\documentclass[aps,prl,twocolumn,reprint,superscriptaddress,amsmath,amssymb,floatfix,longbibliography]{revtex4-2}

\usepackage{graphicx}
\usepackage{bm}
\usepackage{mathtools}
\usepackage{amsthm}
\usepackage{xcolor}
\usepackage{braket}
\usepackage{booktabs}

\newcommand{\nop}{\hat n}
\newcommand{\aop}{\hat a}
\newcommand{\ad}{\hat a^{\dagger}}

\usepackage[colorlinks=true,linkcolor=blue,citecolor=blue,urlcolor=blue]{hyperref}

\begin{document}

\title{Time- and Photon-Optimal Paths for the Preparation and Logical Flip\\of Gottesman--Kitaev--Preskill States}

\author{Aidan Curtis}
\thanks{These authors contributed equally to this work.}
\affiliation{Thayer School of Engineering, Dartmouth College, Hanover, New Hampshire 03755, USA}

\author{Juan S. Salcedo-Gallo}
\thanks{These authors contributed equally to this work.}
\affiliation{Thayer School of Engineering, Dartmouth College, Hanover, New Hampshire 03755, USA}

\author{Sergei Leonov}
\affiliation{Thayer School of Engineering, Dartmouth College, Hanover, New Hampshire 03755, USA}
\affiliation{Department of Computer Science, Amherst College, Amherst, Massachusetts 01002, USA}

\author{Kevin Limanta}
\author{Peter Chin}
\author{Salil Bedkihal}
\author{Mattias Fitzpatrick}
\affiliation{Thayer School of Engineering, Dartmouth College, Hanover, New Hampshire 03755, USA}

\date{\today}

\begin{abstract}
Gottesman–Kitaev–Preskill (GKP) states are highly non-Gaussian, making their control challenging, particularly in the presence of photon loss. Here, we use the brachistochrone, the fastest path at bounded speed, to construct ideal nonlinear Hamiltonians for state preparation and logical gates, and compare it with the photon-optimal path, which minimizes the first-order infidelity from photon loss. We find that the two paths nearly coincide for state preparation but differ markedly for logical operations, namely the brachistochrone preserves the encoded qubit during a bit flip, whereas the photon-optimal path flips a known state through the vacuum. Since these Hamiltonians contain multiphoton terms of arbitrarily high order, we then introduce the tangential capacity to quantify how effectively gate sets that control the oscillator through an ancilla qubit track these paths. Using this analysis, we find that conditional squeezing achieves the highest capacity, preparing GKP states in 824 ns with a fidelity of 0.99 and a minimum closeness to the path of 0.90, compared with 0.98 and 0.49 for conditional displacements and 0.86 and 0.19 for SNAP gates.
\end{abstract}

\maketitle

\noindent\textit{Introduction.}---The brachistochrone problem, posed by Bernoulli in 1696, asks for the curve along which a body falls between two points in the shortest time~\cite{Bernoulli1696}. Bernoulli solved it with Fermat's principle, treating the body as a light ray that follows the path of least time through a medium of varying refractive index~\cite{Bernoulli1697,BornWolf1999},  ultimately leading to the development of the calculus of variations. In quantum mechanics, speed limits bound how fast a state can evolve~\cite{MandelstamTamm1945,Margolus1998,Deffner2017},  but the quantum brachistochrone  has been shown to reach this bound~\cite{Carlini2006,Carlini2007,yang2022minimumtimequantumcontrolquantum}. For an oscillator, photon loss adds a second cost, and the path of least photon exposure obeys Fermat's principle with the mean photon number as the refractive index.  Here, we show that these two paths provide a common framework for preparing and manipulating bosonic states.

Among bosonic codes~\cite{Weedbrook2012,PhysRevLett.82.1784,Terhal2020,Joshi_2021}, which include cat~\cite{PhysRevA.59.2631,Mirrahimi2014} and binomial codes~\cite{Michael2016,CAI202150}, Gottesman--Kitaev--Preskill (GKP) codes protect a qubit against small displacements in both quadratures with a grid of phase-space states~\cite{Gottesman2001,grimsmo2021Quantum}. GKP error correction has been demonstrated in circuit QED~\cite{CampagneIbarcq2020,Blais2021}, beyond break-even~\cite{Sivak2023RealTime}, autonomously~\cite{PhysRevLett.132.150607}, and in a fully planar circuit~\cite{Jha2026}, as well as in the motion of a trapped ion~\cite{Fluhmann2019,deNeeve2022}. Other hardware-efficient platforms, such as the gridium~\cite{nguyen2025superconductinggridstatesqubit}, encode the grid in the circuit Hamiltonian itself, and two Josephson junctions coupled by a gyrator, an element recently demonstrated on chip~\cite{Tu2026}, would encode a GKP qubit in their degenerate ground space~\cite{Rymarz2021}.

The ideal GKP states form a lattice of delta functions in phase space, their Pauli operators are displacements, the Hadamard gate is a phase-space rotation that can be applied in software, and the phase gate is a shear, so Clifford operations on that code space can be performed with Gaussian operations~\cite{Gottesman2001}. A logical rotation by a generic angle, in contrast, superposes the identity and a displacement and is non-Gaussian, and such rotations can be teleported into the code during error correction~\cite{Singh2026}. In practice, the code states are commonly approximated by applying a Gaussian envelope operator to the ideal states. This approximation makes the logical states slightly non-orthogonal and the logical operations non-Gaussian~\cite{grimsmo2021Quantum}, since a displacement also moves the envelope. At $\Delta=0.4$, a displacement flips $\ket{0_L}$ to $\ket{1_L}$ with a fidelity of only 0.78 (Supplementary Information (SI)~\cite{SM}, Sec.~I.4). Non-Gaussian control is thus needed to prepare finite-energy GKP states and to apply logical gates to them, which is often provided by the non-linear Hamiltonian~\cite{PhysRevLett.82.1784} of an ancilla qubit in circuit QED. Gate sets such as echoed conditional displacements~\cite{Eickbusch2022} and selective number-dependent arbitrary phase (SNAP) gates and displacements~\cite{Krastanov2015,Heeres2015} have been used to perform preparation and logical operations of GKP states in superconducting circuits with other protocols for trapped ions and photonic systems~\cite{Fluhmann2019,deNeeve2022,PhysRevLett.133.050602,Shi2019,Hastrup2021,Hastrup2022,Winnel2024,Kolesnikow2024,Gorgichuk2026,Lin2026,Larsen2025}. Conditional displacements and qubit rotations also prepare GKP states deterministically from a squeezed vacuum, with an envelope close to the Gaussian one~\cite{Hastrup2021,Singh2026}, and Rabi interactions can prepare that squeezed vacuum itself~\cite{HastrupPRL2021}. Many of the gate-based schemes for preparing GKP states last microseconds~\cite{Eickbusch2022,Sivak2023RealTime}, which may suffer from ancilla bit-flip errors and oscillator photon loss. Optimal control and reinforcement learning reduce these errors by optimizing the gates at the pulse level and time-optimal control has been recently applied to Fock and cat states in certain qubit--oscillator models~\cite{Garcia2022,Edmunds2026}. To our knowledge, the quantum brachistochrone equation has never been used for controlling and preparing GKP states or providing a reference path for optimization. For GKP states, which contain only even photon numbers, this requires generators that change the photon number in pairs, such as conditional squeezing~\cite{DelGrosso2025}, rather than the single-photon generators of displacements.

In this Letter, we derive the time- and photon-optimal paths for finite-energy GKP state preparation and logical operations and show that conditional squeezing~\cite{DelGrosso2025,Ayyash2024} follows the time-optimal preparation path closely. Using the quantum brachistochrone formalism, we obtain the Hamiltonian generating the time-optimal trajectory explicitly as an infinite series of higher order wave-mixing processes, which no polynomial of finite order in the creation and annihilation operators reproduces. Minimizing the photon exposure instead of the duration gives the photon-optimal path. For state preparation, the two paths nearly coincide. For logical operations, the photon-optimal path leaves the GKP code space by detouring through the vacuum. Since real gate sets cannot implement these Hamiltonians, we introduce the tangential capacity, the largest speed at which a gate set controlled by an ancilla qubit moves a state directly along a path. For conditional squeezing (CS), conditional displacement (CD), and SNAP--displacement control (SNAP) at equal drive strength,  we prove that only CS has a nonzero tangential capacity along the GKP preparation brachistochrone, so that only CS can follow it directly. Finally, we test this prediction with pulses optimized by gradient-ascent pulse engineering (GRAPE) and evaluated with  the decoherence rates and the cavity self-Kerr measured in the GKP error-correction experiment of Ref.~\cite{Sivak2023RealTime}, and extend the comparison in the SI to conditional displacements acting on a squeezed vacuum. In 824~ns, CS prepares $\ket{0_L}$ with a fidelity of 0.99 and a minimum closeness to the brachistochrone of 0.90, against 0.98 and 0.49 for CD.

\begin{figure}[t]
\includegraphics[width=\columnwidth,height=1.050130\columnwidth]{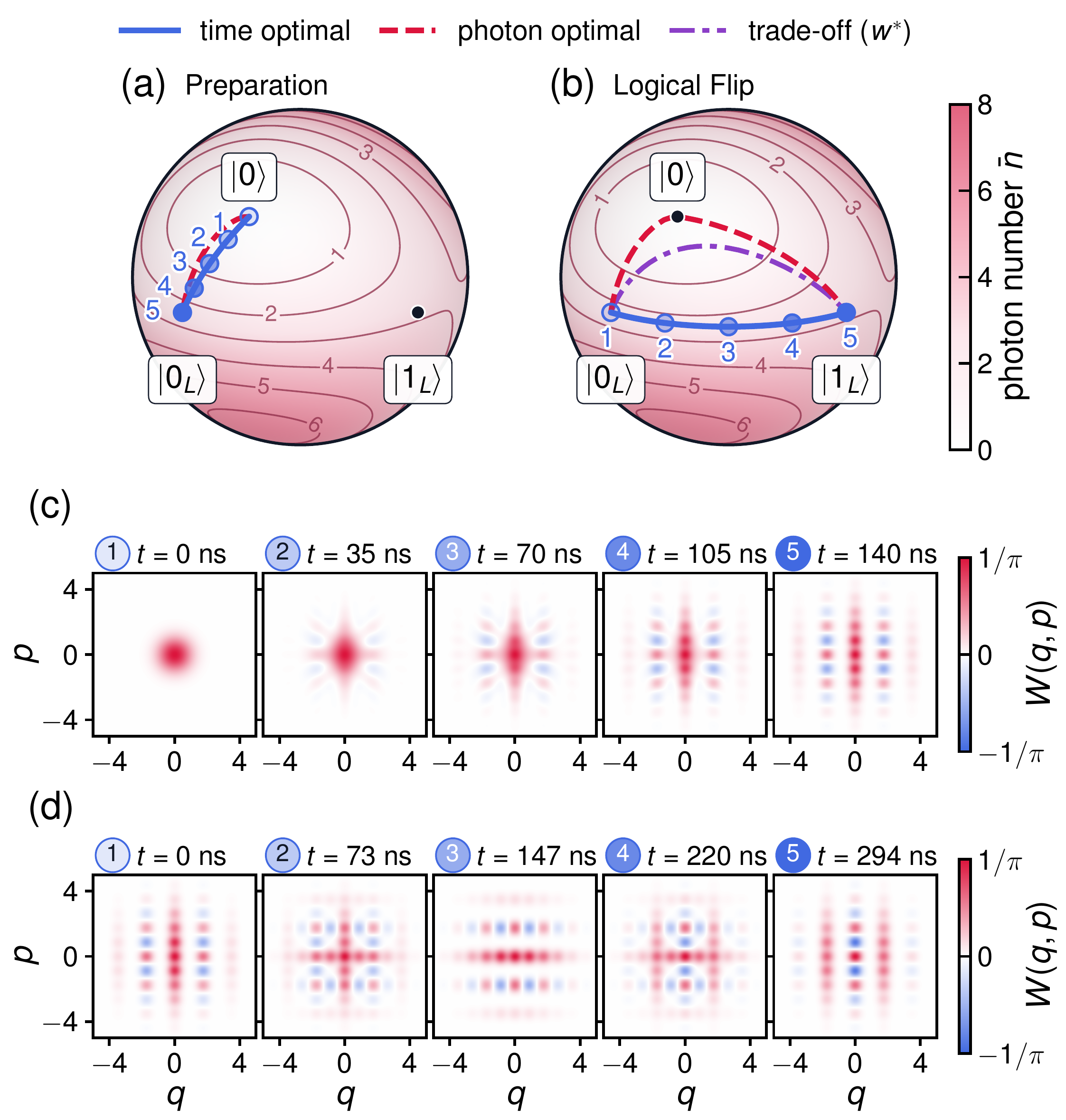}
\caption{Optimal paths for GKP preparation (a,~c) and between the logical states (b,~d) at $\Delta=0.4$. (a,~b)~Brachistochrone (solid), photon-optimal path (dashed), and, in (b), the trade-off path $w^*$ (dash-dotted), which minimizes the first-order error of Eq.~\eqref{eq:eps} without cavity dephasing at the noise rates of Ref.~\cite{Sivak2023RealTime}. Both panels show all paths projected onto the same sphere of real states, spanned by $\ket0$, $\ket{0_L}$, and $\ket{1_L}$ and colored by the mean photon number $\bar n$.  The numbered points mark the times of the Wigner functions in (c,~d), at $\omega=5.3~\mu\mathrm{s}^{-1}$ ($\omega/2\pi=0.84$~MHz). (c,~d)~Wigner functions $W(q,p)$ over the phase space of the quadratures $\hat q=(\aop+\ad)/\sqrt2$ and $\hat p=i(\ad-\aop)/\sqrt2$.}
\label{fig:paths}
\end{figure}

\textit{GKP states.}---Ideal GKP states are $+1$ eigenstates of stabilizer displacement operators $\hat{S}_X = \exp(-i2\sqrt{\pi}\hat{P})$ and $\hat{S}_Z = \exp(i2\sqrt{\pi}\hat{Q})$, where $\hat{Q} = i(\beta^* \hat{a} - \beta \hat{a}^\dagger)/\sqrt{\pi}$ and $\hat{P} = -i(\alpha^* \hat{a} - \alpha \hat{a}^\dagger)/\sqrt{\pi}$ are generalized quadratures parameterized by $\alpha, \beta \in \mathbb{C}$~\cite{Gottesman2001, grimsmo2021Quantum} and $\hat{a}^{\dagger}$ ($\hat{a}$) is the creation (annihilation) operator. It is required that $\beta \alpha^* - \beta^* \alpha = i\pi$ for $\hat{Q}$ and $\hat{P}$ to satisfy the canonical commutation relation $[\hat{Q}, \hat{P}] = i$. The logical $\hat{Z}$ states are the $\pm 1$ eigenstates of the  $\hat{Z}=\hat{D}(\beta)$ operator and can be written as
\begin{equation}
    \ket{\mu_L} \propto \sum_{k,l \in \mathbb{Z}} \hat{X}^{\mu}\hat{S}_X^k \hat{Z}^l \ket{0}
    \label{eq:ideal_gkp}
\end{equation}
where $\mu \in \{0,1\}$ and $\hat{X} = \hat{D}(\alpha)$. The states in Eq.~\eqref{eq:ideal_gkp} are infinite superpositions of coherent states and are non-normalizable. In this work, we apply the Gaussian envelope operator $\hat{E}_{\Delta} = \exp(-\Delta^2\hat{a}^{\dagger}\hat{a})$ to yield a finite-energy approximation~\cite{grimsmo2021Quantum, Royer2020}
\begin{equation}
    \ket{\mu_L^\Delta} = \mathcal{N}_{\mu, \Delta} \hat{E}_{\Delta}\ket{\mu_L}
    \label{eq:finite-energy-gkp}
\end{equation}
where $\mathcal{N}_{\mu, \Delta}$ is the normalization constant. Thus, in this formalism, a finite-energy GKP code can be parameterized by three parameters, $\alpha, \beta$, and $\Delta$. The envelope commutes with the quarter rotation $e^{-i\pi\nop/2}$, the logical Hadamard of the ideal square code, whose eigenstates in the code space are supported on the photon numbers $n\equiv0$ and $n\equiv2\ (\mathrm{mod}\ 4)$. Applied to these two eigenstates, the envelope gives them different norms, and this imbalance accounts exactly for the small overlap $\langle0_L^\Delta|1_L^\Delta\rangle\approx2e^{-\pi/(4\Delta^2)}$ of the code words. Rescaling the two sectors to equal norms makes the code words exactly orthogonal and the quarter rotation an exact logical Hadamard on them, and changes each code word by an infidelity of only $5\times10^{-5}$ at $\Delta=0.4$ (SI~Sec.~I.2). We keep the code words of Eq.~\eqref{eq:finite-energy-gkp} and write them $\ket{\mu_L}$ below.

\textit{Time-optimal paths.}---The distance between two states $\ket{\psi_i}$ and $\ket{\psi_f}$ in Hilbert space is the angle $\theta$, with $\cos\theta=\langle\psi_i|\psi_f\rangle\geq0$, and a state moves at a speed set by the energy uncertainty of the system Hamiltonian, $\Delta H$~\cite{Anandan1990}. At speeds up to $\omega$, reaching $\ket{\psi_f}$ takes at least $T^*=\theta/\omega$~\cite{MandelstamTamm1945,Deffner2017}, a limit reached only by the brachistochrone, the great circle between the two states traversed at full speed. The Hamiltonian that generates the brachistochrone evolution is known for any two states~\cite{Carlini2006, yang2022minimumtimequantumcontrolquantum} (SI~\cite{SM}, Sec.~I),
\begin{equation}
H_{\rm opt}=\frac{i\omega}{\sin\theta}\big(\ket{\psi_f}\bra{\psi_i}-\ket{\psi_i}\bra{\psi_f}\big),
\label{eq:Hopt-general}
\end{equation}
which rotates $\ket{\psi_i}$ into $\ket{\psi_f}$ through the path states $\ket{p(t)}=[\sin(\theta-\omega t)\ket{\psi_i}+\sin(\omega t)\ket{\psi_f}]/\sin\theta$, a coherent Rabi oscillation between $\ket{\psi_i}$ and the part of $\ket{\psi_f}$ orthogonal to it. For an oscillator, $H_{\rm opt}$ is an infinite series of multiphoton terms.

We now consider the time-optimal paths for preparation and logical flip of the square finite-energy GKP code ($\alpha = \sqrt{\pi/2}$, $\beta = i\sqrt{\pi/2}$), with $\Delta = 0.4$. Unless stated otherwise, all results of this Letter are for this value, at which $\ket{0_L}$ has the mean photon number $\bar n=2.44$, and we discuss other values at the end. Time-optimal preparation of $\ket{0_L}$ from the vacuum state takes $T^*=0.74/\omega$, whereas the flip between the nearly orthogonal logical states takes $T^*=1.56/\omega$. To make Eq.~\eqref{eq:Hopt-general} explicit, we expand the state in Eq.~\eqref{eq:finite-energy-gkp} as a series of coherent states, $\ket{\mu_L^\Delta}=\sum_{m\equiv\mu \text{(mod 2)},\,l}c^\mu_{ml}\,\exp({\zeta^\Delta_{ml}\ad)}\ket0$, with $\zeta_{ml}=m\alpha+l\beta$, $\zeta^\Delta_{ml}=\exp({-\Delta^2})\zeta_{ml}$, $c^\mu_{ml}=\mathcal N_{\mu, \Delta} \exp({-i\pi ml/2-|\zeta_{ml}|^2/2})$. Inserting these states into Eq.~\eqref{eq:Hopt-general} and writing the vacuum projector in normal order gives the brachistochrone Hamiltonians of preparation and of the logical flip in explicit normal-ordered form (SI~Secs.~I.3 and I.4),

\begin{subequations}\label{eq:hseries}
\begin{align}
H_{0\to\mu_L}&=\frac{i\omega}{\sin\theta_\mu}\sum_{m\equiv\mu,\,l}c^\mu_{ml}\sum_{n\geq1,\,r\geq0}\frac{(-1)^r}{n!\,r!}\nonumber\\
&\qquad\times(\zeta^\Delta_{ml})^n\,\hat a^{\dagger\,n+r}\aop^{r}+\mathrm{h.c.},\label{eq:Hprep}\\
H_{0_L\to1_L}&=\frac{i\omega}{\sin\theta_L}\sum_{\substack{m'\equiv1,\,l'\\ m\equiv0,\,l}}c^1_{m'l'}c^{0*}_{ml}\sum_{n,n',r\geq0}\frac{(-1)^r}{n!\,n'!\,r!}\nonumber\\
&\qquad\times(\zeta^\Delta_{m'l'})^n(\zeta^{\Delta*}_{ml})^{n'}\,\hat a^{\dagger\,n+r}\aop^{n'+r}+\mathrm{h.c.},\label{eq:Hflip}
\end{align}
\end{subequations}
with $\cos\theta_\mu=\langle0|\mu_L^{\Delta}\rangle$ and $\cos\theta_L=\langle0_L^{\Delta}|1_L^{\Delta}\rangle$. Their lowest terms are squeezers, and the higher terms of Eq.~\eqref{eq:Hprep} are multiphoton drives confined to the vacuum. This confinement comes from the factors $\hat a^{\dagger r}\aop^r=\nop(\nop-1)\cdots(\nop-r+1)$ of the sum over $r$, which for $r=2$ is the self-Kerr operator, so the time-optimal Hamiltonians already contain the self-Kerr operator and its higher-order analogs as photon-number-dependent factors of their drives. Neither expansion has a term that conserves the photon number, however, so neither contains a static Kerr term (SI~Secs.~I.3 and I.4). Because the vacuum and both logical states contain only even photon numbers, both Hamiltonians conserve photon-number parity.  Each Hamiltonian maps every state into the plane of its initial and final states, whereas every nonzero polynomial in $\aop$ and $\ad$ maps distant Fock states onto orthogonal nonzero states, so no truncation of Eq.~\eqref{eq:hseries} at a finite order reproduces it (SI~Secs.~I.1 and I.5). The logical flip Hamiltonian rotates the $\ket{0_L}$ state around the Y axis to $\ket{1_L}$ and couples only Fock states whose photon numbers differ by 2, 6, 10, $\ldots$. Because the finite-energy code words overlap, with $\cos\theta_L=0.014$, the flip brachistochrone rotates the logical Bloch sphere about the Y axis by $2\theta_L=0.991\pi$ rather than $\pi$ and carries $\ket{1_L}$ to $-\ket{0_L}$ with fidelity $(1-2\cos^2\theta_L)^2=0.9992$. This small defect comes entirely from the overlap, which falls from 0.081 at $\Delta=0.5$ to $3\times10^{-4}$ at $\Delta=0.3$, and between exactly orthogonal code words the flip brachistochrone implements the logical Pauli $Y$ exactly, up to a global phase (SI~Secs.~I.2 and I.4). In circuit QED, such multiphoton generators are commonly engineered with an ancilla qubit. Conditional squeezing interleaved with qubit rotations generates qubit-dressed multiphoton terms of the same even orders through nested commutators (SI~Sec.~V), and driven multiphoton qubit--oscillator interactions can supply such terms directly~\cite{Ayyash2024,Gorgichuk2026}.

Figures~\ref{fig:paths}(a) and \ref{fig:paths}(b) show all paths on the same sphere of real states, spanned by $\ket0$, $\ket{0_L}$, and $\ket{1_L}$. Both brachistochrones are great-circle arcs (solid lines). During preparation, the grid grows out of the vacuum as the target fidelity $\mathcal F_T$ rises from 0.54 to 1 [Fig.~\ref{fig:paths}(c)]. During the flip, the state crosses the equator of the logical Bloch sphere at $t=T^*/2$, where it equals  $\ket{+_L}\propto\ket{0_L}+\ket{1_L}$ [Fig.~\ref{fig:paths}(d)].

\textit{Photon-optimal paths.}---The brachistochrone minimizes the duration, but photon loss also depends on the path, because the probability of losing a photon grows with the photon number. To first order in the photon loss rate $\kappa$, a path of even parity loses the fidelity $\kappa\int\bar n\,dt=\kappa J/\omega$ at $\Delta H=\omega$, with $\bar n=\langle\nop\rangle$ and the photon exposure (SI~Sec.~II.1)
\begin{equation}
J=\int\bar n\,d\vartheta ,\qquad d\vartheta=\Delta H\,dt ,
\label{eq:J}
\end{equation}
where $\vartheta$ is the distance traveled in state space. Photon-optimal paths minimize $J$, so they obey Fermat's principle with $\bar n$ as the refractive index~\cite{Brekhovskikh1960} and detour through states with fewer photons [dashed lines in Figs.~\ref{fig:paths}(a) and \ref{fig:paths}(b)] (SI~Sec.~II.2). Because every segment of a shortest path is itself a shortest path~\cite{Bellman1957}, a photon-optimal path through the vacuum splits into two preparations of least exposure (SI~Sec.~II.4).

In state preparation, the photon-optimal path stays close to the brachistochrone $\cos\varphi\ket0+\sin\varphi\ket\chi$, with $\varphi\in[0,\theta]$ the angle along it and $\ket\chi\propto\ket{\psi_f}-\cos\theta\ket0$ [Fig.~\ref{fig:paths}(a)]. To second order in the departure $\delta(\varphi)$, minimizing $J$ is a convex quadratic problem (SI~Sec.~II.5) with the unique closed-form solution
\begin{equation}
\begin{gathered}
\ket{\psi(\varphi)}=\cos|\delta|\big(\cos\varphi\ket0+\sin\varphi\ket\chi\big)+\sin|\delta|\,\delta/|\delta|,\\
\delta=\textstyle\sum_kA_k\Big[\sin\varphi-\dfrac{\sin^2\theta\,\Phi_k(\varphi)}{\sin\varphi\,\Phi_k(\theta)}\Big]\ket{e_k},\\
\Phi_k(\varphi)=\sin^{\nu_k}\!\varphi\;{}_2F_1\big(\tfrac{\nu_k+2}{2},\tfrac{\nu_k-2}{2};\nu_k+\tfrac12;\sin^2\varphi\big),
\end{gathered}
\label{eq:closed}
\end{equation}
where $\ket{e_k}$ and $n_k$ are the eigenstates and eigenvalues of $\nop$ restricted to the states orthogonal to $\ket0$ and $\ket\chi$, the directions in which the path can depart, $A_k=2\langle e_k|\nop|\chi\rangle/[n_\chi(2-\eta_k)]$ and $\nu_k=\tfrac12+(\tfrac14+\eta_k)^{1/2}$ with $\eta_k=2n_k/n_\chi$ and $n_\chi=\langle\chi|\nop|\chi\rangle$, and ${}_2F_1$ is the Gauss hypergeometric function. For a transfer through the vacuum, Eq.~\eqref{eq:closed} gives both preparations. The exposure of every path is also bounded from below. Since every state within a distance $s$ of the target has fidelity at least $\cos^2s$, and hence at least $N(\cos^2s)$ photons, with $N(F)=\min\{\langle\psi|\nop|\psi\rangle:|\langle\psi_f|\psi\rangle|^2\geq F\}$ (SI~Sec.~II.3), every path obeys
\begin{equation}
J\geq B=\int_0^{\theta}N(\cos^2s)\,ds,\qquad\cos\theta=|\langle\psi_i|\psi_f\rangle| .
\label{eq:bound}
\end{equation}

\begin{figure}[t]
\includegraphics[width=\columnwidth]{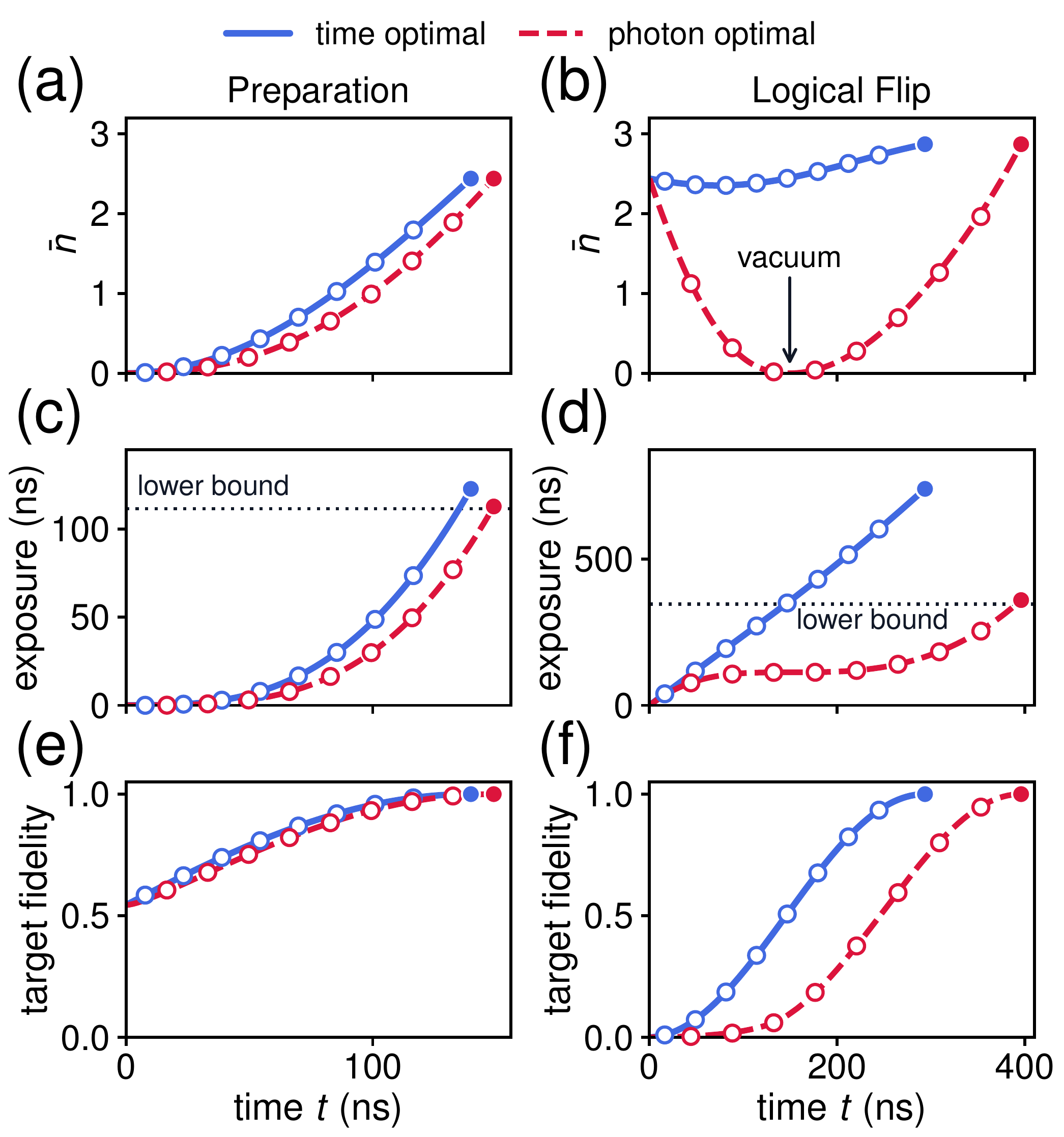}
\caption{Photon number (a,~b), accumulated photon exposure (c,~d), and target fidelity $\mathcal F_T$ (e,~f) along the brachistochrone (blue) and the photon-optimal path (red), for preparation (left) and between the logical states $\ket{0_L}\to\ket{1_L}$ (right), at $\omega=5.3~\mu\mathrm{s}^{-1}$ ($\omega/2\pi=0.84$~MHz). Lines show the analytic paths,  open circles show the numerical optima, and dotted lines the lower bound of photon exposure.}
\label{fig:photon}
\end{figure}

Figure~\ref{fig:photon} compares the two paths. The analytic paths match the numerical optima within 1\% (SI~Secs.~II.2 and II.5), and the photon optima lie within a few percent of their lower bounds [Figs.~\ref{fig:photon}(c) and \ref{fig:photon}(d)], $B$ for preparation and its two-sided form between the logical states (SI~Sec.~II.3). In state preparation, the photon optimum holds slightly fewer photons [Fig.~\ref{fig:photon}(a)] and lowers the exposure to photon loss by 8\% [Fig.~\ref{fig:photon}(c)] at a 7\% longer duration [Fig.~\ref{fig:photon}(e)]. The saving vanishes for Fock and binomial-code targets and stays below 26\% for the squeezed, cat, and GKP targets we study (SI~Secs.~II.3 and II.6). Between the logical states, by contrast, the photon-optimal path returns $\ket{0_L}$ to the vacuum before preparing $\ket{1_L}$ [Fig.~\ref{fig:photon}(b)], so the target fidelity rises mainly in the second stage [Fig.~\ref{fig:photon}(f)]. This detour takes 397~ns instead of 294~ns but halves the exposure [Fig.~\ref{fig:photon}(d)]. Fewer photons also reduce the error of a cavity self-Kerr $(K/2)\hat a^{\dagger2}\aop^2$, which shifts the energy of each Fock state $\ket n$ by $Kn(n-1)/2$ and vanishes on the vacuum and on a single photon. When the generator of a path does not compensate it, its infidelity grows as $K^2$, and the photon-optimal path is 2.2 times less sensitive to it than the brachistochrone for preparation and 10.5 times less sensitive between the logical states, although it lasts 7\% and 35\% longer. At the self-Kerr of the device of Ref.~\cite{Sivak2023RealTime}, $K/2\pi=-4.8$~Hz, this error stays below $10^{-7}$ on every path (SI~Sec.~XII).

A longer path, however, accumulates more errors of the ancilla that controls the oscillator. With cavity dephasing at rate $\kappa_\phi$, whose error grows with the photon-number variance, a path of duration $T$ has the first-order error (SI~Sec.~III)
\begin{equation}
\epsilon_1=\Gamma_qT+\kappa\!\int_0^T\!\bar n\,dt+\kappa_\phi\!\int_0^T\!\mathrm{Var}(\nop)\,dt ,
\label{eq:eps}
\end{equation}
with $\Gamma_q$ the ancilla error rate. Without cavity dephasing, $\epsilon_1=(\Gamma_q/\omega)\int(1+w^*\bar n)\,d\vartheta$ weighs the duration against the photon exposure with the weight $w^*=\kappa/\Gamma_q=0.79$ at the noise rates of Ref.~\cite{Sivak2023RealTime}. Its minimizer, which we call the trade-off path, therefore lies between the two optimal paths and lasts 350~ns between the logical states [dash-dotted line in Fig.~\ref{fig:paths}(b)].

\begin{figure}[t]
\centering
\includegraphics[width=\columnwidth]{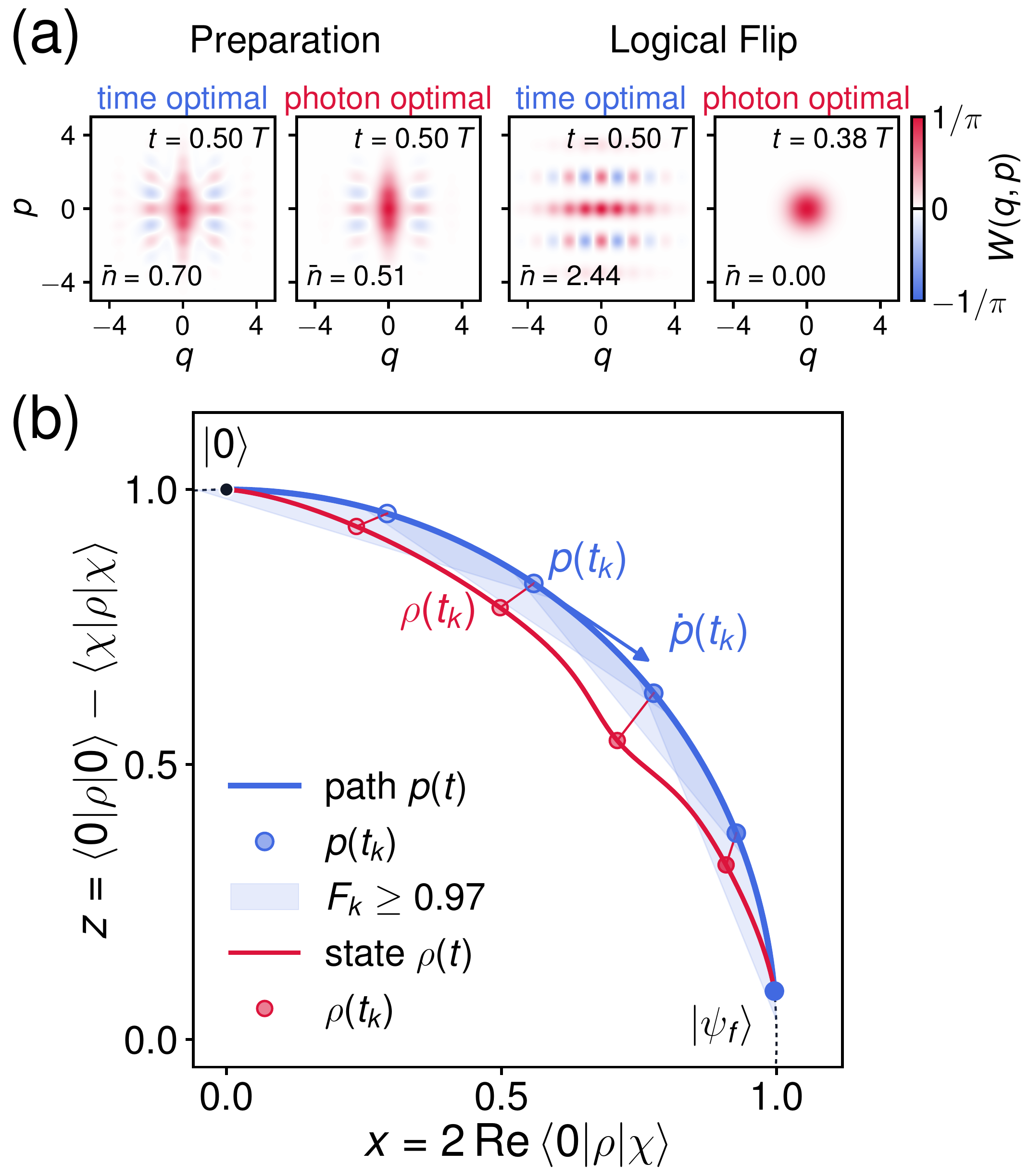}
\caption{(a)~Wigner functions of the path states halfway through preparation (left) and along a path between the logical states $\ket{0_L}\to\ket{1_L}$ (right), labeled by the photon number $\bar n$. (b)~Bloch plane of the brachistochrone path with axes $x=2\,\mathrm{Re}\langle0|\rho|\chi\rangle$ and $z=\langle0|\rho|0\rangle-\langle\chi|\rho|\chi\rangle$, in which pure real states of the plane lie on the unit circle. The reduced oscillator state $\rho(t_k)$ of the pulse (red) is penalized outside the shaded region $F_k=F(t_k)\geq0.97$ around the path state $p(t_k)$ (blue), whose unit tangent is $\dot p(t_k)$.}
\label{fig:policy}
\end{figure}

\textit{Logical operations and state-to-state transfers.}---The detour through the vacuum suits a state-to-state transfer, which moves one known input, and lowers its first-order error (SI~Sec.~III). A logical flip, however, must act correctly on every encoded state, and only one logical state can pass through the vacuum, since unitary evolution keeps them nearly orthogonal. Near the middle of the paths [Fig.~\ref{fig:policy}(a)], the two preparations pass through similar states, whereas between the logical states the brachistochrone passes through $\ket{+_L}$ and the photon-optimal path through the vacuum, which has no grid and no stabilizers~\cite{Pantaleoni2020,Shaw2024}. The photon-optimal path reaches the vacuum at $t=0.38\,T$ rather than at $T/2$, because the vacuum lies closer to $\ket{0_L}$ than to $\ket{1_L}$. The generators of the photon-optimal path therefore carry $\ket{1_L}$ to $\ket{0_L}$ with a fidelity of only 0.60 (SI~Sec.~IV.4). The brachistochrone instead rotates every encoded state, at about twice the exposure to photon loss [Fig.~\ref{fig:photon}(d)]. Among flips that keep the qubit encoded, the exposure averaged over the code space depends only on the duration, so the brachistochrone also minimizes it (SI~Secs.~IV.2 and XI).

\textit{Tangential capacity.}---Real gate sets cannot implement $H_{\rm opt}$, but they can follow its path. Here, we measure how closely a pulse does so by the closeness $F(t)=\langle p(t)|\rho(t)|p(t)\rangle$ of the reduced oscillator state $\rho(t)$ to the path state $\ket{p(t)}$, the state on the reference path at time $t$ [Fig.~\ref{fig:policy}(b)]. A pulse that does nothing keeps the closeness above 0.54 during GKP preparation. Whether a gate set can follow a path directly is set by the tangential capacity of its oscillator generators $ u\cdot K$, with Hermitian $K$ and amplitudes $|u|\leq g$, the largest speed they give an unentangled state along the unit tangent $\ket{\dot p}=d\ket p/d\vartheta$ of the path,
\begin{equation}
v=\max_{|u|\leq g}\mathrm{Re}\,\langle\dot p|{-i}\,u\cdot K|p\rangle .
\label{eq:vtan}
\end{equation}
On a path faster than $v$, an unentangled state loses closeness at second order in time (SI~Sec.~VI.2). Generators that change the photon number by an odd amount, such as displacements, have $v=0$ on every even-parity path (SI~Sec.~VI.3). Generators at most quadratic in $\aop$ and $\ad$, even when conditioned on a qubit, must entangle it with the oscillator to create a non-Gaussian state (SI~Sec.~VI.1), so $v$ bounds the direct motion rather than the pulse duration.

\begin{figure*}[t]
\includegraphics[width=\textwidth,height=0.428255\textwidth]{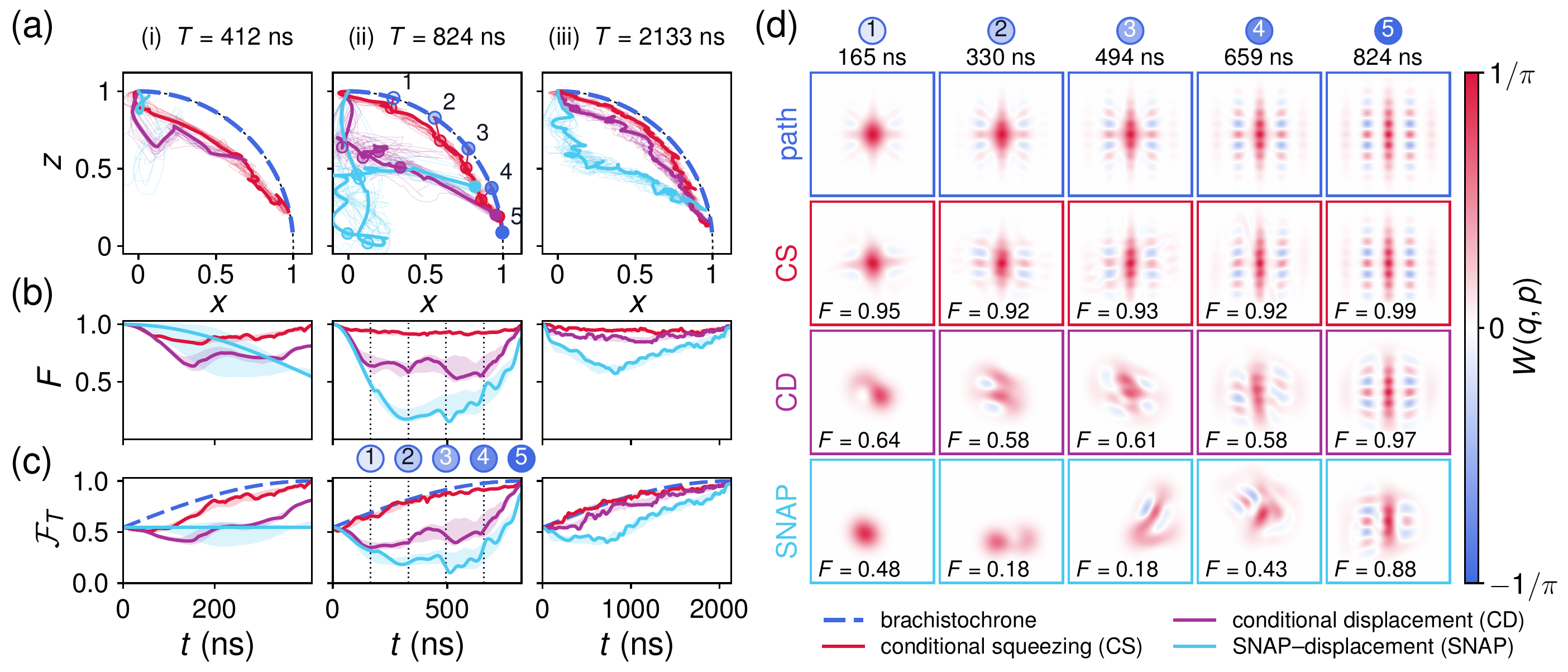}
\caption{Following the GKP brachistochrone (dashed) with CS, CD, and SNAP at $T=412$, 824, and 2133~ns [(i)--(iii)], 20 starts each, at the largest tracking weight with $\mathcal F_T\geq0.99$ under noise (else the best $\mathcal F_T$). Noisy evolutions solve the Lindblad master equation~\cite{Lindblad1976} with photon loss, cavity dephasing, and ancilla decay and dephasing at the rates of Ref.~\cite{Sivak2023RealTime}, $T_{1c}=606~\mu$s and $T_{\phi c}=5.1$~ms for the cavity and $T_{1q}=280~\mu$s and $T_{\phi q}=414~\mu$s for the ancilla, at Fock cutoff 90. Every optimization and evaluation includes the cavity self-Kerr $K/2\pi=-4.8$~Hz of the same device (SI~Secs.~III and VII). Thin lines show every start, thick lines the start closest to the mean (the mean for SNAP in (a)(i)), and bands one standard deviation. (a)~State in the plane of Fig.~\ref{fig:policy}(b). (b)~Closeness $F$ to the path state. (c)~Target fidelity $\mathcal F_T$. (d)~Wigner functions and closeness at the numbered points of (a).}
\label{fig:benchmark}
\end{figure*}

\textit{Qubit-controlled gate sets.}---We compare three gate sets. Each combines a qubit drive $\tfrac12(\Omega_x\sigma_x+\Omega_y\sigma_y)$, $|\Omega|=|\Omega_x+i\Omega_y|\leq10g$, with one bounded oscillator generator,
\begin{align}
H_{\rm CS}&=\textstyle\sum_j|j\rangle\langle j|\otimes\tfrac{i}{2}\big(\xi_j^*\aop^2-\xi_j\hat a^{\dagger 2}\big),\nonumber\\
H_{\rm CD}&=\textstyle\sum_j|j\rangle\langle j|\otimes\tfrac{i}{\sqrt2}\big(\varepsilon_j^*\aop-\varepsilon_j\ad\big),\label{eq:gates}\\
H_{\rm SNAP}&=-\chi_d|\mathrm e\rangle\langle\mathrm e|\otimes\nop+\tfrac{i}{\sqrt2}\big(\varepsilon^*\aop-\varepsilon\ad\big),\nonumber
\end{align}
with $|\xi_j|,|\varepsilon_j|,|\varepsilon|\leq g$, where $j\in\{\mathrm g,\mathrm e\}$ labels the ground and excited qubit states. CS squeezes and CD displaces the oscillator by an amount that depends on the qubit state, and SNAP combines a dispersive shift $\chi_d=g$ with a displacement drive. CD underlies echoed conditional-displacement gates~\cite{Eickbusch2022}. CS can be realized with two drive tones on the circuit of Ref.~\cite{DelGrosso2025} or by driving a qubit near the two-photon cross resonance with a resonator~\cite{Ayyash2024}. At their bounds, all three move the vacuum at the same speed, $\omega=g/\sqrt2=5.3~\mu\mathrm{s}^{-1}$ for the proposed $g=7.5~\mu\mathrm{s}^{-1}$~\cite{DelGrosso2025}, although they can move other states faster. In every optimization and evaluation of the benchmark, the oscillator also carries the static self-Kerr $(K/2)\hat a^{\dagger2}\aop^2$ that its coupling to the ancilla induces in the device of Ref.~\cite{Sivak2023RealTime}, $K/2\pi=-4.8$~Hz. Stronger couplings induce larger Kerr terms, which the optimization absorbs when they are modeled (SI~Secs.~VIII and XII). This common bound favors CD, since experiments displace the oscillator at up to $3.8~\mu\mathrm{s}^{-1}$~\cite{Eickbusch2022,Sivak2023RealTime}. Along a brachistochrone from the vacuum, CS has $v=g|\langle2|\chi\rangle|/\sqrt2=0.51\,\omega$ for GKP preparation, above the $0.34\,\omega$ that the path requires at 412~ns, whereas CD and SNAP have $v=0$ (SI~Sec.~VI.3). With qubit rotations, CS also generates even-order terms like those of Eq.~\eqref{eq:hseries} (SI~Sec.~V). The capacity therefore predicts that only CS can follow the GKP brachistochrone directly.

\begin{table}[t]
\caption{Errors and closeness of the GKP preparation runs of Fig.~\ref{fig:benchmark} at the durations (i)--(iii). $\epsilon_1$ is the first-order error of the path itself [Eq.~\eqref{eq:eps}], and $1-\mathcal F_T$ and $C$ are the final infidelity and the minimum closeness of each gate set, as mean (s.d.) over 20 starts.}
\label{tab:tracking}
\setlength{\tabcolsep}{3pt}
\begin{ruledtabular}
\begin{tabular}{lcccccc}
 & \multicolumn{3}{c}{$1-\mathcal F_T$ (\%)} & \multicolumn{3}{c}{$C$}\\\cline{2-4}\cline{5-7}
$T$ (ns) & 412 & 824 & 2133 & 412 & 824 & 2133\\\hline
$\epsilon_1$ (\%) & 0.23 & 0.45 & 1.2 & & & \\
CS & 0.97(2) & 0.99(4) & 1.4(1) & 0.82(2) & 0.90(1) & 0.91(1)\\
CD & 19(1) & 2.3(5) & 1.6(1) & 0.62(4) & 0.49(5) & 0.84(1)\\
SNAP & 43(4) & 14(2) & 2.1(1) & 0.51(7) & 0.19(6) & 0.56(2)\\
\end{tabular}
\end{ruledtabular}
\end{table}

\textit{GKP state preparation.}---To test the prediction that only CS can follow the GKP brachistochrone directly, we optimize GRAPE pulses~\cite{Khaneja2005} on 50 time slices (100 at 2133~ns). Their objective penalizes a closeness below a threshold at each slice [shaded region in Fig.~\ref{fig:policy}(b)], with a tracking weight $W$ that balances following the path against reaching the target, and a final fidelity below a stricter threshold (SI~Sec.~VII). Figure~\ref{fig:benchmark} and Table~\ref{tab:tracking} show the results for 20 random starts at three durations [(i)--(iii)]. As predicted, CS follows the brachistochrone closely, with a minimum closeness $C=\min_tF(t)$ of 0.82 to 0.91 on average [Figs.~\ref{fig:benchmark}(a) and \ref{fig:benchmark}(b)], and its Wigner functions follow those of the path state as the grid grows out of the vacuum [Fig.~\ref{fig:benchmark}(d)]. CD leaves the path, with on average at least half of the lost closeness in odd parity (SI~Sec.~VI.4), and its grid appears only near the end [Figs.~\ref{fig:benchmark}(c) and \ref{fig:benchmark}(d)]. At 824~ns, it still reaches $\mathcal F_T=0.98$, but along a route far from the path, with $C={}$0.49, below the 0.54 of a pulse that does nothing. SNAP follows least, and at 412~ns most of its starts never leave the zero-drive point, where parity makes the objective stationary (SI~Secs.~VII and VIII). At 2133~ns, all three come within a factor of two of the first-order error $\epsilon_1$ of the path, and CD follows the path more closely than at shorter durations ($C=0.84$), so the advantage of CS lies at short durations. These results include the self-Kerr of the device, which changes the fidelity of every stored pulse by less than $10^{-4}$. A Kerr term that the model omits costs the 412-ns CS pulses 0.012 at $K/2\pi=-4.8$~kHz, a thousand times the device value, and the 2133-ns pulses of every gate set 0.15--0.17. A Kerr term in the model is absorbed by the optimization, and CS keeps $\mathcal F_T=0.990$ at 412 and 824~ns for each tested $K$ up to $K/2\pi=-0.43$~MHz, the Kerr term of the proposed circuit, although its minimum closeness to the path then falls to 0.65 at 412~ns and 0.64 at 824~ns (SI~Sec.~XII). Conditional displacements that start from a squeezed vacuum with the peak width of the target, the starting point of deterministic protocols~\cite{Hastrup2021,Singh2026}, do not follow the brachistochrone from that state either, with a minimum closeness of 0.23 at 824~ns (SI~Sec.~XIII).

 Strikingly, this ranking, with CS closest to the path, followed by CD and then SNAP, holds at every tracking weight (Supplementary Fig.~9) and in the stabilizers of the final states (Supplementary Table~7). CS also follows the paths to Fock $\ket2$, an even cat state of amplitude 2, and the binomial-code state most closely and prepares them with a mean $\mathcal F_T\geq{}$0.98 at every duration, whereas CD needs 824~ns and SNAP reaches 0.98--0.99 only at 2133~ns (Supplementary Table~10 and Supplementary Figs.~10--12). The ranking also holds at half the squeezing rate and qubit-drive bound, where CS uses less energy uncertainty than CD and still reaches $\mathcal F_T=0.988$ in 824~ns, against 0.977 for CD at the full rate, so the advantage does not come from a larger energy uncertainty alone (SI~Sec.~VIII). Parity also offers an error check. Since the CS drive, cavity dephasing, and ancilla errors conserve parity, an ideal parity measurement after the pulse~\cite{Sun2014} can detect every odd number of photon losses, as in erasure conversion~\cite{Wu2022,Kubica2023,Teoh2023} (SI~Sec.~VIII). As with dual-rail qubits~\cite{Teoh2023,Koottandavida2024}, heralding photon loss in this way holds promise for realizing quantum error-correction schemes with enhanced thresholds and improved scaling.

\textit{Discussion and outlook.}---Closed-form time-optimal and, to second order, photon-optimal paths describe the preparation of bosonic states, state-to-state transfers, and logical operations within one framework. A known input can save photons by detouring through the vacuum. A logical gate, in contrast, must act correctly on every encoded state, and among the flips that keep the qubit in the code space throughout, the brachistochrone is both the fastest and, averaged over the code space, the least exposed to photon loss. The tangential capacity identifies which hardware can follow these paths directly. At equal drive strength, experimentally motivated noise rates, and the measured self-Kerr, CS follows the GKP brachistochrone most closely and prepares $\ket{0_L}$ with the highest fidelity in 412~ns, and CS pulses optimized for the gate implement the logical flip with a gate fidelity of 0.987 at 824~ns, whereas a numerical search finds no Gaussian unitary above 0.78 (SI~Sec.~IX and Supplementary Table~11). No gate set, however, follows the flip brachistochrone closely for all encoded inputs at once. The smallest closeness over three tested inputs stays below 0.40 in every start and at every tracking weight (Supplementary Table~11), so the flip brachistochrone serves as a reference for the duration and photon cost of the gate rather than as a path to track. The CS results assume independent squeezing for both qubit states, which requires two drive tones on the proposed circuit~\cite{DelGrosso2025}, and the coherence of the device of Ref.~\cite{Sivak2023RealTime}, and they motivate an experimental test of that circuit. Larger GKP states, with smaller $\Delta$ and narrower peaks, lengthen the preparation brachistochrone and lower the tangential capacity of CS at the vacuum, from $0.64\,\omega$ at $\Delta=0.5$ to $0.38\,\omega$ at $\Delta=0.3$, so following the path directly needs longer pulses there (SI~Secs.~II.7 and XII). In the benchmark at $\Delta=0.3$, the 412-ns CS pulses chosen as in Fig.~\ref{fig:benchmark} indeed leave the path ($C=0.45$) but still reach $\mathcal F_T=0.985$, against 0.58 for CD and 0.35 for SNAP. Because a cavity self-Kerr shifts the energy of each Fock state quadratically in its photon number, photon-optimal paths are also less sensitive to a Kerr term that the control omits or miscalibrates, which makes them attractive for oscillators with stronger nonlinearities. Beyond GKP codes, the paths, the bound, and the capacity provide a common reference for state preparation and control, from model-free reinforcement learning~\cite{PhysRevX.12.011059,Sivak2023RealTime} to the design of new qubit--oscillator interactions that can follow these paths directly, a promising route for bosonic error correction and, more broadly, quantum information science.

\section*{Data availability}
The data and code that support the findings of this article are openly available at \url{https://github.com/jussalcedoga/GKP_brachistochrone}.

\begin{acknowledgments}
We thank Steven Girvin for valuable feedback on this manuscript. This work was supported by the U.S. National Science Foundation under Awards 2540460 (CAREER), 2412555 (with Lorenza Viola), and 2553633 (Quantum Leap Challenge Institute); by the Defense Advanced Research Projects Agency (DARPA) under contract W912CG23C0031; and by startup funds from the Thayer School of Engineering at Dartmouth College. The authors acknowledge the use of Claude Opus 5.5 (Anthropic) to check derivations and numerical values and to edit the text. The authors reviewed all AI-assisted content and take full responsibility for it. All scientific ideas came from the authors.
\end{acknowledgments}

\section*{Author Contributions}

Mattias Fitzpatrick (MF) and Salil Bedkihal (SB) jointly conceived and supervised the project and formulated the research problem. MF provided experimental guidance and acquired funding, while SB provided theoretical guidance and contributed to the theoretical interpretation.
Aidan Curtis (AC) assisted in preparing the manuscript and derived the brachistochrone Hamiltonian. Juan S. Salcedo-Gallo (JSSG) derived the photon-optimal path, performed simulations using experimentally motivated noise rates, and led manuscript preparation and visualization. Sergei Leonov (SL) derived the logical bit-flip Hamiltonian.
SB derived the conditional-squeezing Hamiltonian using the Baker--Campbell--Hausdorff expansion, developed the GRAPE optimization framework with analytical gradients, introduced tangential capacity, and derived the photon-number evolution along the time-optimal path. Kevin Limanta (KL) and Peter Chin (PC) guided the optimization calculations without tracking weights.
All authors contributed to validation and interpretation of the results and approved the manuscript.

\end{document}


\title{Supplementary Information for\\ Time- and Photon-Optimal Paths for the Preparation and Logical Flip\\of Gottesman--Kitaev--Preskill States}

\author{Aidan Curtis}
\thanks{These authors contributed equally to this work.}
\affiliation{Thayer School of Engineering, Dartmouth College, Hanover, New Hampshire 03755, USA}

\author{Juan S. Salcedo-Gallo}
\thanks{These authors contributed equally to this work.}
\affiliation{Thayer School of Engineering, Dartmouth College, Hanover, New Hampshire 03755, USA}

\author{Sergei Leonov}
\affiliation{Thayer School of Engineering, Dartmouth College, Hanover, New Hampshire 03755, USA}
\affiliation{Department of Computer Science, Amherst College, Amherst, Massachusetts 01002, USA}

\author{Kevin Limanta}
\author{Peter Chin}
\author{Salil Bedkihal}
\author{Mattias Fitzpatrick}
\affiliation{Thayer School of Engineering, Dartmouth College, Hanover, New Hampshire 03755, USA}

\maketitle

{\makeatletter\def\l@subsection#1#2{}\makeatother
\tableofcontents}

\bigskip \noindent In this Supplementary Information, we derive the time-optimal paths between bosonic states, whose Hamiltonians contain multiphoton terms of arbitrarily high order (Sec.~\ref{app:derivation}), and the photon-optimal paths, which minimize the first-order infidelity from photon loss (Sec.~\ref{app:paths}), and evaluate them for GKP states. We then give the noise model (Sec.~\ref{app:noise}), separate a logical bit flip from a state-to-state transfer (Sec.~\ref{app:logical}), show how conditional squeezing generates the even-order terms that these paths require (Sec.~\ref{app:cs}), and define how closely and how directly a gate set follows a path (Sec.~\ref{app:capacity}). Sections~\ref{app:numerics}--\ref{app:pulses} give the numerical methods, the benchmarks of conditional squeezing (CS), conditional displacement (CD), and SNAP--displacement control (SNAP), Eq.~(10), all optimized with gradient-ascent pulse engineering (GRAPE), and the optimized pulses. Section~\ref{app:flipopt} solves the photon optimization of logical flips that keep the qubit encoded. Section~\ref{app:kerr} quantifies the self-Kerr of the cavity, which every benchmark includes, and repeats the benchmark at other envelope widths, and Sec.~\ref{app:sqcd} compares the benchmark with conditional displacements that start from a squeezed vacuum. Supplementary Table~\ref{tab:params} collects the physical parameters. Equations, figures, and tables without the prefix S or ``Supplementary'' refer to the main text.

\section{\texorpdfstring{Time-Optimal Paths and Brachistochrone Hamiltonians for GKP}{Time-Optimal Paths and Brachistochrone Hamiltonians for GKP}}
\label{app:derivation}

In this section, we derive the brachistochrone Hamiltonian, Eq.~(3) of the main text, write it in normal order, and evaluate it for GKP preparation and for the logical flip. We prove that its normal-ordered series never terminates and count how many orders a target requires.

\subsection{\texorpdfstring{General Form}{General Form}}
\label{sec:hopt-general}
We consider an initial state $\ket{\psi_i}$ and a target $\ket{\psi_f}$ with $\langle\psi_i|\psi_f\rangle=c\,e^{i\phi}$ and $c=\cos\theta$, where $\theta\in[0,\pi/2]$ is the Fubini--Study distance between them, the length of the shortest path in projective Hilbert space. For every transfer of this work, $\phi=0$. The normalized component of the target orthogonal to $\ket{\psi_i}$ is
\begin{equation}
\ket{\psi_f^{\perp}}=\frac{\ket{\psi_f}-\langle\psi_i|\psi_f\rangle\ket{\psi_i}}{\big\|\ket{\psi_f}-\langle\psi_i|\psi_f\rangle\ket{\psi_i}\big\|},
\label{eq:psi_perp}
\end{equation}
and
\begin{align}
\sigma^x_{\rm eff}&=\ket{\psi_i}\bra{\psi_f^\perp}+\ket{\psi_f^\perp}\bra{\psi_i},\label{eq:sigma_x}\\
\sigma^y_{\rm eff}&=-i\big(\ket{\psi_i}\bra{\psi_f^\perp}-\ket{\psi_f^\perp}\bra{\psi_i}\big)\label{eq:sigma_y}
\end{align}
act, together with $\sigma^z_{\rm eff}=\ket{\psi_i}\bra{\psi_i}-\ket{\psi_f^\perp}\bra{\psi_f^\perp}$, as Pauli operators on the plane of $\ket{\psi_i}$ and $\ket{\psi_f^\perp}$ and vanish outside it. A state moves at the speed $\Delta H$, its energy uncertainty~\cite{S-Anandan1990}, and we bound it by $\Delta H\leq\omega$, with $\omega=g/\sqrt2=5.303~\mu\mathrm{s}^{-1}$ the speed at which a squeezer of rate $g=7.5~\mu\mathrm{s}^{-1}$, the bound on every oscillator generator, moves the vacuum (Supplementary Table~\ref{tab:params}). The optimal paths are therefore references at a given energy uncertainty, not hardware schedules. The brachistochrone Hamiltonian at $\Delta H=\omega$~\cite{S-Carlini2006},
\begin{equation}
H_{\rm opt}=\omega\big(\sin\phi\,\sigma^x_{\rm eff}+\cos\phi\,\sigma^y_{\rm eff}\big)=i\omega\big(\ket\chi\bra{\psi_i}-\ket{\psi_i}\bra\chi\big),
\label{eq:Hopt-sigma}
\end{equation}
with $\ket\chi=e^{-i\phi}\ket{\psi_f^\perp}$ the direction in which the target lies, generates
\begin{equation}
e^{-iH_{\rm opt}t}\ket{\psi_i}=\cos(\omega t)\ket{\psi_i}+e^{-i\phi}\sin(\omega t)\ket{\psi_f^\perp},
\label{eq:path-general}
\end{equation}
which reaches $e^{-i\phi}\ket{\psi_f}$ at the minimal time $T^*=\theta/\omega$. We write $s=t/T$ for the fraction of a path of duration $T$. For $0<c<1$ the minimizing path is unique, because a path of length $\theta$ is a great circle $\cos\omega t\ket{\psi_i}+\sin\omega t\ket v$ with $\ket v\perp\ket{\psi_i}$, which reaches the target only for $\ket v=\ket\chi$. Since $\langle p|H_{\rm opt}|p\rangle=0$ and $H_{\rm opt}^2=\omega^2$ on the plane, $\Delta H=\|H_{\rm opt}\|=\omega$ at every instant. The path fixes the generator only through its action on the path state, and $H_{\rm opt}$ is the unique generator of the path with the least Hilbert--Schmidt norm. For the flip, the statements below on the code space refer to this generator.

\emph{Compact form.} With the global phase of the target chosen such that $\langle\psi_i|\psi_f\rangle=c\geq0$ ($\ket{\psi_f}\to e^{-i\phi}\ket{\psi_f}$), $\ket\chi=(\ket{\psi_f}-c\ket{\psi_i})/\sin\theta$. Inserting it in Eq.~\eqref{eq:Hopt-sigma}, where the terms $\propto c\ket{\psi_i}\bra{\psi_i}$ cancel, gives Eq.~(3),
\begin{equation}
\begin{gathered}
H_{\rm opt}=\frac{i\omega}{\sin\theta}\big(\ket{\psi_f}\bra{\psi_i}-\ket{\psi_i}\bra{\psi_f}\big),\\
\ket{p(t)}=\frac{\sin(\theta-\omega t)\ket{\psi_i}+\sin(\omega t)\ket{\psi_f}}{\sin\theta}.
\end{gathered}
\label{eq:Hopt-compact}
\end{equation}
The same expression covers preparation from the vacuum and the logical flip.

\emph{Normal order.} A product of ladder operators is normally ordered if every $\ad$ stands to the left of every $\aop$. The normal-ordered monomials have the Fock matrix elements
\begin{equation}
\langle m|\hat a^{\dagger j}\aop^k|n\rangle=\begin{cases}\delta_{m-j,\,n-k}\,\dfrac{\sqrt{m!\,n!}}{(n-k)!}, & n\geq k,\\[6pt] 0, & n<k.\end{cases}
\label{eq:no-matrix}
\end{equation}
With $\hat a^{\dagger r}\aop^r\ket n=[n!/(n-r)!]\ket n$ and $\sum_{r=0}^n(-1)^r\binom nr=\delta_{n0}$, the vacuum projector is, with ${:}f{:}$ the normally ordered form of a series $f$,
\begin{equation}
\ket0\bra0={:}e^{-\ad\aop}{:}=\sum_{r\geq0}\frac{(-1)^r}{r!}\,\hat a^{\dagger r}\aop^r .
\label{eq:no-vac}
\end{equation}
With $\ket k=\hat a^{\dagger k}\ket0/\sqrt{k!}$ and its adjoint $\bra n=\bra0\aop^n/\sqrt{n!}$, every outer product of two Fock states is normally ordered as
\begin{equation}
\ket k\bra n=\frac{\hat a^{\dagger k}\ket0\bra0\,\aop^{n}}{\sqrt{k!\,n!}}=\sum_{r\geq0}\frac{(-1)^r}{r!\sqrt{k!\,n!}}\,\hat a^{\dagger\,k+r}\aop^{n+r}.
\label{eq:no-kn}
\end{equation}
We call this series the tower of $\ket k\bra n$. Its term $r$ has the order $k+n+2r$, so every tower contains terms of arbitrarily high order. On a state with finitely many photons only finitely many terms act, and we use such series only through their Fock matrix elements, which are finite sums. An operator with the Fock matrix elements $A_{mn}$ has a unique normal-ordered expansion $A=\sum_{j,k\geq0}h_{jk}\hat a^{\dagger j}\aop^k$, since $\hat a^{\dagger j}\aop^k$ arises only from the towers of $\ket{j-r}\bra{k-r}$,
\begin{equation}
h_{jk}=\sum_{r=0}^{\min(j,k)}\frac{(-1)^r\,A_{j-r,\,k-r}}{r!\sqrt{(j-r)!\,(k-r)!}} .
\label{eq:no-coeff}
\end{equation}

\emph{A finite series has infinite rank.} Let $A=\sum_{j+k\leq K}h_{jk}\hat a^{\dagger j}\aop^k$ be a polynomial of finite order $K$ whose coefficients do not all vanish. For $n\geq K$, Eq.~\eqref{eq:no-matrix} gives
\begin{equation}
A\ket n=\sum_{j+k\leq K}h_{jk}\,\frac{n!}{(n-k)!}\,\sqrt{\frac{(n+j-k)!}{n!}}\,\ket{n+j-k}.
\label{eq:no-rank}
\end{equation}
For a shift $j-k$ with a nonzero coefficient, the component of $A\ket n$ along $\ket{n+j-k}$ is a polynomial in $n$ that does not vanish identically, because the falling factorials $n(n-1)\cdots(n-k+1)$ are linearly independent, so $A\ket n\neq0$ for all large $n$. Images of Fock states more than $2K$ photons apart have disjoint supports, so $A$ has infinite rank. A brachistochrone Hamiltonian with $\theta>0$ maps every state into the plane of $\ket{\psi_i}$ and $\ket{\psi_f}$ and has rank two. Its normal-ordered series therefore has nonzero terms of arbitrarily high order, and no polynomial in $\aop$ and $\ad$ of finite order, in any ordering, equals it. A truncation, quadratic (Gaussian) or not, is a different operator (Supplementary Table~\ref{tab:order}). Other generators of the same path can be polynomials, as for the path from $\ket0$ to $\ket2$, but for GKP preparation no generator of the path is a polynomial of finite order at its start (Sec.~\ref{sec:vac-gkp}).

\subsection{\texorpdfstring{GKP Logical States}{GKP Logical States}}
\label{sec:gkp-words}
We write $\hat D(\zeta)=\exp(\zeta\ad-\zeta^*\aop)$. A GKP code is defined by two lattice vectors whose displacements anticommute~\cite{S-Gottesman2001,S-grimsmo2021Quantum}, here $\alpha=\sqrt{\pi/2}$ and $\beta=i\alpha$. The ideal logical states are combs of position eigenstates, $\sum_j\ket{(2j+\mu)\sqrt\pi}_{\hat q}$ with $\hat q=(\aop+\ad)/\sqrt2$ and $\mu\in\{0,1\}$. Their logical $X$ and $Z$ are the displacements $\hat D(\alpha)$ and $\hat D(\beta)$~\cite{S-Gottesman2001}, which are Gaussian unitaries~\cite{S-Weedbrook2012}, and their stabilizers are $\hat S_X=\hat D(2\alpha)$ and $\hat S_Z=\hat D(2\beta)$. The Wigner functions $W(q,p)$ of the figures use $\hat q$ and $\hat p=i(\ad-\aop)/\sqrt2$. Projecting the vacuum onto the $+1$ eigenspace of $\hat S_X$ and $\hat D(\beta)$ gives $\ket{0_L}$, and $\hat D(\alpha)$ maps it to $\ket{1_L}$, so that, with $\zeta_{m,l}\equiv m\alpha+l\beta$,
\begin{equation}
  \ket{\mu_L;\alpha,\beta} \propto \sum_{m\equiv\mu,\,l} e^{-i\pi ml/2}\,\ket{\zeta_{m,l}},
  \qquad \mu\in\{0,1\},
  \label{eq:param_gkp}
\end{equation}
where $\sum_{m\equiv\mu,\,l}$ runs over $m,l\in\mathbb{Z}$ with $m\equiv\mu\ (\mathrm{mod}\ 2)$. These states are not normalizable. The finite-energy states $\ket{\mu_L}=\mathcal N^\Delta_\mu e^{-\Delta^2\nop}\ket{\mu_L;\alpha,\beta}$ carry a Gaussian envelope of width $\Delta$~\cite{S-Tzitrin2020,S-Royer2020}, and with $\hat D(\zeta)=e^{-|\zeta|^2/2}e^{\zeta\ad}e^{-\zeta^*\aop}$ and $e^{-\Delta^2\nop}e^{\zeta\ad}=e^{\zeta^\Delta\ad}e^{-\Delta^2\nop}$ they are
\begin{equation}
\ket{\mu_L}=\mathcal{N}_{\mu}^{\Delta}\sumGKP e^{-i\pi ml/2}e^{-\frac12|\zeta_{m,l}|^2}e^{\zeta_{m,l}^{\Delta}\hat a^\dagger}\ket{0},
\label{eq:gkp-main}
\end{equation}
with $\zeta^\Delta_{m,l}\equiv e^{-\Delta^2}\zeta_{m,l}$ and $c^\mu_{m,l}\equiv\mathcal N^\Delta_\mu e^{-i\pi ml/2}e^{-\frac12|\zeta_{m,l}|^2}$. For the square lattice the Fock amplitudes are real. Both logical states are parity even, because $(-1)^{\nop}$ fixes the vacuum, commutes with the envelope, and maps $\hat D(\zeta)$ to $\hat D(-\zeta)$, while $(m,l)\to(-m,-l)$ maps the lattice sum onto itself. So is every state on the brachistochrones of preparation and of the flip [Eq.~\eqref{eq:Hopt-compact}], which therefore have $\langle\aop\rangle=0$. Unless stated otherwise, $\Delta=0.4$ and the lattice sums run over $|m|,|l|\leq10$ at the Fock cutoff 80 (Sec.~\ref{app:numerics}). The logical states then have $\langle\hat D(\beta)\rangle=\pm0.882$ and stabilizer expectation values 0.59--0.62, below the ideal value 1 because of the finite envelope. At finite energy $\hat D(\beta)$ does not map the code space onto itself, so the logical operators $X_L$, $Y_L$, $Z_L$ below are defined on $\mathcal C=\mathrm{span}\{\ket{0_L},\ket{1_L}\}$.

\emph{Orthogonality of the code words.} The quarter rotation $R=e^{-i\pi\nop/2}$ maps the lattice of Eq.~\eqref{eq:param_gkp} onto itself and acts on the ideal code as the logical Hadamard. Its eigenstates in the ideal code space, $\ket{h_\pm}$ with eigenvalues $\pm1$, give $\ket{0_L;\alpha,\beta}=\cos\frac\pi8\ket{h_+}+\sin\frac\pi8\ket{h_-}$ and $\ket{1_L;\alpha,\beta}=\sin\frac\pi8\ket{h_+}-\cos\frac\pi8\ket{h_-}$, two code words normalized alike since $\ket{1_L;\alpha,\beta}=\hat D(\alpha)\ket{0_L;\alpha,\beta}$. Because $R=(-1)^{\nop/2}$ on even Fock states, $\ket{h_+}$ is supported on the photon numbers $n\equiv0$ and $\ket{h_-}$ on $n\equiv2\ (\mathrm{mod}\ 4)$. The envelope commutes with $R$, so $e^{-\Delta^2\nop}\ket{h_+}$ and $e^{-\Delta^2\nop}\ket{h_-}$ stay exactly orthogonal, but their norms $\nu_+$ and $\nu_-$ differ. With $\ket a$ and $\ket b$ the normalized components of $\ket{0_L}$ on the two sectors,
\begin{equation*}
\ket{0_L}\propto\nu_+\cos\tfrac\pi8\ket a+\nu_-\sin\tfrac\pi8\ket b,\qquad\ket{1_L}\propto\nu_+\sin\tfrac\pi8\ket a-\nu_-\cos\tfrac\pi8\ket b,
\end{equation*}
so that $\langle0_L|1_L\rangle\propto(\nu_+^2-\nu_-^2)\sin\frac\pi8\cos\frac\pi8$. This imbalance of the norms, $\nu_-/\nu_+=0.980$ at $\Delta=0.4$, accounts exactly for the overlap of the finite-energy code words, and the sector norms reproduce $c=0.014160$ of Eq.~\eqref{eq:S01} to machine precision. Rescaling the two sectors to equal norms,
\begin{equation*}
\ket{\tilde0_L}=\cos\tfrac\pi8\ket a+\sin\tfrac\pi8\ket b,\qquad\ket{\tilde1_L}=\sin\tfrac\pi8\ket a-\cos\tfrac\pi8\ket b,
\end{equation*}
gives exactly orthogonal code words on which $R$ is an exact logical Hadamard. At $\Delta=0.4$ each differs from $\ket{0_L}$ or $\ket{1_L}$ by an infidelity of $5\times10^{-5}$, and the overlap of the code words of Eq.~\eqref{eq:gkp-main} decreases as $c\approx2e^{-\pi/(4\Delta^2)}$ to leading order, from 0.081 at $\Delta=0.5$ to $3.2\times10^{-4}$ at $\Delta=0.3$ (Supplementary Table~\ref{tab:delta}). We keep the code words of Eq.~\eqref{eq:gkp-main} throughout.

\subsection{Vacuum to GKP}
\label{sec:vac-gkp}
For preparation from the vacuum, $\ket{\psi_i} = \ket{0}$ and $\ket{\psi_f} = \ket{\mu_L}$. Since $\langle 0|e^{\zeta\hat{a}^\dagger}|0\rangle = 1$, the vacuum overlap is
\begin{equation}
  c_\mu \equiv \langle 0|\mu_L\rangle
  = \mathcal{N}^{\Delta}_{\mu}\sum_{m\equiv\mu,\,l}
    e^{-i\pi ml/2}\,e^{-\frac{1}{2}|\zeta_{m,l}|^2},
  \label{eq:cmu}
\end{equation}
which is real for the square lattice. With the global phase of each logical state chosen such that $c_\mu>0$, $c_0=0.7376$ and $c_1=0.3099$. The Fock amplitudes are
\begin{equation}
\langle k|\mu_L\rangle=\frac{1}{\sqrt{k!}}\sum_{m\equiv\mu,\,l}c^\mu_{m,l}\big(\zeta^\Delta_{m,l}\big)^k .
\label{eq:fock-amp}
\end{equation}
Since $c_\mu$ is real, Eq.~\eqref{eq:Hopt-compact} applies with $\theta_\mu=\arccos c_\mu$, and preparation takes $T^*=140$~ns for $\ket{0_L}$ and 237~ns for $\ket{1_L}$. The target lies in the direction $\ket{\mu_L^\perp}=(\ket{\mu_L}-c_\mu\ket0)/\sin\theta_\mu$, whose Fock amplitudes $\langle n|\mu_L\rangle/\sin\theta_\mu$ ($n\geq1$) are real and vanish for odd $n$.

\emph{Fock-basis form.} For $\phi=0$, Eq.~\eqref{eq:Hopt-sigma} is $H_{\rm opt}=i\omega(\ket{\mu_L^\perp}\bra0-\ket0\bra{\mu_L^\perp})$, and the Fock expansion of $\ket{\mu_L^\perp}$ gives
\begin{equation}
H_{0\to\mu_L}=i\omega\sum_{n=2,4,6,\ldots}\langle n|\mu_L^\perp\rangle\big(\ket n\bra0-\ket0\bra n\big),
\label{eq:vac-fock}
\end{equation}
a generator of rank two. It couples the vacuum to every even Fock state below the cutoff, each coupling $\ket0\leftrightarrow\ket n$ being an $n$-photon transition, and couples no two excited Fock states.

\emph{Normal-ordered form.} Replacing each $\ket n\bra0$ and $\ket0\bra n$ in Eq.~\eqref{eq:vac-fock} by its tower, Eq.~\eqref{eq:no-kn}, gives
\begin{equation}
H_{0\to\mu_L}=i\omega\sum_{n=2,4,6,\ldots}h_n\sum_{r\geq0}\frac{(-1)^r}{r!}\big(\hat a^{\dagger\,n+r}\aop^r-\hat a^{\dagger r}\aop^{n+r}\big),
\label{eq:vac-series}
\end{equation}
with $h_n=\langle n|\mu_L\rangle/(\sqrt{n!}\,\sin\theta_\mu)$. With Eq.~\eqref{eq:fock-amp}, $h_n=\sum_{m\equiv\mu,\,l}c^\mu_{m,l}(\zeta^\Delta_{m,l})^n/(n!\sin\theta_\mu)$, which turns Eq.~\eqref{eq:vac-series} into the lattice form, Eq.~(4a) of the main text. There the sum runs over all $n\geq1$, because the lattice sums vanish for odd $n$.
The term $(n,r)$ changes the photon number by $\pm n$ and has the total order $n+2r$ in $\aop$ and $\ad$. Collecting the terms of equal total order gives, through order six,
\begin{equation}
\begin{split}
H_{0\to\mu_L}=i\omega\Big[&\,h_2\big(\hat a^{\dagger2}-\aop^2\big)\\
&+h_4\big(\hat a^{\dagger4}-\aop^4\big)-h_2\big(\hat a^{\dagger3}\aop-\ad\aop^3\big)\\
&+h_6\big(\hat a^{\dagger6}-\aop^6\big)-h_4\big(\hat a^{\dagger5}\aop-\ad\aop^5\big)+\tfrac12h_2\big(\hat a^{\dagger4}\aop^2-\hat a^{\dagger2}\aop^4\big)+\ldots\Big],
\end{split}
\label{eq:vac-orders}
\end{equation}
where the three lines are the orders two, four, and six. At $\Delta=0.4$, $h_2=-0.3595$ for $\ket{0_L}$ and $0.6252$ for $\ket{1_L}$. The coefficients are small at high order, but the matrix elements of the terms grow factorially with the photon number, so the size of a coefficient does not measure the weight of its term.

\emph{Physical content.} The order-two term is a squeezer, the only Gaussian part of the series~\cite{S-Weedbrook2012}. Its rate, $2\omega|h_2|=g|\langle2|\mu_L^\perp\rangle|$, lies within the bound $g$, and alone it reaches the target fidelity 0.730 for $\ket{0_L}$ (Supplementary Table~\ref{tab:order}) and 0.331 for $\ket{1_L}$. The terms $\hat a^{\dagger n}-\aop^n$ with $n\geq4$ are $n$-photon drives, or generalized squeezers~\cite{S-Ashhab2026}, and the terms with $r\geq1$ make these drives depend on the photon number, since $\hat a^{\dagger\,n+r}\aop^r=\hat a^{\dagger n}\nop(\nop-1)\cdots(\nop-r+1)$. Summed over $r$, they give $\hat a^{\dagger n}\ket0\bra0$ and so confine each $n$-photon drive to the vacuum. The term $r=2$ multiplies the $n$-photon drive by the self-Kerr operator $\hat a^{\dagger2}\aop^2=\nop(\nop-1)$, as in $\tfrac12h_2(\hat a^{\dagger4}\aop^2-\hat a^{\dagger2}\aop^4)$ of Eq.~\eqref{eq:vac-orders}, and the terms $r>2$ by $\hat a^{\dagger r}\aop^r=\nop(\nop-1)\cdots(\nop-r+1)$. The series has no term that conserves the photon number, however, so it contains no static detuning or Kerr term, and a static self-Kerr of the device is an additional term that the control must compensate (Sec.~\ref{app:kerr}). Every term changes the photon number by an even amount, so $H_{0\to\mu_L}$ conserves parity, whereas a displacement changes the photon number by one. Under photon loss, an ideal parity measurement after the path therefore flags every trajectory with an odd number of loss events, as for conditional squeezing (Sec.~\ref{app:robust}).

\emph{Nonlinearity.} Every $n$ with $h_n\neq0$ contributes a term at each order $n+2r$, so the series does not terminate, even for Fock $\ket2$. For GKP preparation no generator $H'$ of the path is a polynomial of finite order at its start. Indeed, $H'(0)\ket0=i\omega\ket{\mu_L^\perp}+\lambda\ket0$ with a real $\lambda$, a polynomial of order $K$ maps $\ket0$ into the span of $\ket0,\ldots,\ket K$, and the Fock amplitudes of $\ket{\mu_L^\perp}$ do not vanish beyond any photon number (at the cutoff 80, $H'(0)$ has at least the order 78). Single terms of the finite sums reach $2\times10^{25}\omega$ below 80 photons and cancel to matrix elements of at most $\omega$, so we evaluate the series in 90-digit arithmetic, where it agrees with Eq.~\eqref{eq:vac-fock} to $10^{-65}\omega$.

\subsection{\texorpdfstring{Logical Flip}{Logical Flip}}
\label{sec:flip-hopt}
Both end states now carry photons. With $\langle0|e^{y^*\aop}e^{z\ad}|0\rangle=e^{y^*z}$ and Eq.~\eqref{eq:gkp-main}, the overlap of the logical states is
\begin{equation}
c=\langle0_L|1_L\rangle=\sum_{m\equiv0,\,l}\;\sum_{m'\equiv1,\,l'}c^{0*}_{m,l}\,c^1_{m',l'}\,
\exp\big(\zeta^{\Delta*}_{m,l}\zeta^\Delta_{m',l'}\big).
\label{eq:S01}
\end{equation}
It is real because the Fock amplitudes are real, and positive because both logical states have positive position wave functions (by Poisson summation~\cite{S-SteinShakarchi2003}, and by Mehler's formula for the envelope~\cite{S-Mehler1866}). Numerically, $c=0.014160$ and the angle between the logical states is $\theta_L=\arccos c=0.991\,\pi/2$, so the finite-energy logical states are nearly, but not exactly, orthogonal. With $\ket{1_L^\perp}=(\ket{1_L}-c\ket{0_L})/\sin\theta_L$, the orthonormal basis $\{\ket{0_L},\ket{1_L^\perp}\}$ spans the code space $\mathcal C$, and Eq.~\eqref{eq:Hopt-compact} gives the flip brachistochrone,
\begin{equation}
H_{0_L\to1_L}=\frac{i\omega}{\sin\theta_L}\big(\ket{1_L}\bra{0_L}-\ket{0_L}\bra{1_L}\big)=\omega Y_L,
\label{eq:H01-main}
\end{equation}
with $Y_L\equiv i(\ket{1_L^\perp}\bra{0_L}-\ket{0_L}\bra{1_L^\perp})$, a rotation about the logical $Y$ axis that reaches $\ket{1_L}$ at $T^*=294$~ns.

\emph{The flip brachistochrone is a logical gate.} Equation~\eqref{eq:H01-main} acts only on $\mathcal C$, so $U(t)=e^{-i\omega tY_L}$ maps $\mathcal C$ onto itself and rotates every encoded state about the logical $Y$ axis. Halfway, $\ket{0_L}$ passes through $\ket{+_L}\propto\ket{0_L}+\ket{1_L}$. Let $X_L$ be the unitary on $\mathcal C$ that swaps the finite-energy logical states. In the basis $\{\ket{0_L},\ket{1_L^\perp}\}$ it reads $\bigl(\begin{smallmatrix}\cos\theta_L&\sin\theta_L\\\sin\theta_L&-\cos\theta_L\end{smallmatrix}\bigr)$, while $U(T^*)=\bigl(\begin{smallmatrix}\cos\theta_L&-\sin\theta_L\\\sin\theta_L&\cos\theta_L\end{smallmatrix}\bigr)$. Hence $U(T^*)=X_LZ_L$ exactly, with $Z_L\equiv\ket{0_L}\bra{0_L}-\ket{1_L^\perp}\bra{1_L^\perp}$ a frame update, and $U(T^*)$ maps $\ket{1_L}$ to $\ket{0_L}$ with fidelity $(1-2c^2)^2=0.9992$. Since the logical states are not orthogonal, $X_L$ is not exactly the Pauli $X$ of the orthonormal basis. In the gate metric of Sec.~\ref{app:logical}, the gate fidelity of $U(T^*)$ with $X_L$ is 0.9997 (closed evolution). Both defects come only from the overlap $c$ of the finite-energy code words. For the orthogonal code words of Sec.~\ref{sec:gkp-words}, $\theta_L=\pi/2$, the flip brachistochrone lasts $T^*=\pi/(2\omega)=296$~ns, and $U(T^*)$ is exactly the logical Pauli $Y$ up to a global phase. With $c$ from Supplementary Table~\ref{tab:delta}, the defect $1-(1-2c^2)^2\approx4c^2$ falls from $2.6\times10^{-2}$ at $\Delta=0.5$ to $8.0\times10^{-4}$ at $\Delta=0.4$ and $4.0\times10^{-7}$ at $\Delta=0.3$.

\emph{Fock-basis form.} On the oscillator $\omega Y_L$ is a nonlinear Hamiltonian with the Fock matrix elements
\begin{equation}
\langle j|H_{0_L\to1_L}|k\rangle=\omega\langle j|Y_L|k\rangle=\frac{i\omega}{\sin\theta_L}\big(\langle j|1_L\rangle\langle0_L|k\rangle-\langle j|0_L\rangle\langle1_L|k\rangle\big),\qquad j,k\geq0.
\label{eq:flip-fock}
\end{equation}
It is imaginary, antisymmetric, of rank two, and couples only even Fock states. The quarter rotation $e^{-i\pi\nop/2}$, the logical Hadamard of the ideal code~\cite{S-Gottesman2001,S-grimsmo2021Quantum}, maps $\ket{0_L}$ to $\ket{+_L}$ to within $5\times10^{-5}$ in fidelity and equals $(-1)^{\nop/2}$ on even Fock states. Since the envelope commutes with this rotation, the sector decomposition of Sec.~\ref{sec:gkp-words} gives exactly $\langle n|1_L\rangle=(\mathcal N^\Delta_1/\mathcal N^\Delta_0)\big[\sqrt2(-1)^{n/2}-1\big]\langle n|0_L\rangle$ for even $n$, and Eq.~\eqref{eq:flip-fock} becomes proportional to $\big[(-1)^{j/2}-(-1)^{k/2}\big]\langle j|0_L\rangle\langle k|0_L\rangle$. The flip generator thus couples $\ket j$ and $\ket k$ only if $j-k\equiv2\pmod4$, including pairs of excited Fock states such as $\ket2$ and $\ket4$, which the generator of preparation does not couple.

\emph{Normal-ordered form.} By Eqs.~\eqref{eq:no-coeff} and \eqref{eq:flip-fock}, $H_{0_L\to1_L}=\sum_{j,k\geq0}h_{jk}\hat a^{\dagger j}\aop^k$ with
\begin{equation}
h_{jk}=\omega\sum_{r=0}^{\min(j,k)}\frac{(-1)^r}{r!}\,\frac{\langle j-r|Y_L|k-r\rangle}{\sqrt{(j-r)!\,(k-r)!}}.
\label{eq:flip-hjk}
\end{equation}
The coefficients inherit four selection rules from the Fock matrix. They vanish unless $j-k$ is even, so the series conserves parity. They vanish for $j=k$, so the series has no detuning or Kerr term. They obey $h_{kj}=-h_{jk}=h_{jk}^*$, so each pair of conjugate terms is Hermitian. They vanish unless $j-k\equiv2\pmod4$, so every term changes the photon number by $\pm2$, $\pm6$, $\pm10$, \dots. Through the total order six,
\begin{equation}
\begin{split}
H_{0_L\to1_L}={}&h_{2,0}\Big[\hat a^{\dagger2}-\aop^2-\big(\hat a^{\dagger3}\aop-\ad\aop^3\big)+\tfrac12\big(\hat a^{\dagger4}\aop^2-\hat a^{\dagger2}\aop^4\big)\Big]\\
&+\frac{\omega\langle4|Y_L|2\rangle}{4\sqrt3}\big(\hat a^{\dagger4}\aop^2-\hat a^{\dagger2}\aop^4\big)+h_{6,0}\big(\hat a^{\dagger6}-\aop^6\big)+\dots,
\end{split}
\label{eq:flip-low}
\end{equation}
with $h_{2,0}=0.5138\,i\omega$ and $h_{6,0}=-0.01081\,i\omega$ at $\Delta=0.4$. The squeezer alone reaches the target fidelity 0.27--0.30 only (Supplementary Table~\ref{tab:order}), because it moves both logical states almost entirely out of the code space, and its tangential capacity along the flip is $0.0061\,g$ (Sec.~\ref{app:capacity}). Like the generator of preparation, $\omega Y_L$ has rank two, so no polynomial of finite order equals it.

\emph{Lattice form.} Normal ordering the vacuum projector of Eq.~\eqref{eq:no-vac} between two exponentials gives
\begin{equation*}
e^{\zeta'\ad}\ket0\bra0\,e^{\zeta^*\aop}=\sum_{n,n',r\geq0}\frac{(-1)^r\,\zeta'^{\,n}\zeta^{*n'}}{n!\,n'!\,r!}\,\hat a^{\dagger\,n+r}\aop^{n'+r},
\end{equation*}
so that, with Eq.~\eqref{eq:gkp-main} for both logical states, $\ket{1_L}\bra{0_L}$ becomes a double lattice sum of such terms and Eq.~\eqref{eq:H01-main} becomes the lattice form, Eq.~(4b) of the main text. Its coefficients agree with Eq.~\eqref{eq:flip-hjk}.

\emph{Contrast with the ideal code.} For ideal logical states the logical $X$ is the Gaussian displacement $\hat D(\alpha)$. At $\Delta=0.4$, however, $\hat D(\alpha)$ also displaces the envelope and maps $\ket{0_L}$ to $\ket{1_L}$ with fidelity 0.778 only, and a numerical search finds no Gaussian unitary with a gate fidelity above 0.780 (Sec.~\ref{app:flip}). A gate set whose oscillator generators are at most quadratic therefore needs entanglement to exceed this numerical bound (Proposition~\ref{thm:nogo}). The generator $\omega Y_L$ instead is nonlinear to all orders and conserves parity, so the parity flag of Sec.~\ref{sec:vac-gkp} applies to the flip as well.

\subsection{\texorpdfstring{How Many Nonlinear Terms}{How Many Nonlinear Terms}}
\label{sec:orders}
Here we count how many orders a target requires. We replace $H_{\rm opt}$ by the Hermitian part of $\sum_{j+k\leq K}h_{jk}\hat a^{\dagger j}\aop^k$ [Eq.~\eqref{eq:no-coeff}], assembled in 60-digit arithmetic, and give the target fidelity at $T^*$ in Supplementary Table~\ref{tab:order}. Beyond $K=2$ the series contain generalized-squeezing terms whose evolution in a truncated Fock space depends on the truncation~\cite{S-Ashhab2026}, so each entry is the range over the cutoffs 40, 60, and 80. Two conclusions do not depend on the cutoff. Squeezing-order terms ($K=2$) are far from sufficient for every target. A fidelity of at least 0.98 at every cutoff is first reached at $K=6$ for Fock $\ket2$, 12 for Fock $\ket4$, 24 for the even cat $\ket{C_\gamma^+}\propto\ket\gamma+\ket{-\gamma}$ ($\gamma=2$), and 40 for GKP preparation and the flip. Multiphoton qubit--oscillator interactions can supply some of these terms directly~\cite{S-Gorgichuk2026}, and conditional squeezing with qubit rotations generates them through nested commutators (Sec.~\ref{app:cs}).

\begin{table}[h!]
\caption{Target fidelity at $T^*$ with the normal-ordered expansion of $H_{\rm opt}$ truncated at total order $K$, given as the range over the Fock cutoffs 40, 60, and 80 (operator assembled and exponentiated in 60-digit arithmetic). A single value means that the three cutoffs agree to the digits shown.}
\label{tab:order}
\footnotesize
\setlength{\tabcolsep}{3pt}
\begin{ruledtabular}
\begin{tabular}{lccccccccc}
$K$ & 2 & 4 & 6 & 8 & 12 & 16 & 24 & 40 & 56\\\hline
Fock $\ket2$ & 0.102--0.104 & 0.93--0.94 & 1.000 & 1.000 & 1.000 & 1.000 & 1.000 & 1.000 & 1.000\\
Fock $\ket4$ & 0.000 & 0.02--0.78 & 0.002--0.966 & 0.001--0.997 & 1.000 & 1.000 & 1.000 & 1.000 & 1.000\\
cat $\gamma=2$ & 0.530 & 0.69 & 0.14--0.53 & 0.32--0.64 & 0.83--0.98 & 0.93--0.99 & 1.000 & 1.000 & 1.000\\
GKP $\ket{0_L}$ & 0.730 & 0.76--0.77 & 0.43--0.76 & 0.51--0.60 & 0.80--0.84 & 0.81--0.82 & 0.97--0.98 & 0.990--0.992 & 0.999\\
flip & 0.27--0.30 & 0.17--0.26 & 0.58--0.72 & 0.31--0.53 & 0.56--0.82 & 0.61--0.70 & 0.64--0.97 & 0.984--0.996 & 0.995--0.999\\
\end{tabular}
\end{ruledtabular}
\end{table}

\section{\texorpdfstring{Photon-Optimal Paths}{Photon-Optimal Paths}}
\label{app:paths}

In this section, we support the statements of the main text on photon-optimal paths. Section~\ref{sec:exposure} shows that the photon exposure $J$ of Eq.~(5) sets the first-order loss of a path and evaluates it along the brachistochrones. Section~\ref{sec:fermat} proves that photon-optimal paths exist and obey Fermat's principle with $\bar n$ as the refractive index, which is why they detour through states with fewer photons, and describes the numerical optima of Fig.~2. Section~\ref{sec:bounds} proves the lower bound of Eq.~(7), Sec.~\ref{sec:vacuum} explains why photon-optimal transfers pass through the vacuum, Sec.~\ref{sec:closed} derives the closed form of Eq.~(6), and Sec.~\ref{sec:gallery} compares the two optimal paths for further targets.

\subsection{\texorpdfstring{Photon Exposure and the Photon Number along the Brachistochrones}{Photon Exposure and the Photon Number along the Brachistochrones}}
\label{sec:exposure}
\begin{lemma}[first-order error of a unitary path]\label{lem:loss}
Let $\dot\rho=-i[H,\rho]+\sum_j\gamma_j\mathcal D[\hat L_j]\rho$, with the dissipator $\mathcal D[\hat L]\rho=\hat L\rho\hat L^\dagger-\tfrac12\{\hat L^\dagger\hat L,\rho\}$ and rates $\gamma_j$, $\rho(0)=\ket{\psi(0)}\bra{\psi(0)}$, and let $\ket{\psi(t)}$ be the unitary trajectory of $H$. Then $1-\langle\psi(T)|\rho(T)|\psi(T)\rangle=\sum_j\gamma_j\int_0^T\big(\langle\hat L_j^\dagger\hat L_j\rangle-|\langle\hat L_j\rangle|^2\big)dt+O(\gamma^2)$, with expectation values in $\ket{\psi(t)}$.
\end{lemma}
\begin{proof}
Along the unitary trajectory, $\ket{\dot\psi}=-iH\ket\psi$ cancels the Hamiltonian part of $\dot\rho$ in $\frac{d}{dt}\langle\psi|\rho|\psi\rangle$, so only the dissipators change the overlap. Since $\rho=\ket\psi\bra\psi+O(\gamma)$, each dissipator contributes $\gamma_j\langle\psi|\mathcal D[\hat L_j](\ket\psi\bra\psi)|\psi\rangle=\gamma_j(|\langle\hat L_j\rangle|^2-\langle\hat L_j^\dagger\hat L_j\rangle)$ up to order $\gamma^2$. Integrating from the initial value 1 gives the lemma.
\end{proof}
For photon loss at the rate $\kappa$, $\hat L=\aop$, and the lemma gives the loss term
\begin{equation}
\kappa\int_0^T\big(\langle\nop\rangle-|\langle\aop\rangle|^2\big)\,dt .
\label{eq:nbar-loss}
\end{equation}
The annihilation operator changes the parity, so on every state of definite parity the integrand is $\bar n=\langle\nop\rangle$. This holds on the brachistochrones of GKP preparation and of the flip (Sec.~\ref{sec:gkp-words}). For cavity dephasing, $\hat L=\nop$ and the integrand is $\mathrm{Var}(\nop)$, and for an equatorial ancilla the two ancilla dissipators give $\Gamma_q$, which together give Eq.~(8). The first-order loss term agrees with a direct Lindblad integration of the brachistochrones to a relative $4\times10^{-4}$ (preparation) and $1.1\times10^{-3}$ (flip). The loss cost of a path is therefore its photon exposure $J=\int\bar n\,d\vartheta$, Eq.~(5), with $\vartheta$ the Fubini--Study arclength. At $\Delta H=\omega$, $J/\omega=\int\bar n\,dt$, and 1~rad of exposure is $10^3/\omega=188.6$~photon~ns. Subscripts geo and ph denote the brachistochrone and the photon optimum.

\emph{Preparation.} Along $\ket{p(s)}=\cos(s\theta_0)\ket0+\sin(s\theta_0)\ket\chi$, with $\ket\chi=\ket{0_L^\perp}$ and $\theta_0=\arccos c_0$, only the term $n_\chi\sin^2(s\theta_0)$ of $\bar n$ survives, since $\nop\ket0=0$. With $n_\chi=n_0/\sin^2\theta_0$ and $n_0=\langle0_L|\nop|0_L\rangle$,
\begin{equation}
\bar n_{\rm geo}(s)=n_0\,\frac{\sin^2(s\theta_0)}{\sin^2\theta_0},
\label{eq:nbar-prep}
\end{equation}
and, with $\int_0^1\sin^2(s\theta_0)\,ds=\tfrac12-\sin2\theta_0/(4\theta_0)$,
\begin{equation}
\int_0^T\bar n_{\rm geo}\,dt=T\,\frac{n_0}{\sin^2\theta_0}\Big(\frac12-\frac{\sin2\theta_0}{4\theta_0}\Big).
\label{eq:nbar-prep-exposure}
\end{equation}
At $\Delta H=\omega$ this is $J_{\rm geo}/\omega$ with $J_{\rm geo}=n_\chi(\theta_0/2-\sin2\theta_0/4)$. For $\Delta=0.4$, $\theta_0=0.7412$, $n_0=2.439$, and $n_\chi=5.350$, so the time-averaged photon number is 0.878, about a third of that of the target, and the exposure at $T^*=140$~ns is 122.7~photon~ns.

\emph{Logical flip.} Along the path of Eq.~\eqref{eq:Hopt-compact}, $\ket{p(s)}=\{\sin[(1-s)\theta_L]\ket{0_L}+\sin(s\theta_L)\ket{1_L}\}/\sin\theta_L$, the real cross term $n_{01}=\langle0_L|\nop|1_L\rangle$ must be kept, and with $n_1=\langle1_L|\nop|1_L\rangle$,
\begin{equation}
\bar n_{\rm geo}(s)=\frac{\sin^2[(1-s)\theta_L]\,n_0+\sin^2(s\theta_L)\,n_1+2\sin[(1-s)\theta_L]\sin(s\theta_L)\,n_{01}}{\sin^2\theta_L}.
\label{eq:nbar-flip}
\end{equation}
The diagonal terms integrate to $\tfrac12-\sin2\theta_L/(4\theta_L)$ each and the cross term to $\sin\theta_L/\theta_L-\cos\theta_L$, so
\begin{equation}
\int_0^T\bar n_{\rm geo}\,dt=\frac{T}{\sin^2\theta_L}\Big[(n_0+n_1)\Big(\frac12-\frac{\sin2\theta_L}{4\theta_L}\Big)+n_{01}\Big(\frac{\sin\theta_L}{\theta_L}-\cos\theta_L\Big)\Big].
\label{eq:nbar-flip-exposure}
\end{equation}
For $\Delta=0.4$, $\theta_L=1.5566$, $n_1=2.870$, and $n_{01}=-0.178$. The photon number then has its minimum, 2.353, at $s=0.248$ and the time average 2.519, below the mean 2.655 of the end values mainly because of the cross term. Both closed forms agree with a direct evaluation on the state vectors to $5\times10^{-15}$. At equal duration the flip brachistochrone thus carries 2.87 times the time-integrated photon number of preparation, and at the speed limit, where it lasts 2.10 times as long, 6.0 times its exposure (739.5 against 122.7~photon~ns). The shortest path need not have the least exposure. A path through the vacuum is longer ($\arccos c_0+\arccos c_1=1.997$ against 1.557) but passes through $\bar n=0$, and the photon-optimal transfer from $\ket{0_L}$ to $\ket{1_L}$ takes this route (Sec.~\ref{sec:vacuum}).

\subsection{\texorpdfstring{Fermat's Principle for Photon-Optimal Paths}{Fermat's Principle for Photon-Optimal Paths}}
\label{sec:fermat}
In geometric optics, a light ray between two points makes its optical path length stationary, the geometric length of the ray weighted by the refractive index of the medium, which is Fermat's principle~\cite{S-BornWolf1999}. In Bernoulli's solution of the brachistochrone problem, the refractive index is the inverse speed of the falling body, so that the optical path length is the travel time~\cite{S-Bernoulli1697}. The photon exposure of Eq.~(5), $J=\int\bar n\,d\vartheta$, has the same form, with the Fubini--Study arclength $\vartheta$ as the geometric length and the mean photon number $\bar n$ as the refractive index, and we call $\bar n\,d\vartheta$ the photon metric. At the fixed speed $\Delta H=\omega$ a path of length $L=\int d\vartheta$ lasts $L/\omega$, so the brachistochrone, the shortest path, is a ray of a uniform medium, whereas a photon-optimal path is a ray of the medium whose index is $\bar n$. This subsection makes the analogy exact. It shows that photon-optimal paths exist, reduces their search for preparation to real states of even parity, derives their ray equation, and explains from it the detour of the photon-optimal paths in Figs.~1(a,~b) and 2(a,~b).

\emph{Existence.} Let $J^*(\varphi,\psi)$ be the least exposure of any path from $\ket\varphi$ to $\ket\psi$. It is symmetric and obeys the triangle inequality, since $J$ is invariant under reversal of the direction of travel and additive under concatenation. Every state has $\bar n\geq1-|\langle0|\psi\rangle|^2$, so $\bar n$ vanishes only at the vacuum and $J^*$ is positive for distinct states. At a finite Fock cutoff $\bar n$ is also bounded, so $J^*$ is a distance that induces the usual topology on the compact space of states (rays), and by the Hopf--Rinow--Cohn-Vossen theorem for length spaces~\cite[Chap.~I.3]{S-BridsonHaefliger1999} any two states are joined by a path of least exposure, a photon-optimal path.

\emph{Reduction to real, even paths.} For preparation from the vacuum, the search can be restricted to real amplitudes of even parity.
\begin{lemma}[real, even paths suffice for preparation]\label{lem:reduce}
Let $\ket{\psi_f}$ have even parity. For every path from $\ket0$ to $\ket{\psi_f}$, with complex amplitudes of either parity, there is a path from $\ket0$ to $\ket{\psi_f}$ of no larger exposure whose amplitudes vanish for odd photon numbers and carry the phases of the amplitudes of $\ket{\psi_f}$. For a real target, they are real and carry its signs.
\end{lemma}
\begin{proof}
The diagonal unitary $e^{-i\varphi(\nop)}$, with $e^{i\varphi(n)}=\langle n|\psi_f\rangle/|\langle n|\psi_f\rangle|$ and $\varphi(n)=0$ where $\langle n|\psi_f\rangle=0$, commutes with $\nop$, so it leaves $\bar n$ unchanged, preserves the Fubini--Study metric, and fixes the vacuum up to a phase. It maps $\ket{\psi_f}$ to the state with the amplitudes $|\langle n|\psi_f\rangle|$, so it suffices to prove the lemma for a target with non-negative amplitudes and to apply $e^{i\varphi(\nop)}$ to the path it gives. Let $\psi(\tau)$, $\tau\in[0,1]$, be a piecewise smooth lift of the path to normalized vectors with $\langle\psi|\partial_\tau\psi\rangle=0$, so that its Euclidean speed $|\partial_\tau\psi|$ equals its Fubini--Study speed, and write $\psi_n=\langle n|\psi\rangle$. The path $\tilde\psi$ with the real amplitudes $\tilde\psi_{2k}=(|\psi_{2k}|^2+|\psi_{2k+1}|^2)^{1/2}$ and $\tilde\psi_{2k+1}=0$ is normalized, starts at the vacuum, and ends at $\ket{\psi_f}$, whose odd amplitudes vanish. At every $\tau$ its photon number is $\bar n(\tilde\psi)=\bar n(\psi)-\sum_k|\psi_{2k+1}|^2\leq\bar n(\psi)$, and its speed is at most that of $\psi$, because $|\partial_\tau\tilde\psi_{2k}|\leq(|\partial_\tau\psi_{2k}|^2+|\partial_\tau\psi_{2k+1}|^2)^{1/2}$ almost everywhere by the Cauchy--Schwarz inequality, and a real normalized path has equal Euclidean and Fubini--Study speeds. Hence $J[\tilde\psi]=\int_0^1\bar n(\tilde\psi)\,|\partial_\tau\tilde\psi|\,d\tau\leq\int_0^1\bar n(\psi)\,|\partial_\tau\psi|\,d\tau=J[\psi]$.
\end{proof}
The inequality is strict wherever the path moves while it has odd weight, so every photon-optimal preparation is even. Applied to a photon-optimal preparation, the lemma gives a real one of the same exposure, so the least exposure over real even paths equals that over all paths.

\emph{Real even paths.} In this subsection and in Sec.~\ref{sec:closed}, a state of the real even sector is a real unit vector $\psi$ with the components $\psi_n=\langle n|\psi\rangle$ for even $n$ below the Fock cutoff (40 components at the cutoff 80), with ${}^\top$ the transpose and $|\cdot|$ the Euclidean norm. Its photon number is $\bar n=\psi^\top\nop\psi=\sum_nn\,\psi_n^2$, whose gradient in this space of real vectors is $\nabla\bar n=2\nop\psi$. Since $\psi^\top d\psi=0$ for real unit vectors, the Fubini--Study line element is the Euclidean one, $d\vartheta=|d\psi|$, so the real even sector is a unit sphere with its usual geometry, on which $\psi$ and $-\psi$ are the same state. The projector $P_\psi=1-\psi\psi^\top$ maps onto the directions tangent to the sphere at $\psi$. Unless stated otherwise, a path $\psi(\vartheta)$ is parametrized by its arclength $\vartheta$, a prime denotes $d/d\vartheta$, and $\psi'$ is the unit tangent, $|\psi'|=1$.

\begin{proposition}[Fermat's principle for the photon exposure]\label{prop:el}
On an interval where $\bar n>0$, a real even path makes $J$ stationary under variations that vanish at the ends of the interval if and only if
\begin{equation}
P_\psi\big[(\bar n\,\psi')'-\nabla\bar n\big]=0,
\label{eq:ray}
\end{equation}
or, equivalently,
\begin{equation}
\psi''=-\psi+2\Big[\frac{\nop\psi}{\bar n}-\psi-\frac{\psi'^\top\nop\psi}{\bar n}\,\psi'\Big].
\label{eq:EL}
\end{equation}
\end{proposition}
\begin{proof}
For a variation $\delta\psi$ tangent to the sphere, $\psi^\top\delta\psi=0$, which keeps $\psi$ normalized to first order, and vanishing at the ends of the interval, $\delta(\bar n\,|\psi'|)=\nabla\bar n^\top\delta\psi+\bar n\,\psi'^\top\delta\psi'$ at $|\psi'|=1$. An integration by parts gives $\delta J=-\int\delta\psi^\top[(\bar n\,\psi')'-\nabla\bar n]\,d\vartheta$, which vanishes for every such $\delta\psi$ if and only if Eq.~\eqref{eq:ray} holds. To write it out, we differentiate $\psi^\top\psi'=0$ and use $|\psi'|=1$, which gives $\psi^\top\psi''=-1$ and hence $P_\psi\psi''=\psi''+\psi$. With $P_\psi\psi'=\psi'$, $\bar n'=2\psi'^\top\nop\psi$, and $P_\psi\nabla\bar n=2(\nop\psi-\bar n\psi)$, Eq.~\eqref{eq:ray} reads $\bar n'\psi'+\bar n(\psi''+\psi)-2(\nop\psi-\bar n\psi)=0$, which is Eq.~\eqref{eq:EL}. Conversely, a solution of Eq.~\eqref{eq:EL} that starts with $|\psi|=1$, $\psi^\top\psi'=0$, and $|\psi'|=1$ keeps these values, because Eq.~\eqref{eq:EL} gives $\psi^\top\psi''=-1$ and $\psi'^\top\psi''=0$ for them, so it satisfies Eq.~\eqref{eq:ray}.
\end{proof}
Equation~\eqref{eq:ray} is the ray equation of geometric optics, which states that the derivative of the refractive index times the unit tangent of a ray equals the gradient of the index~\cite{S-BornWolf1999,S-Brekhovskikh1960}, here restricted by $P_\psi$ to the sphere of states, with $\bar n$ as the index. Three consequences connect it to the main text. First, the projector $P_\perp=P_\psi-\psi'\psi'^\top$ onto the directions normal to the path picks out the geodesic curvature $P_\perp\psi''$, the part of the acceleration of the path that is not needed to keep it on the sphere, and Eq.~\eqref{eq:EL} gives $P_\perp\psi''=P_\perp\nabla\ln\bar n$. For a constant index, as for $\nop$ replaced by the identity, the curvature vanishes and the ray is a great circle, the brachistochrone of Eq.~(3). Second, the curvature points along the component of $\nabla\ln\bar n$ normal to the path, toward higher photon number, as a light ray bends toward higher refractive index. Between fixed end points a photon-optimal path therefore bows away from the great circle into states with fewer photons, as a light ray above a hot road dips into the air of lower index, which gives the detour of the dashed lines in Figs.~1(a,~b) and the lower photon number of the photon optimum in Fig.~2(a). Third, Eq.~\eqref{eq:EL} holds only where $\bar n>0$. At the vacuum, where the index vanishes, a photon-optimal path can change its direction, which is why photon-optimal transfers can pass through it (Sec.~\ref{sec:vacuum}). Section~\ref{sec:closed} solves the variational problem to second order in the departure from the brachistochrone, which gives the closed form of Eq.~(6).

\emph{Numerical optima.} The open circles of Fig.~2 minimize $J$ directly. A path in the real even sector at the Fock cutoff 80 is represented by $M=30$--240 free unit vectors $\psi^{(1)},\dots,\psi^{(M)}$ between the fixed end points, joined by great-circle arcs. Along the arc between consecutive vectors $\psi^{(a)}$ and $\psi^{(b)}$, of length $\vartheta_{ab}=\arccos(\psi^{(a)\top}\psi^{(b)})$, the photon number is a trigonometric polynomial in the arclength, so the exposure of the arc is exact,
\begin{equation}
J_{\rm arc}=\frac{\big(\psi^{(a)\top}\nop\psi^{(a)}+\psi^{(b)\top}\nop\psi^{(b)}\big)\big(\tfrac{\vartheta_{ab}}2-\tfrac{\sin2\vartheta_{ab}}4\big)+\psi^{(a)\top}\nop\psi^{(b)}\,(\sin\vartheta_{ab}-\vartheta_{ab}\cos\vartheta_{ab})}{\sin^2\vartheta_{ab}}.
\label{eq:arc}
\end{equation}
The total exposure is minimized over the free vectors by L-BFGS-B~\cite{S-Byrd1995}, converged to $3\times10^{-5}$ for preparation and $1.1\times10^{-4}$ for the transfer $\ket{0_L}\to\ket{1_L}$. Neither end state of this transfer is the vacuum, so the lemma does not apply directly, but complex paths of either parity in the full Fock space reach the same optimum (Sec.~\ref{sec:vacuum}). With $\nop$ replaced by the identity, the method recovers the great circle of Eq.~(3) to $7\times10^{-6}$~rad, and the photon optima satisfy Eq.~\eqref{eq:EL} with a median relative residual of $4\times10^{-4}$ (preparation) and $7\times10^{-4}$ (transfer).

\subsection{\texorpdfstring{Lower Bounds}{Lower Bounds}}
\label{sec:bounds}
\begin{lemma}[minimal photon number at given fidelity]\label{lem:m}
Let $N(F)=\min\{\bar n:|\langle\psi_f|\psi\rangle|^2\geq F\}$ and $f_n=\langle n|\psi_f\rangle$. Then $N$ is non-decreasing, $N(F)=\sup_{\lambda>0}[F/S_1(\lambda)-\lambda]$ with $S_1(\lambda)=\sum_n|f_n|^2/(n+\lambda)$ for $F>|f_0|^2$, and $N(F)=0$ for $F\leq|f_0|^2$. For $f_0\neq0$ the supremum is attained at the $\lambda$ with $S_1(\lambda)^2/S_2(\lambda)=F$, $S_2(\lambda)=\sum_n|f_n|^2/(n+\lambda)^2$, and the minimum by $\ket{\psi_\lambda}\propto(\nop+\lambda)^{-1}\ket{\psi_f}$. For $f_0=0$ the same holds for $F>S_1(0)^2/S_2(0)$, and up to that value the supremum is the limit $\lambda\to0^+$, $N(F)=F/S_1(0)$, attained by $\sqrt{1-x}\,\ket0+\sqrt x\,\nop^{-1}\ket{\psi_f}/S_2(0)^{1/2}$ with $x=FS_2(0)/S_1(0)^2$.
\end{lemma}
\begin{proof}
For $\lambda>0$, Cauchy--Schwarz gives $|\langle\psi_f|\psi\rangle|^2\leq\sum_n\frac{|f_n|^2}{n+\lambda}\sum_n(n+\lambda)|\psi_n|^2=S_1(\lambda)(\bar n+\lambda)$, so $\bar n\geq F/S_1(\lambda)-\lambda$ for every admissible $\psi$, with equality iff $\psi\propto(\nop+\lambda)^{-1}\psi_f$. The fidelity of $\psi_\lambda$ is $S_1(\lambda)^2/S_2(\lambda)$, which for $f_0\neq0$ covers every $F\in(|f_0|^2,1)$. The vacuum has $\bar n=0$ and fidelity $|f_0|^2$. For $f_0=0$ the same inequality at $\lambda\to0^+$ gives $\bar n\geq F/S_1(0)$. For $F\leq S_1(0)^2/S_2(0)$ the superposition of the statement has fidelity $xS_1(0)^2/S_2(0)=F$ and $\bar n=xS_1(0)/S_2(0)=F/S_1(0)$, and for larger $F$ the argument for $f_0\neq0$ applies, since $S_1(\lambda)^2/S_2(\lambda)$, the fidelity of $\psi_\lambda$, covers $(S_1(0)^2/S_2(0),1)$ for $\lambda>0$.
\end{proof}

\begin{proposition}[lower bound]\label{prop:bound}
Every path from $\ket{\psi_i}$ to $\ket{\psi_f}$ traversed at $\Delta H=\omega$, with complex amplitudes and components of either parity, has $J\geq B=\int_0^\theta N(\cos^2s)\,ds$ [Eq.~(7)].
\end{proposition}
\begin{proof}
Parametrize the last $\theta$ of the path by the arclength $s\in[0,\theta]$ still to go ($L\geq\theta$). A state with $s\leq\pi/2$ still to go has $|\langle\psi_f|\psi\rangle|\geq\cos s$, so $\bar n\geq N(\cos^2s)$ by Lemma~\ref{lem:m}. The rest of the path contributes $\geq0$.
\end{proof}
For the transfer $\ket{0_L}\to\ket{1_L}$ each end bounds a different part of the path. With $N_0$ and $N_1$ the functions $N$ for the targets $\ket{0_L}$ and $\ket{1_L}$,
\begin{equation}
\begin{split}
J\geq B_2=\min_{L\geq\theta_L}\int_0^L\max\{&N_0(\cos^2\vartheta),\\
&N_1(\cos^2(L-\vartheta))\}\,d\vartheta
\end{split}
\label{eq:b2}
\end{equation}
($\cos^2$ set to 0 beyond $\pi/2$), minimal at $L=\theta_L$. This two-sided bound, 345.6~photon~ns, lies 3.8\% below the numerical optimum (359.2~photon~ns), whereas the one-sided bound (238.3~photon~ns) is loose.

\begin{proposition}[$\ket\chi$ a Fock state]\label{thm:fock}
If $\ket{\psi_f}=\cos\theta\ket0+\sin\theta\ket{k}$ with a Fock state $\ket k$, $k\geq1$, $0<\theta\leq\pi/2$, every path from $\ket0$ to $\ket{\psi_f}$ has $J\geq k(\theta/2-\sin2\theta/4)=J_{\rm geo}$, with equality only for the brachistochrone.
\end{proposition}
\begin{proof}
At fixed $|\psi_k|$, the overlap $|\cos\theta\,\psi_0+\sin\theta\,\psi_k|$ is largest when all remaining weight sits in $\ket0$, which costs no photons, so the minimizer of Lemma~\ref{lem:m} lies in $\mathrm{span}\{\ket0,\ket{k}\}$. It is $\psi=\cos\vartheta\ket0+\sin\vartheta\ket{k}$ with $\vartheta=\theta-\arccos\sqrt F$, a state of the brachistochrone, so $N(F)=k\sin^2(\theta-\arccos\sqrt F)$. Hence $B=\int_0^{\theta}k\sin^2(\theta-s)\,ds=J_{\rm geo}$, and equality in Proposition~\ref{prop:bound} requires the minimizer at every arclength and $L=\theta$, that is, the great circle.
\end{proof}
This covers Fock $\ket n$ ($\theta=\pi/2$) and the binomial-code state $(\ket0+\ket4)/\sqrt2$ ($k=4$, $\theta=\pi/4$), for which the brachistochrone is also the photon-optimal path, as the numerical optima confirm.

\subsection{\texorpdfstring{Transfers through the Vacuum}{Transfers through the Vacuum}}
\label{sec:vacuum}
The photon metric $\bar n\,d\vartheta$ of Sec.~\ref{sec:fermat} vanishes only at the vacuum. On the even sector $\bar n=\sum_nn|\psi_n|^2\geq2(1-|\psi_0|^2)$, with equality only for states in $\mathrm{span}\{\ket0,\ket2\}$. This single feature shapes every photon-optimal path below.
\begin{proposition}[photon-optimal transfers]\label{prop:vac}
Let $J^*(\varphi,\psi)$ be the least exposure of any path from $\ket\varphi$ to $\ket\psi$, as in Sec.~\ref{sec:fermat}. (i) For all $\varphi$, $\psi$, $J^*(\varphi,\psi)\leq J^*(0,\varphi)+J^*(0,\psi)$. (ii) If a minimizing path from $\ket\varphi$ to $\ket\psi$ passes through the vacuum, equality holds, and the path is the time-reversed photon-optimal preparation of $\ket\varphi$ followed by the photon-optimal preparation of $\ket\psi$. (iii) A minimizer is smooth wherever $\bar n>0$, so it can change its direction only at the vacuum.
\end{proposition}
\begin{proof}
(i) $J$ is additive, invariant under reparametrization and under reversal of the direction of travel, so the reversed minimizer from $\ket0$ to $\ket\varphi$ followed by the minimizer from $\ket0$ to $\ket\psi$ is a path from $\ket\varphi$ to $\ket\psi$ with exposure $J^*(0,\varphi)+J^*(0,\psi)$. (ii) Every sub-arc of a minimizer is a minimizer between its end points (otherwise replacing it would lower $J$), the principle of optimality~\cite{S-Bellman1957}. Applied to the two halves on either side of the vacuum, it gives the decomposition and the equality. (iii) Where $\bar n>0$ the photon metric is a Riemannian metric, conformal to the Fubini--Study metric, so a shortest path is a smooth geodesic there, in the real even sector a solution of Eq.~\eqref{eq:EL}. Equivalently, at a corner of a piecewise smooth minimizer the momentum $\bar n\,\psi'$ of Eq.~\eqref{eq:ray}, with $\psi'$ the unit tangent, is continuous in its component tangent to the sphere, the Weierstrass--Erdmann condition~\cite{S-GelfandFomin1963}. Where $\bar n>0$ this condition fixes the unit tangent, whereas at the vacuum the momentum vanishes on both sides for any pair of directions.
\end{proof}
\emph{The vacuum as a cone point.} On the great circle $\cos\vartheta\ket0+\sin\vartheta\ket e$ that leaves the vacuum along a direction $\ket e\perp\ket0$ of the even sector, $\bar n=n_e\sin^2\vartheta$, with $n_e=\langle e|\nop|e\rangle\geq2$. Near the vacuum, the states $\ket0+x\ket{e_1}+y\ket{e_2}$, with orthonormal directions $\ket{e_1}$ and $\ket{e_2}$ of equal photon number $n_e$, form a surface on which, with $z=x+iy$, the photon metric is $n_e|z|^2|dz|=|d(n_ez^3/3)|$ to leading order. This metric is flat, with the vacuum as the apex of a cone of total angle $6\pi$, so a shortest path between two directions that differ by at least $\pi/3$ runs exactly through the vacuum~\cite[Chap.~I.5]{S-BridsonHaefliger1999}. The model is exact to leading order for $\ket{e_1}=\ket2$ and $\ket{e_2}=i\ket2$, whereas the real even sector is anisotropic.

\emph{Numerical evidence.} That a minimizer runs through the vacuum is a numerical finding, which the cone model explains but does not prove. For the transfers of Supplementary Table~\ref{tab:vacuum} and for $\ket{0_L}\to\ket{1_L}$ at $\Delta=0.5$, every start we tried, including the brachistochrone and starts that avoid the vacuum, passes through the vacuum (weight $\geq1-2\times10^{-4}$) and, except one start $1.1\times10^{-3}$ above, attains the sum of the two photon-optimal preparations to $10^{-4}$ or better. Complex paths of either parity in the full Fock space reach the same optima to $4\times10^{-4}$ in the two cases tested, the preparation of $\ket{0_L}$ and $\ket{0_L}\to\ket{1_L}$ at $\Delta=0.4$.

\emph{At the vacuum the optimum does not turn.} Near the vacuum a real even state is $\ket0+x\ket2+\sum_{n\geq4}y_n\ket n$, and the linearization of Eq.~\eqref{eq:EL} has the regular solutions $y_n\propto x^{b_n}$ with $b_n=(\sqrt{1+4n}-1)/2>1$. A minimizer therefore generically reaches the vacuum tangent to Fock $\ket2$, the direction of lowest photon number. The two halves do so with opposite signs, so the path crosses the vacuum straight along the squeezing axis, un-squeezing the initial state and squeezing toward the target. Two brachistochrones through the vacuum, which leave it along $\ket\chi$, cost 410.4~photon~ns, 14\% above the optimum.

\begin{table*}[tbp]
\caption{Photon-optimal preparations of the logical states and transfers through the vacuum (exposure in photon~ns). The columns give the brachistochrone, the numerical optimum, the sum of the two photon-optimal preparations [Proposition~\ref{prop:vac}(ii)], the closed form, Eq.~(6), and its one-mode form, Eq.~\eqref{eq:closed1}, both applied to each half of a transfer [Eq.~\eqref{eq:transfer}], and the lower bound ($B$ of Eq.~(7), or $B_2$ of Eq.~\eqref{eq:b2}). Every optimum leaves or crosses the vacuum nearly tangent to $\ket2$. The row $\ket{0_L}\to\ket{+_L}$ illustrates a transfer that hardware applies as a frame update, since $\ket{+_L}=e^{-i\pi\nop/2}\ket{0_L}$ to within $5\times10^{-5}$ in fidelity.}
\label{tab:vacuum}
\begin{ruledtabular}
\begin{tabular}{lcccccc}
 & $J_{\rm geo}$ & optimum & sum & closed form & one-mode & bound\\\hline
preparation of $\ket{0_L}$ & 122.7 & 112.6 & -- & 112.7 & 114.3 & 111.3\\
preparation of $\ket{1_L}$ & 287.7 & 246.6 & -- & 246.9 & 253.1 & 238.3\\
$\ket{0_L}\to\ket{1_L}$, $\Delta=0.4$ & 739.5 & 359.2 & 359.2 & 359.6 & 367.4 & 345.6\\
$\ket{0_L}\to\ket{1_L}$, $\Delta=0.3$ & 1498.7 & 722.0 & 722.0 & 725.8 & 763.0 & --\\
$\ket{0_L}\to\ket{+_L}$ & 349.7 & 225.8 & 225.8 & 226.0 & 229.2 & --\\
cat $\gamma=2\to\ket{0_L}$ & 799.2 & 536.5 & 536.5 & 536.8 & 543.5 & --\\
\end{tabular}
\end{ruledtabular}
\end{table*}

\subsection{\texorpdfstring{Closed Form}{Closed Form}}
\label{sec:closed}
\emph{Closed form for preparation.} The photon-optimal path of preparation departs from the brachistochrone by at most 0.30~rad for every preparation we consider, so we expand $J$ to second order in this departure and minimize the expansion exactly, in nine steps. By Lemma~\ref{lem:reduce} we work in the real even sector with the conventions of Sec.~\ref{sec:fermat}, except that a nearby path is parametrized by the angle $\varphi$ along the brachistochrone rather than by its own arclength, so that $|\psi'|\neq1$ in general.

\emph{Step~1 (nearby paths).} Let $\psi_{\rm geo}(\varphi)=\cos\varphi\ket0+\sin\varphi\ket\chi$, $\varphi\in[0,\theta]$, be the brachistochrone to $\ket{\psi_f}$, with $\ket\chi=(\ket{\psi_f}-\cos\theta\ket0)/\sin\theta$. Its normal space is the same at every $\varphi$, $V=\mathrm{span}\{\ket0,\ket\chi\}^\perp$, with the projector $P$, and a nearby path is described by a departure $\delta(\varphi)\in V$ with $\delta(0)=\delta(\theta)=0$,
\begin{equation}
\psi(\varphi)=\cos|\delta|\,\psi_{\rm geo}(\varphi)+\sin|\delta|\,\frac{\delta}{|\delta|},\qquad \delta(\varphi)\in V,
\label{eq:cf-path}
\end{equation}
normalized for every $\delta$. Its exposure is $J=\int_0^\theta\bar n\,|\psi'|\,d\varphi$ with $\bar n=\psi^\top\nop\psi$.

\emph{Steps~2--4 (second-order functional).} To second order in $\delta$ and $\delta'$, with $\nop\ket0=0$, $P\nop\ket\chi=(\nop-n_\chi)\ket\chi\equiv\sigma_\chi\ket d$ ($\sigma_\chi^2$ the photon-number variance of $\ket\chi$, $\ket d\in V$ normalized), and $M=P\nop P$, the photon number is $\bar n=(1-|\delta|^2)\,n_\chi\sin^2\varphi+2\sigma_\chi\sin\varphi\,\langle d|\delta\rangle+\delta^\top M\delta$. Since $\psi=(1-\tfrac12|\delta|^2)\,\psi_{\rm geo}+\delta$ and $\psi_{\rm geo}'$, $\psi_{\rm geo}$, and $\delta'$ are mutually orthogonal, the speed is $|\psi'|=1-\tfrac12|\delta|^2+\tfrac12|\delta'|^2$. Their product, integrated with $J_{\rm geo}=\int_0^\theta n_\chi\sin^2\varphi\,d\varphi$, gives
\begin{equation}
\begin{split}
J={}&J_{\rm geo}+\int_0^\theta\Big[2\sigma_\chi\sin\varphi\,\langle d|\delta\rangle+\tfrac12n_\chi\sin^2\varphi\,|\delta'|^2\\
&+\delta^\top\big(M-\tfrac32n_\chi\sin^2\varphi\big)\delta\Big]d\varphi+O(\delta^3).
\end{split}
\label{eq:J2}
\end{equation}
\emph{Step~5 (Euler--Lagrange equation).} Stationarity under variations of $\delta$ that vanish at both ends gives $(\sin^2\varphi\,\delta')'=(2\sigma_\chi/n_\chi)\sin\varphi\ket d+(2/n_\chi)M\delta-3\sin^2\varphi\,\delta$. The substitution $u=\delta\sin\varphi$, with $(\sin^2\varphi\,\delta')'=(u''+u)\sin\varphi$, turns it into a vector P\"oschl--Teller problem with an inverse-square-sine potential,
\begin{equation}
u''+\Big(4-\frac{2M}{n_\chi\sin^2\varphi}\Big)u=\frac{2\sigma_\chi}{n_\chi}\ket d,\qquad u(0)=u(\theta)=0.
\label{eq:PT-vector}
\end{equation}
\begin{lemma}[convexity of the second-order problem]\label{lem:convex}
For $\theta\leq\pi/2$ the quadratic part of Eq.~\eqref{eq:J2} is positive definite on departures with $\delta(0)=\delta(\theta)=0$, so Eq.~\eqref{eq:PT-vector} has exactly one solution, and it is the unique minimizer of the second-order functional.
\end{lemma}
\begin{proof}
With $u=\delta\sin\varphi$, $\sin^2\varphi\,|\delta'|^2=|u'|^2-2\cot\varphi\,u^\top u'+\cot^2\varphi\,|u|^2$. Integrating the middle term by parts (the boundary term $|\delta|^2\sin\varphi\cos\varphi$ vanishes at both ends) turns it into $-\csc^2\varphi\,|u|^2$, so $\int_0^\theta[\tfrac12n_\chi\sin^2\varphi|\delta'|^2+\delta^\top(M-\tfrac32n_\chi\sin^2\varphi)\delta]\,d\varphi =\tfrac{n_\chi}2\int_0^\theta[|u'|^2-4|u|^2+\tfrac{2}{n_\chi}u^\top Mu/\sin^2\varphi]\,d\varphi$. Wirtinger's inequality for $u(0)=u(\theta)=0$, $\int_0^\theta|u'|^2\geq(\pi/\theta)^2\int_0^\theta|u|^2\geq4\int_0^\theta|u|^2$ for $\theta\leq\pi/2$, and $M\geq2$ on $V$ make it positive for every $u\neq0$.
\end{proof}
\emph{Step~6 (decoupling).} The operator $M$ is real and symmetric on $V$, with eigenvectors $M\ket{e_k}=n_k\ket{e_k}$ and $n_k\geq2$, since every vector in $V$ is even and orthogonal to $\ket0$. With $\sigma_\chi\langle e_k|d\rangle=\langle e_k|\nop|\chi\rangle$, each component $w_k=\langle e_k|u\rangle$ obeys
\begin{equation}
w_k''+\Big(4-\frac{\eta_k}{\sin^2\varphi}\Big)w_k=\frac{2\langle e_k|\nop|\chi\rangle}{n_\chi},\qquad\eta_k=\frac{2n_k}{n_\chi},
\label{eq:PT-scalar}
\end{equation}
with $w_k(0)=w_k(\theta)=0$.

\emph{Steps~7--9 (solution).} Since $(\sin^2\varphi)''+4\sin^2\varphi=2$, $A_k\sin^2\varphi$ with $A_k=2\langle e_k|\nop|\chi\rangle/[n_\chi(2-\eta_k)]$ is a particular solution. In the variable $y=\sin^2\varphi$, the homogeneous equation has near the vacuum the exponents $\nu$ with $\nu(\nu-1)=\eta_k$, of which only $\nu_k=\tfrac12+(\tfrac14+\eta_k)^{1/2}>1$ gives a solution regular at the vacuum, and $w=y^{\nu_k/2}f(y)$ turns it into the hypergeometric equation of Gauss. The regular homogeneous solution is therefore
\begin{equation}
\Phi_k(\varphi)=\sin^{\nu_k}\varphi\;{}_2F_1\big(\tfrac{\nu_k+2}2,\tfrac{\nu_k-2}2;\nu_k+\tfrac12;\sin^2\varphi\big)\simeq\varphi^{\nu_k}\quad(\varphi\to0),
\label{eq:cf-uk}
\end{equation}
which holds on all of $[0,\pi/2]$, because the series converges at $y=1$. The boundary condition $w_k(\theta)=0$ fixes
\begin{equation}
w_k(\varphi)=A_k\Big[\sin^2\varphi-\frac{\sin^2\theta\,\Phi_k(\varphi)}{\Phi_k(\theta)}\Big],
\label{eq:cf-wk}
\end{equation}
and $\delta=\sum_kw_k\ket{e_k}/\sin\varphi$ is Eq.~(6), the exact minimizer of the second-order functional (Lemma~\ref{lem:convex}), which we call the closed form. For $\eta_k\to2$ the bracket and $2-\eta_k$ vanish together and $w_k$ has a finite limit, and $\Phi_k>0$ on $(0,\theta]$ by Lemma~\ref{lem:convex}. For GKP $\ket{0_L}$ at $\Delta=0.4$ ($n_\chi=5.350$, $\sigma_\chi=3.116$), only the lowest of the 38 modes at the Fock cutoff 80 has $\eta_k<2$. Evaluated exactly, Eq.~(6) has the exposure 112.7~photon~ns (Supplementary Table~\ref{tab:vacuum}) and lasts 149~ns at $\Delta H=\omega$. Modes with $\eta_k<2$ have a large $\delta'$ near the vacuum, so the closed form leaves the vacuum $35^\circ$ from $\ket2$ for $\ket{0_L}$, whereas the exact minimizer leaves it along $\ket2$. The error this causes is weighted by $\bar n$, which vanishes quadratically there, and it is small (below).

\emph{One-mode form.} Keeping only the source direction, $\delta=\delta_d\ket d$, gives one term with $\eta=2n_d/n_\chi$, $n_d=\langle d|\nop|d\rangle$, $\nu=\tfrac12+(\tfrac14+\eta)^{1/2}$, $A=2\sigma_\chi/[n_\chi(2-\eta)]$, and the function $\Phi_\nu$ of Eq.~\eqref{eq:cf-uk} with $\nu_k$ replaced by $\nu$,
\begin{equation}
\delta_d=A\Big[\sin\varphi-\frac{\sin^2\theta\,\Phi_\nu(\varphi)}{\sin\varphi\,\Phi_\nu(\theta)}\Big],\qquad
\psi=\cos \delta_d\,\psi_{\rm geo}(\varphi)+\sin \delta_d\ket d,
\label{eq:closed1}
\end{equation}
which is simpler but less accurate.

\emph{Closed form for any transfer through the vacuum.} By Proposition~\ref{prop:vac}, the same expression approximates every photon-optimal transfer that runs through the vacuum,
\begin{equation}
\ket{\psi(\varphi)}=\cos|\delta|\big[\cos\varphi\ket0+\sin\varphi\ket{\chi}\big]+\sin|\delta|\,\delta/|\delta|,
\label{eq:transfer}
\end{equation}
traversed from $\varphi=\theta_i$ to 0 with the parameters of $\ket{\psi_i}$ and then from 0 to $\theta_f$ with those of $\ket{\psi_f}$. For $\ket{0_L}\to\ket{1_L}$ it gives 359.6~photon~ns in 396~ns.

\begin{figure}[tbp]
\includegraphics[width=0.72\textwidth,height=0.42\textheight,keepaspectratio]{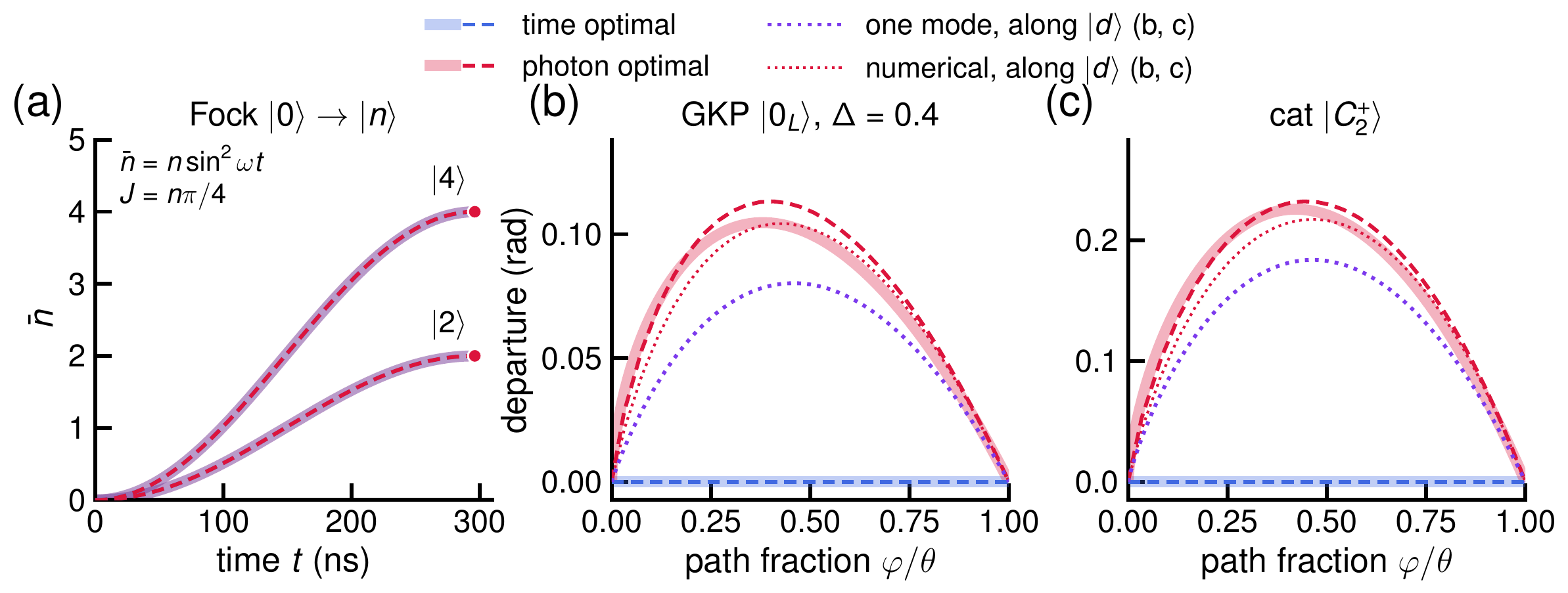}
\caption{Photon-optimal path (crimson) against the brachistochrone (blue). Bands show the analytic paths of Eqs.~(3) and (6) (minimizer of the second-order expansion of $J$, Eq.~\eqref{eq:J2}, in all transverse directions), and dashed lines the numerical optimum. (a)~Photon number at $\Delta H=\omega$ for Fock $\ket2$ and $\ket4$, where all three paths coincide. (b,~c)~Departure from the plane of the brachistochrone for GKP $\ket{0_L}$ ($\Delta=0.4$) and the even cat $\ket{C_2^+}$, against the path fraction $\varphi/\theta$, with $\varphi$ the angle along the brachistochrone from which the path departs, Eq.~(6). Dotted lines show the component of the numerical optimum along $\ket d$ (crimson) and the one-mode form along $\ket d$, Eq.~\eqref{eq:closed1} (purple).}
\label{fig:analytic}
\end{figure}

\emph{Accuracy of the closed forms.} The closed form reproduces the departure of the numerical optimum from the brachistochrone [Supplementary Fig.~\ref{fig:analytic}(b,~c)]. In exposure, it lies at most 0.14\% above the numerical optimum for every preparation we computed, except $\ket{1_L}$ at $\Delta=0.3$ (0.82\%), and at most 0.52\% above it for the transfers of Supplementary Table~\ref{tab:vacuum}, whereas the one-mode form lies up to 7.7\% above it. The numerical optima lie 1.2\% (GKP) and 2.6\% (cat) above the bound $B$.

\subsection{\texorpdfstring{State Gallery and the Price of Saving Photons}{State Gallery and the Price of Saving Photons}}
\label{sec:gallery}
\begin{table*}[tbp]
\caption{State gallery of Supplementary Fig.~\ref{fig:S01} (cutoff 80, $\Delta H=\omega$). The columns give the Bures angle $\theta$ of the brachistochrone ($T^*=\theta/\omega$), the mean $n_\chi$ and standard deviation $\sigma_\chi$ of $\nop$ in its direction $\ket\chi$, $n_d=\langle d|\nop|d\rangle$, and $\eta=2n_d/n_\chi$ of the one-mode form, Eq.~\eqref{eq:closed1}. The exposures of the brachistochrone ($J_{\rm geo}$) and of the numerical photon optimum ($J_{\rm ph}$) and the bound $B$ are in rad (multiply by 188.6 for photon~ns). The columns ``closed form'' and ``one-mode'' give the exposure of Eqs.~(6) and \eqref{eq:closed1} above $J_{\rm ph}$, and ``time'' and ``saving'' give $L_{\rm ph}/\theta-1$ and $1-J_{\rm ph}/J_{\rm geo}$. Entries ``--'' mark targets whose $\ket\chi$ is a Fock state, for which the two paths coincide (Proposition~\ref{thm:fock}).}
\label{tab:gallery}
\scriptsize
\setlength{\tabcolsep}{2.6pt}
\begin{ruledtabular}
\begin{tabular}{lcccccccccccc}
target & $\theta$ & $n_\chi$ & $\sigma_\chi$ & $n_d$ & $\eta$ & $J_{\rm geo}$ & $J_{\rm ph}$ & $B$ & closed form (\%) & one-mode (\%) & time (\%) & saving (\%)\\\hline
squeezed, $r=1$ & 0.6351 & 3.924 & 2.951 & 11.111 & 5.663 & 0.3091 & 0.2785 & 0.2742 & 0.078 & 1.7 & 5.3 & 9.9\\
Fock $\ket2$ & 1.5708 & 2.000 & 0.000 & -- & -- & 1.5708 & 1.5708 & 1.5708 & -- & -- & 0.0 & 0.0\\
Fock $\ket4$ & 1.5708 & 4.000 & 0.000 & -- & -- & 3.1416 & 3.1416 & 3.1416 & -- & -- & 0.0 & 0.0\\
Fock $\ket6$ & 1.5708 & 6.000 & 0.000 & -- & -- & 4.7124 & 4.7124 & 4.7124 & -- & -- & 0.0 & 0.0\\
$(\ket0+\ket4)/\sqrt2$ & 0.7854 & 4.000 & 0.000 & -- & -- & 0.5708 & 0.5708 & 0.5708 & -- & -- & 0.0 & 0.0\\
cat, $\gamma=1.5$ & 1.0966 & 2.780 & 1.203 & 4.499 & 3.237 & 0.9597 & 0.9100 & 0.9000 & 0.008 & 0.3 & 2.8 & 5.2\\
cat, $\gamma=2.5$ & 1.5086 & 6.274 & 2.474 & 7.380 & 2.352 & 4.5381 & 4.1254 & 3.9855 & 0.079 & 1.8 & 9.0 & 9.1\\
4-component cat, $\gamma=2$ & 1.2933 & 4.550 & 1.411 & 7.799 & 3.428 & 2.3432 & 2.2769 & 2.2589 & 0.000 & 0.0 & 1.2 & 2.8\\
GKP $\ket{0_L}$, $\Delta=0.3$ & 0.9569 & 7.566 & 5.284 & 19.391 & 5.126 & 1.8387 & 1.6298 & 1.5996 & 0.132 & 2.9 & 9.0 & 11.4\\
GKP $\ket{0_L}$, $\Delta=0.5$ & 0.5425 & 4.053 & 2.335 & 6.813 & 3.363 & 0.2034 & 0.1890 & 0.1872 & 0.038 & 0.6 & 4.9 & 7.0\\
GKP $\ket{+_L}$, $\Delta=0.4$ & 0.7444 & 5.321 & 3.122 & 9.768 & 3.672 & 0.6546 & 0.6003 & 0.5932 & 0.064 & 1.5 & 6.7 & 8.3\\
\end{tabular}
\end{ruledtabular}
\end{table*}

\begin{figure}[tbp]
\includegraphics[width=0.72\textwidth,height=0.42\textheight,keepaspectratio]{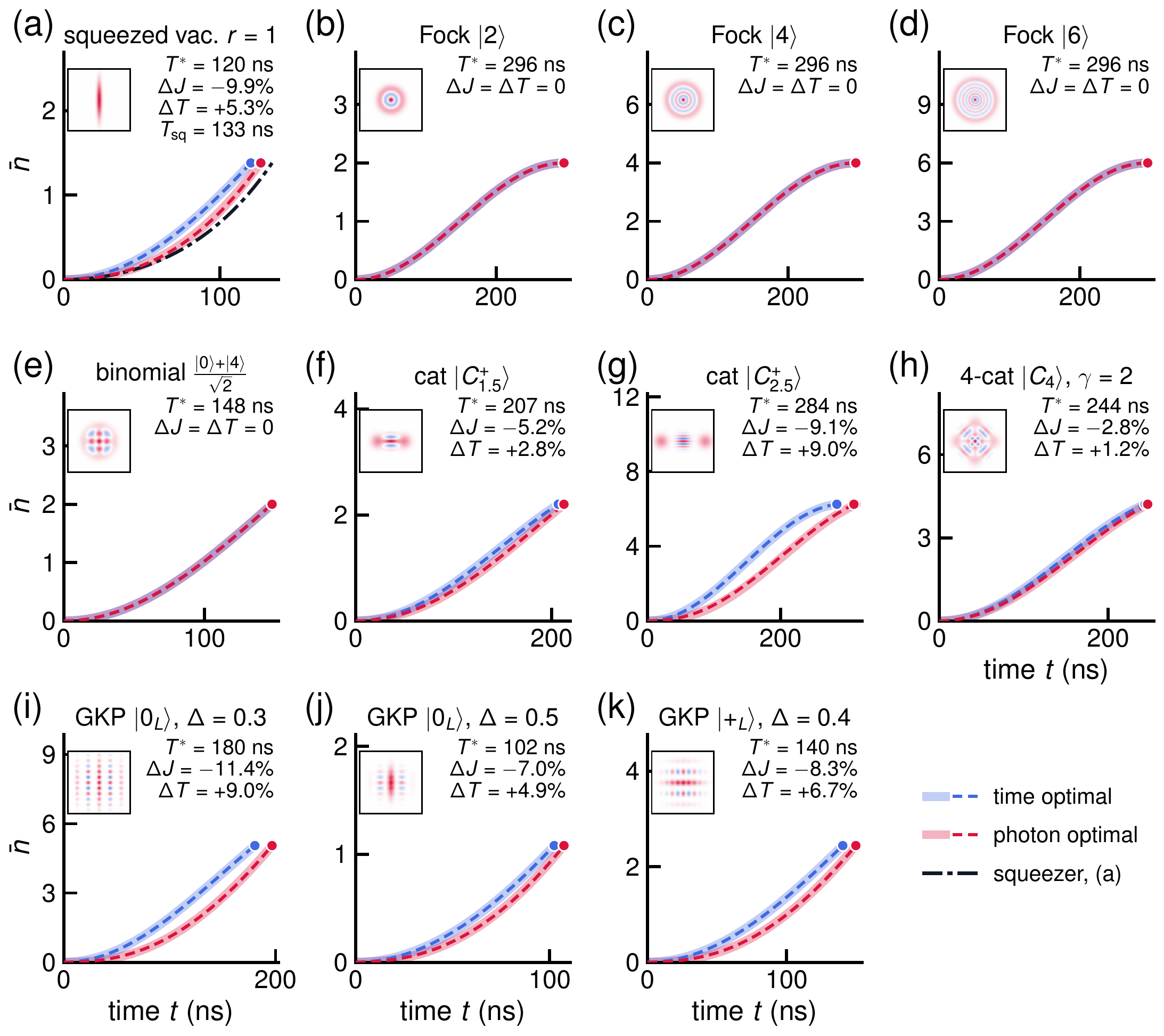}
\caption{Photon number along the time-optimal (blue) and photon-optimal (crimson) paths at $\Delta H=\omega$ for eleven targets prepared from the vacuum. Bands show the analytic paths of Eqs.~(3) and (6), dashed lines the numerical optimum, and the dash-dotted line in (a) the native squeezer. For Fock targets and the binomial-code state (b--e) the curves coincide. The numbers give $T^*$, the changes $\Delta J$ in exposure and $\Delta T$ in duration of the numerical optimum, and the duration $T_{\rm sq}$ of the squeezer. Insets show the Wigner functions of the targets.}
\label{fig:S01}
\end{figure}

\begin{figure}[tbp]
\includegraphics[width=0.72\textwidth,height=0.42\textheight,keepaspectratio]{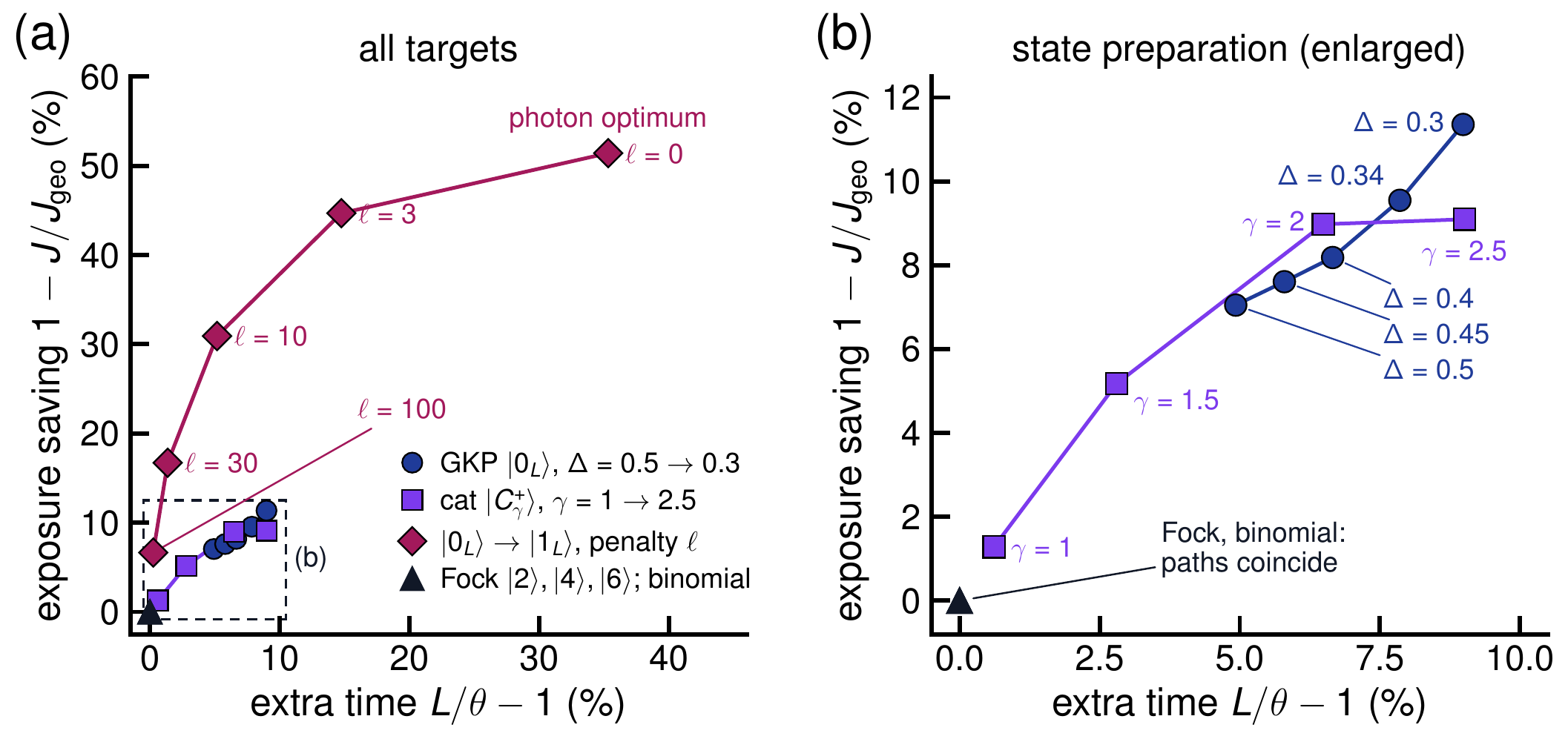}
\caption{Price of saving photons relative to the brachistochrone at $\Delta H=\omega$. The extra time $L/\theta-1$ of a photon-saving path of length $L$ is plotted against its exposure saving $1-J/J_{\rm geo}$, so the brachistochrone lies at the origin. Preparations are shown for GKP $\ket{0_L}$ ($\Delta=0.5$, 0.45, 0.4, 0.34, 0.3), cats $\ket{C_\gamma^+}$ ($\gamma=1$, 1.5, 2, 2.5), Fock $\ket2$, $\ket4$, $\ket6$, and the binomial-code state. The transfer $\ket{0_L}\to\ket{1_L}$ is shown by the minimizers of $J_\ell$, from $\ell=\infty$ to $\ell=0$. (a)~All targets. (b)~The preparations, enlarged (dashed box in a).}
\label{fig:S03}
\end{figure}

Supplementary Figure~\ref{fig:S01} and Supplementary Table~\ref{tab:gallery} compare the two optimal paths for eleven targets prepared from the vacuum, and Supplementary Fig.~\ref{fig:S03} plots the extra duration of each photon-saving path against the exposure it saves. For Fock states and the binomial-code state the two paths coincide (Proposition~\ref{thm:fock}). For the other targets the photon-optimal path detours through states with fewer photons, and within each family the price grows with the photon number of the target, from 0.6\% in time for 1.3\% in exposure (cat, $\gamma=1$) to 9.0\% for 11.4\% (GKP, $\Delta=0.3$). For GKP $\ket{0_L}$ at $\Delta=0.4$ it is 6.7\% for 8.2\% (149 instead of 140~ns). The largest saving among the preparations we computed is 25.7\%, for $\ket{1_L}$ at $\Delta=0.3$ (278 instead of 251~ns). Between the logical states, the minimizers of $J_\ell=\int\langle\nop+\ell(1-\Pi)\rangle\,d\vartheta$, with $\Pi$ the projector on $\mathcal C$, span the range from the brachistochrone ($\ell\to\infty$) to the photon optimum ($\ell=0$), which takes 35\% longer and has 51\% less exposure (Supplementary Table~\ref{tab:route}).

\subsection{\texorpdfstring{Dependence on the Envelope Width}{Dependence on the Envelope Width}}
\label{sec:delta}
Unless stated otherwise, all results of the main text are for $\Delta=0.4$. Supplementary Table~\ref{tab:delta} gives the quantities of the main text for $\Delta=0.3$--0.5. Smaller values of $\Delta$, which approach the ideal grid and hold more photons, lengthen the preparation brachistochrone from 102 to 180~ns and raise the photon saving of the photon-optimal preparation from 7.0 to 11.4\%, whereas the flip brachistochrone stays near $\pi/(2\omega)=296$~ns and the overlap of the code words, with the defect of the flip, vanishes rapidly. The photon-optimal transfer between the logical states saves 41--53\% of the exposure at every $\Delta$. The tangential capacity of conditional squeezing at the vacuum, $v=g|\langle2|\chi\rangle|/\sqrt2$, falls from $0.64\,\omega$ at $\Delta=0.5$ to $0.38\,\omega$ at $\Delta=0.3$, so a direct push along the preparation brachistochrone needs at least $T^*\omega/v=159$~ns at $\Delta=0.5$ but 478~ns at $\Delta=0.3$, longer than the 412~ns of Fig.~4, with the crossover near $\Delta=0.33$. Section~\ref{app:kerr} tests this prediction with the benchmark of Fig.~4 at $\Delta=0.3$ and 0.5.

\section{\texorpdfstring{Noise and the Time--Photon Trade-Off}{Noise and the Time--Photon Trade-Off}}
\label{app:noise}

In this section, we specify the master equation and its rates, and use the first-order error of Eq.~(8) to compare the time-optimal, photon-optimal, and intermediate paths. Saving photons lowers the error when photon loss and cavity dephasing dominate it, and the brachistochrone is better when ancilla errors, which grow with the duration, dominate.

\emph{Master equation.} Every noisy evolution solves the Lindblad equation~\cite{S-Lindblad1976}
\begin{equation}
\dot\rho=-i[H,\rho]+\kappa\mathcal D[\aop]\rho+\kappa_\phi\mathcal D[\nop]\rho+\gamma_1\mathcal D[\sigma_-]\rho
+\tfrac{\gamma_\phi}{2}\mathcal D[\sigma_z]\rho ,
\label{eq:lindblad}
\end{equation}
with $\mathcal D[A]\rho=A\rho A^\dagger-\tfrac12\{A^\dagger A,\rho\}$, $\kappa=1/T_{1c}$, $\kappa_\phi=2/T_{\phi c}$, $\gamma_1=1/T_{1q}$, and $\gamma_\phi=1/T_{\phi q}$, where $T_{1c}$ and $T_{\phi c}$ are the energy-relaxation and pure-dephasing times of the cavity and $T_{1q}$ and $T_{\phi q}$ those of the ancilla (Supplementary Table~\ref{tab:params}).

\emph{Rates and their sources.} The measured rates are those of Table~S1 of Ref.~\cite{S-Sivak2023RealTime}, with $T_{\phi c}$ from its cavity coherence time $T_{2c}=980~\mu$s and $T_{\phi q}=414~\mu$s from its echo time $T_{2E}=238~\mu$s. The shorter Ramsey time, $T_{2R}=62~\mu$s, reflects low-frequency noise, which changes the closed-evolution fidelity of the stored pulses by less than $10^{-4}$ when modeled as a static detuning, so the Markovian Ramsey value $T_{\phi q}=70~\mu$s is a pessimistic bracket. The rates of the conditional-squeezing proposal~\cite{S-DelGrosso2025} are lifetimes of $200~\mu$s for resonator and qubit and a qubit dephasing time of $10~\mu$s. As in the proposal we set cavity dephasing to zero, and we neglect its thermal population.

\emph{Self-Kerr.} The coupling to its ancilla gives the cavity of Ref.~\cite{S-Sivak2023RealTime} the self-Kerr $K=-2\pi\times4.8$~Hz (Table~S1 there, where it enters as $\tfrac12K(\ad\aop)^2$). We add $H_K=\tfrac K2\hat a^{\dagger2}\aop^2$, which differs from that form by the detuning $\tfrac K2\ad\aop$ that a calibrated frame absorbs, to $H$ of Eq.~\eqref{eq:lindblad} for every gate set, in every pulse optimization and evaluation of the benchmark. The second-order dispersive shift of the same device, $\chi'=2\pi\times5.8$~Hz, is of the same size and is not modeled. A static $H_K$ is coherent and dissipates nothing, so it does not enter $\epsilon_1$ of Eq.~(8). Along a path that the control follows exactly, an uncompensated $H_K$ lowers the fidelity at second order in $K$, and the optimization absorbs a modeled $H_K$ (Sec.~\ref{app:kerr}). The Kerr term grows with the square of the dispersive coupling, so gate sets that couple more strongly than this device carry larger Kerr terms, of order $2\pi\times3$~kHz for SNAP with $\chi_d=g$, 26 times the dispersive shift $\chi/2\pi=46.5$~kHz of this device, and about $2\pi\times0.43$~MHz for the circuit of Ref.~\cite{S-DelGrosso2025} (Sec.~\ref{app:robust}).

\begin{table*}[tbp]
\caption{Physical parameters. [LIT] marks a value from a publication, [DEF] a definition of this work, and [ASSUMPTION] a modeling choice. Frequencies ($g$, $\omega$, $\Omega_{\max}$, $\chi_d$, $K$) are angular, and decay rates are inverse lifetimes.}
\label{tab:params}
\small
\begin{ruledtabular}
\begin{tabular}{llp{5.0cm}p{5.2cm}}
symbol & value & role & source\\\hline
$g$ & $7.5~\mu\mathrm{s}^{-1}$ & bound on every oscillator generator and rate of the conditional squeeze $r=gt$ ($r=1.5$ in 200~ns) & [LIT] proposal of Ref.~\cite{S-DelGrosso2025}\\
$\omega$ & $g/\sqrt2=5.303~\mu\mathrm{s}^{-1}$ & energy uncertainty $\Delta H$ of every analytic path, $T=L/\omega$ & [DEF] Fubini--Study speed of each generator on $\ket0$\\
$\Omega_{\max}$ & $10g=75~\mu\mathrm{s}^{-1}$ & qubit-drive bound ($\pi$ rotation in 41.9~ns) & [ASSUMPTION]\\
$\chi_d$ & $g$ (and $4g$) & dispersive shift of SNAP--displacement control & [ASSUMPTION]\\
$T_{1c},T_{\phi c},T_{1q}$ & 606~$\mu$s, 5.1~ms, 280~$\mu$s & Lindblad evolutions at measured rates & [LIT] Table~S1 of Ref.~\cite{S-Sivak2023RealTime}\\
$T_{\phi q}$ & 414~$\mu$s (70~$\mu$s) & ancilla dephasing (Markovian Ramsey bracket in parentheses) & [LIT] from $T_{2E}=238~\mu$s ($T_{2R}=62~\mu$s) and $T_{1q}$ of Ref.~\cite{S-Sivak2023RealTime}\\
$T_{1c},T_{1q},T_{\phi q}$ & 200, 200, 10~$\mu$s & noise of the conditional-squeezing proposal (Supplementary Table~\ref{tab:proposal}) & [LIT] Ref.~\cite{S-DelGrosso2025}\\
$\kappa$, $\kappa_\phi$, $\gamma_1$, $\gamma_\phi$ & from the times above & rates of Eq.~\eqref{eq:lindblad}, $\kappa=1/T_{1c}$, $\kappa_\phi=2/T_{\phi c}$, $\gamma_1=1/T_{1q}$, $\gamma_\phi=1/T_{\phi q}$ & [DEF] $\kappa_\phi$ such that $\rho_{n,n+1}$ decays at $1/T_{\phi c}$\\
$\Gamma_q$ & $1/(2T_{\phi q})+1/(4T_{1q})$ & ancilla error rate in $\epsilon_1$, Eq.~(8), equal to $(476~\mu\mathrm{s})^{-1}$ for $T_{\phi q}=414~\mu$s & [DEF] equatorial ancilla state\\
$K$ & $-2\pi\times4.8$~Hz & static self-Kerr $\tfrac K2\hat a^{\dagger2}\aop^2$ of the oscillator, in every optimization and evaluation of the benchmark & [LIT] Table~S1 of Ref.~\cite{S-Sivak2023RealTime}\\
$\Delta$ & 0.4 & GKP envelope & [ASSUMPTION], with paths for 0.3--0.5 in Supplementary Tables~\ref{tab:gallery} and \ref{tab:delta}, and the benchmark at 0.3 and 0.5 in Sec.~\ref{app:kerr}\\
\end{tabular}
\end{ruledtabular}
\end{table*}

\begin{figure}[tbp]
\includegraphics[width=0.72\textwidth,height=0.42\textheight,keepaspectratio]{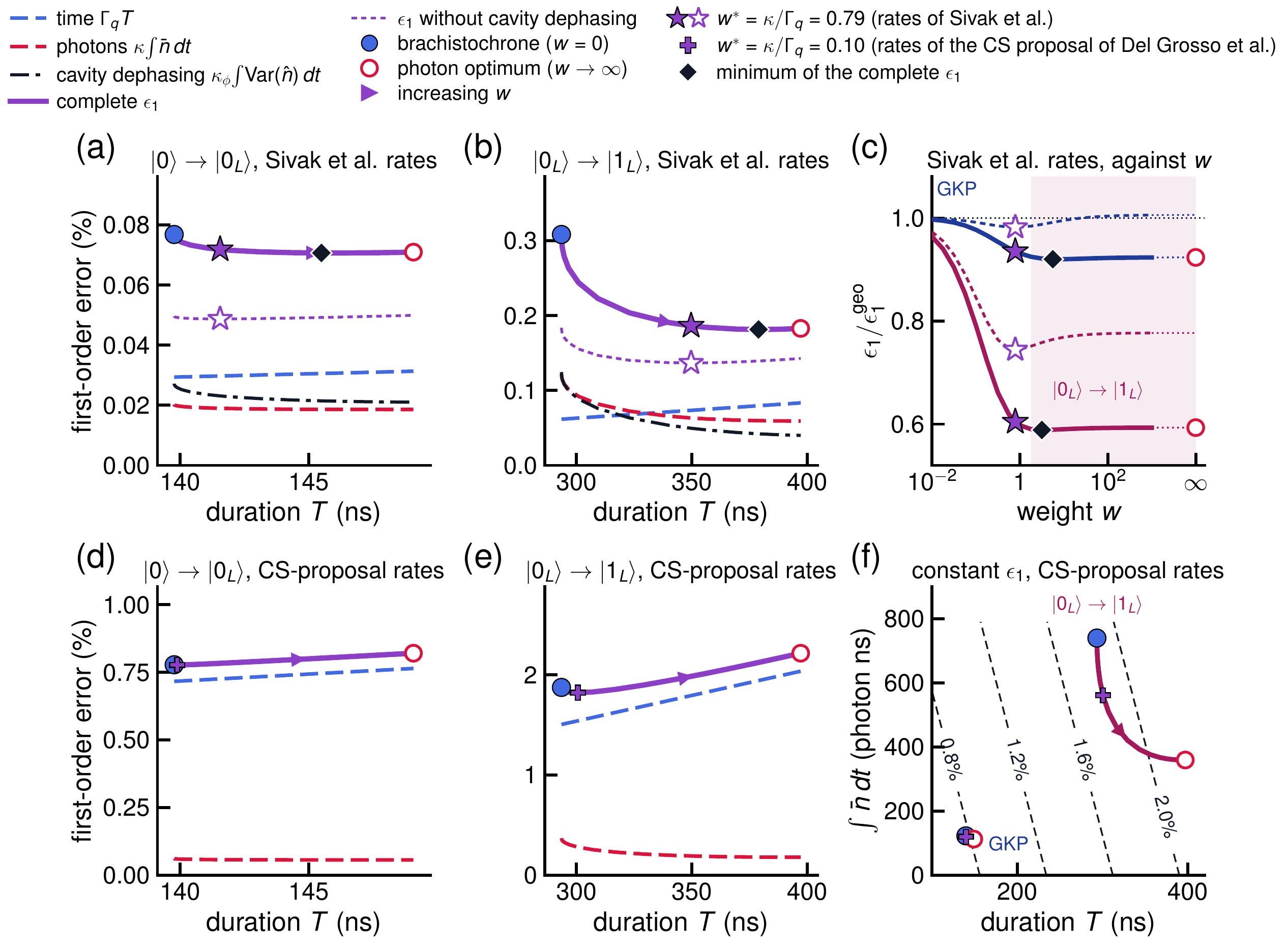}
\caption{First-order error $\epsilon_1$, Eq.~(8), along the trade-off family from the brachistochrone ($w=0$, filled circle) to the photon-optimal path ($w\to\infty$, open circle) at $\Delta H=\omega$, with arrowheads toward increasing $w$. (a--c)~Rates measured by Sivak et al.~\cite{S-Sivak2023RealTime}. (a,~b)~The three terms of $\epsilon_1$, the complete $\epsilon_1$ (solid), and $\epsilon_1$ without cavity dephasing (dotted) against $T$, for GKP preparation and $\ket{0_L}\to\ket{1_L}$. (c)~Both against $w$, relative to their brachistochrone values, with shading within 1\% of the minimum of the complete $\epsilon_1$ of $\ket{0_L}\to\ket{1_L}$, stars at $w^*=\kappa/\Gamma_q=0.79$, and diamonds at the minimum. (d--f)~Rates of the conditional-squeezing proposal of Del Grosso et al.~\cite{S-DelGrosso2025}, without cavity dephasing. (d,~e)~As (a,~b). (f)~The two families in the plane of duration and $\int\bar n\,dt$, with lines of constant $\epsilon_1$ and plus signs at $w^*=0.10$.}
\label{fig:budget}
\end{figure}

\emph{Cavity dephasing and ancilla errors.} The dissipator $\kappa_\phi\mathcal D[\nop]$ is fixed by the measured decay of $\rho_{01}$ and damps $\rho_{nm}$ at $\kappa_\phi(n-m)^2/2$. If the cavity dephasing arises from thermal excitations of the ancilla~\cite{S-Sivak2023RealTime}, the damping of $\rho_{nm}$ is at most $(n-m)^2$ times that of $\rho_{01}$, so our cavity-dephasing terms are upper estimates. Without them, the photon-saving paths lower $\epsilon_1$ of the transfer by 22--26\% instead of about 40\% [Supplementary Fig.~\ref{fig:budget}(c)]. For an equatorial ancilla the ancilla dissipators give $\Gamma_q=(476~\mu\mathrm{s})^{-1}$.

\emph{Trade-off family.} The minimizers of $\int(1+w\bar n)\,d\vartheta$ form a family from the brachistochrone ($w=0$) to the photon-optimal path ($w\to\infty$). Without cavity dephasing, the member $w^*=\kappa/\Gamma_q$ minimizes $\epsilon_1=(\Gamma_q/\omega)\int(1+\kappa\bar n/\Gamma_q)\,d\vartheta$ exactly, with $w^*=0.786$ at the measured rates and 0.098 at those of the proposal. With the measured rates and cavity dephasing [Supplementary Fig.~\ref{fig:budget}(a--c)], $\epsilon_1$ of the transfer $\ket{0_L}\to\ket{1_L}$ is minimal at $w\approx3.2$ ($T=379$~ns, 41\% below the brachistochrone), and for preparation the photon-optimal path gains 7.6\%. At the rates of the proposal [Supplementary Fig.~\ref{fig:budget}(d--f)], time dominates, and saving photons gains at most 3\% for the transfer and 0.1\% for preparation. These statements assume Markovian noise and a fidelity measure. Since a photon loss acts on a GKP state mainly as a small, correctable displacement, whereas an ancilla fault applies the wrong conditional operation, a decoded figure of merit would move the optimum toward the brachistochrone.

\begin{figure}[tbp]
\includegraphics[width=0.72\textwidth,height=0.42\textheight,keepaspectratio]{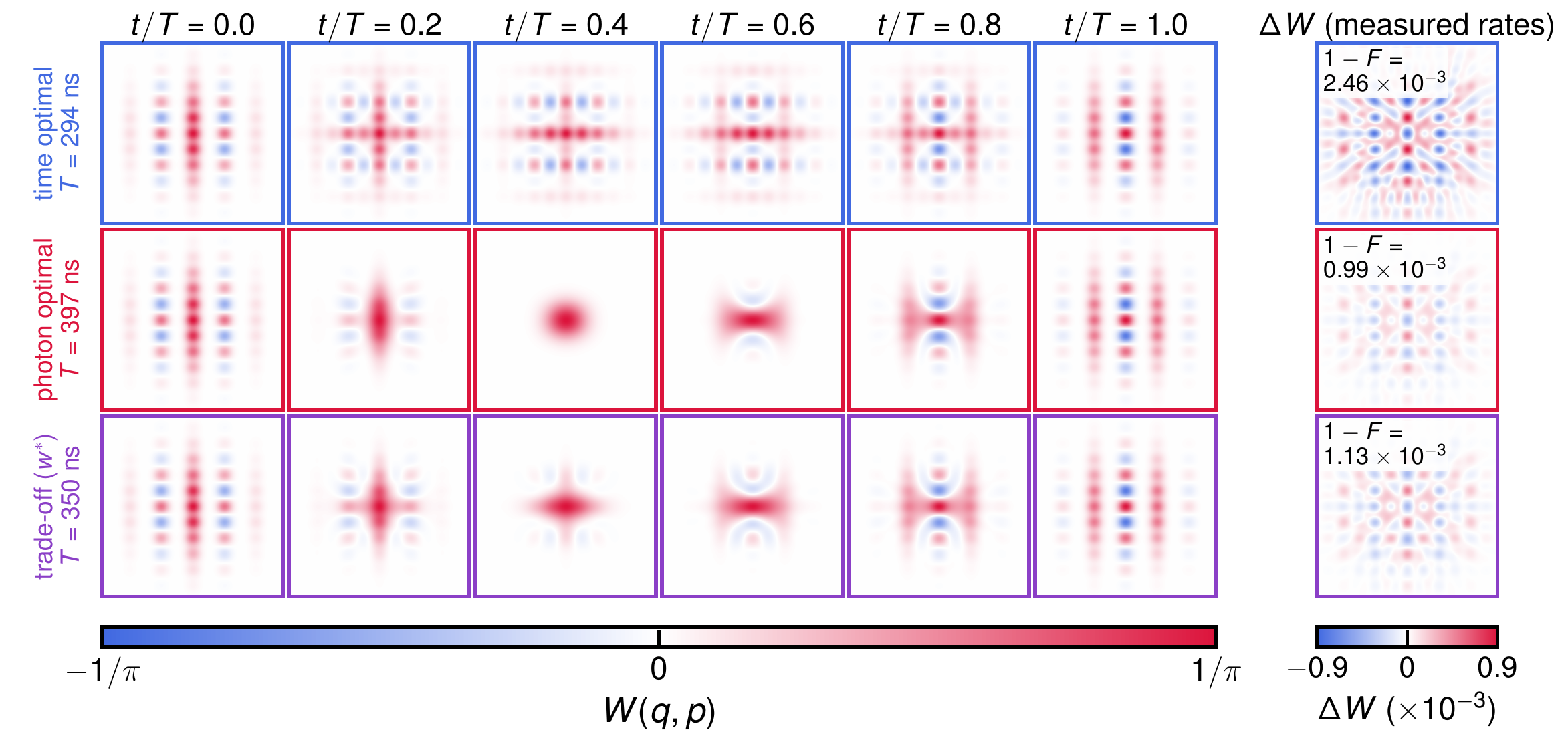}
\caption{Wigner functions along the time-optimal (logical flip), photon-optimal and trade-off paths between the logical states at $t/T=0,0.2,\dots,1$, and (last column) the change $\Delta W$ of the final state when each path is generated under photon loss and cavity dephasing, with the resulting error $1-F$.}
\label{fig:S04}
\end{figure}

\emph{The states along the three paths.} Supplementary Figure~\ref{fig:S04} shows the oscillator along the three optimal paths between $\ket{0_L}$ and $\ket{1_L}$. The brachistochrone, the logical flip, passes through $\ket{+_L}$ at $t=T/2$, the photon-optimal path dissolves the grid into the vacuum near $t=0.4\,T$ and rebuilds it, and the trade-off path passes through a squeezed state with few photons. Generated by their rank-two Hamiltonians under photon loss and cavity dephasing, they end with errors of $2.5$, $1.0$, and $1.1\times10^{-3}$, as Eq.~(8) predicts.

\section{\texorpdfstring{Logical Operations and State-to-State Transfers}{Logical Operations and State-to-State Transfers}}
\label{app:logical}

In this section, we separate a logical operation, which must act on every encoded state, from a state-to-state transfer of one known state. We prove that, among evolutions that keep the qubit encoded, the flip brachistochrone is time optimal and, averaged over the code space, photon and first-order error optimal, and we construct an exact flip that briefly leaves the code space and lowers the average exposure.

\subsection{\texorpdfstring{Why a Logical Flip}{Why a Logical Flip}}

On the ideal GKP code the logical Paulis are displacements~\cite{S-Gottesman2001}, applied directly or tracked in software~\cite{S-Fluhmann2019,S-CampagneIbarcq2020,S-Sivak2023RealTime}. At finite energy a displacement also moves the envelope, which stabilization restores~\cite{S-Royer2020,S-Sellem2025}, and gates designed for the ideal code leave the finite-energy code space~\cite{S-Rojkov2024}. Optimized logical gates have been demonstrated on cat codes~\cite{S-Heeres2017} and GKP qubits~\cite{S-Matsos2025}. We do not propose our paths as a replacement for displacements or frame updates. We study the flip as the simplest operation between two logical states of similar photon number ($\bar n=2.44$ and 2.87).

\subsection{\texorpdfstring{Operations within the Code Space and the Gate Metric}{Operations within the Code Space and the Gate Metric}}
\label{sec:codepres}

Let $\Pi$ be the projector on $\mathcal C$ and $X_L$ the swap of the finite-energy logical states. A protocol implements the logical flip if it sends every encoded state $\ket\psi$ to $X_Le^{-i\varphi Z_L/2}\ket\psi$ for some frame update $\varphi$, and the state-to-state transfer if it only maps $\ket{0_L}$ to $\ket{1_L}$. An oscillator evolution $U(t)$ keeps the qubit encoded if $U(t)\mathcal C=\mathcal C$ for all $t$. We score a logical operation by its gate fidelity with $X_L$, the average fidelity over the six cardinal states $\ket{0_L}$, $\ket{1_L}$, $\ket{\pm_L}\propto\ket{0_L}\pm\ket{1_L}$, and $\ket{\pm i_L}\propto\ket{0_L}\pm i\ket{1_L}$, the eigenstates of the logical Pauli operators, after the best virtual $Z$ rotation, so the identity scores 1/3. This frame coincides with $e^{-i\varphi Z_L/2}$ only up to $O(c)$, so the flip brachistochrone scores 0.9997 rather than 1.

\begin{proposition}[logical operations within the code space]\label{thm:code}
Let $U(t)$ keep the qubit encoded. For every function $f$ of the state, the average over code states drawn uniformly (Haar measure $d\psi$, the uniform measure on the Bloch sphere of $\mathcal C$), $\langle f\rangle_{\mathcal C}(t)=\int f(U(t)\ket\psi)\,d\psi$, is independent of $t$. Consequently, for an operation of duration $T$ that keeps the qubit encoded, the first-order error of Lemma~\ref{lem:loss}, averaged over $\mathcal C$, is
\begin{equation}
\bar\epsilon_1=T\big(\Gamma_q+\kappa\,\bar n_{\mathcal C}+\kappa_\phi V_{\mathcal C}\big),
\label{eq:eps-code}
\end{equation}
with $\bar n_{\mathcal C}=\tfrac12\mathrm{Tr}(\Pi\nop)$ and $V_{\mathcal C}=\langle\mathrm{Var}\,\nop\rangle_{\mathcal C}=\tfrac12\mathrm{Tr}(\Pi\nop^2)-\tfrac16\{(\mathrm{Tr}\,\Pi\nop)^2+\mathrm{Tr}[(\Pi\nop\Pi)^2]\}$.
\end{proposition}
\begin{proof}
For each $t$, $U(t)$ restricted to $\mathcal C$ is a unitary on $\mathcal C$, and the Haar measure on $\mathcal C$ is invariant under it, so $\int f(U(t)\psi)\,d\psi=\int f(\psi)\,d\psi$. Code states are parity even, so $\langle\aop\rangle=0$ and the integrands of Lemma~\ref{lem:loss} are $\bar n$ and $\mathrm{Var}\,\nop$. The Haar moments of a qubit, $\int\langle\psi|A|\psi\rangle\,d\psi=\tfrac12\mathrm{Tr}A$ and $\int\langle\psi|A|\psi\rangle^2d\psi=\tfrac16[(\mathrm{Tr}A)^2+\mathrm{Tr}A^2]$ for $A$ on $\mathcal C$, give $\bar n_{\mathcal C}$ and $V_{\mathcal C}$, and the ancilla term does not depend on the state. Hence $1-\bar\epsilon_1$ is, to first order, the average gate fidelity of the noisy operation with its noiseless version $U(T)$ on $\mathcal C$, the standard figure of merit of a gate~\cite{S-Nielsen2002}.
\end{proof}

\begin{corollary}[the time-optimal logical flip]\label{cor:flip}
Every logical flip whose Hamiltonian has $\Delta H\leq\omega$ in the state that starts as $\ket{0_L}$ takes $T\geq T^*=\theta_L/\omega$, because it transfers $\ket{0_L}$ to $\ket{1_L}$. The flip brachistochrone, Eq.~\eqref{eq:H01-main}, keeps the qubit encoded (Sec.~\ref{app:derivation}), has $\Delta H\leq\omega$ on every code state and attains $T^*$. By Proposition~\ref{thm:code}, it therefore minimizes the duration and, averaged over the code space, the time-integrated photon number $\bar n_{\mathcal C}T$ and the first-order error among all logical flips that keep the qubit encoded and have $\Delta H\leq\omega$ in the state that starts as $\ket{0_L}$.
\end{corollary}
Averaged over the code space, the error of an operation that keeps the qubit encoded is thus proportional to its duration, so the fastest such operation is also the most photon and error efficient. The corollary excludes displacements and every pulse benchmarked here, which leave the code space, and operations that leave the code space and return can do better (Sec.~\ref{sec:exactflip}). For $\Delta=0.4$, $\bar n_{\mathcal C}=2.658$ and $V_{\mathcal C}=9.60$, so the flip brachistochrone has the averaged exposure $\int\bar n\,dt=\bar n_{\mathcal C}T^*=780$~photon~ns and $\bar\epsilon_1=3.0\times10^{-3}$ at the measured rates.

\subsection{\texorpdfstring{The Subsystem Picture}{The Subsystem Picture}}

In the stabilizer subsystem decomposition of GKP codes~\cite{S-Shaw2024}, an alternative to the modular one~\cite{S-Pantaleoni2020}, tracing out the stabilizer subsystem amounts to an ideal decoding. Logical action in this sense and keeping the qubit encoded are distinct notions, since a $\sqrt\pi$ displacement acts logically but moves the envelope out of $\mathcal C$. They agree on the distinction that matters here. Along the flip brachistochrone the stabilizers keep their code values, $\langle S_X\rangle\in[0.589,0.623]$ and $\langle S_Z\rangle\in[0.589,0.606]$, whereas the photon-optimal path drives both to $e^{-\pi}=0.043$ at the vacuum and so dissolves the stabilizer structure.

\subsection{\texorpdfstring{Leaving the Code Space}{Leaving the Code Space}}

Supplementary Table~\ref{tab:route} applies the rank-two Hamiltonians that carry $\ket{0_L}$ along each path of the family $J_\ell$ (Sec.~\ref{sec:gallery}) to $\ket{1_L}$. Along the photon-optimal path, $\ket{0_L}$ leaves the code space by up to 37\%, and the same Hamiltonians map $\ket{1_L}$ to $\ket{0_L}$ with fidelity 0.60 only, so they perform a state-to-state transfer, not a gate. A unitary preserves overlaps, so a gate that carries $\ket{0_L}$ to the vacuum leaves the image of $\ket{1_L}$ a vacuum weight of only $c^2=2\times10^{-4}$. Only one encoded state can pass through the vacuum at a time, and the halving of the exposure belongs to a single known input.

\subsection{\texorpdfstring{An Exact Flip That Leaves the Code Space}{An Exact Flip That Leaves the Code Space}}
\label{sec:exactflip}

A gate can nonetheless lower the averages over the code space by tilting the whole code space toward fewer photons and back. Let $U(t)=e^{\lambda\sin(\pi t/T)\mathcal A}e^{-i\Theta(t)Y_L}$, with tilt amplitude $\lambda$, in-plane angle $\Theta(t)$ from 0 to $\theta_L$, and the anti-Hermitian, parity-preserving tilt generator $\mathcal A=[\Pi,\nop]/\|[\Pi,\nop]\|$, which couples $\mathcal C$ only to its complement. The tilt vanishes at both ends, so $U(T)=X_LZ_L$ on $\mathcal C$ exactly, and the schedule $\dot\Theta^2=\omega^2-(\lambda\pi/T)^2\cos^2(\pi t/T)$ keeps $\Delta H\leq\omega$ on every encoded state. The averaged photon number falls with the tilt, initially at the positive rate $\tfrac12\|[\Pi,\nop]\|_{\rm HS}^2/\|[\Pi,\nop]\|=5.45$, so the flip brachistochrone ($\lambda=0$) is not a local minimizer of the averaged exposure among exact flips. At $\lambda=0.3$ the flip lasts 320~ns ($1.09\,T^*$), keeps at most 7\% of the average weight outside $\mathcal C$, and has an averaged exposure of 602~photon~ns ($-23\%$) and $\bar\epsilon_1=2.29\times10^{-3}$ ($-24\%$, or $-13\%$ without cavity dephasing), whose oscillator part a Lindblad simulation of the six cardinal states reproduces to within 0.3\%. Like $\omega Y_L$, $\mathcal A$ is a reference, not a hardware Hamiltonian, and the family is not optimized. In general, the averaged exposure and error of a gate depend only on its duration and on the path of the code plane $U(t)\mathcal C$, so the rotation within the plane costs time but no photons.

\begin{table}[tbp]
\caption{The Hamiltonians that carry $\ket{0_L}$ along a path, applied to $\ket{1_L}$. The columns give the extra time $L/\theta_L-1$ of the path, the fidelity of $\ket{1_L}\to\ket{0_L}$, the peak weight outside the code space of each input, and the exposure (photon~ns) of each input and their mean.}
\label{tab:route}
\begin{ruledtabular}
\begin{tabular}{lccccccc}
path of $\ket{0_L}$ & time & $F_{1\to0}$ & leak$_0$ & leak$_1$ & $J_0$ & $J_1$ & mean\\\hline
brachistochrone & 0 & 0.9992 & 0 & 0 & 739 & 822 & 781\\
$\ell=100$ & 0.3\% & 0.9976 & 0.1\% & 0.2\% & 690 & 803 & 747\\
$\ell=30$ & 1.4\% & 0.9870 & 0.9\% & 1.2\% & 616 & 779 & 698\\
$\ell=10$ & 5.2\% & 0.9436 & 4.6\% & 5.6\% & 511 & 761 & 636\\
$\ell=3$ & 14.8\% & 0.8258 & 15\% & 17\% & 409 & 790 & 600\\
photon optimal & 35.3\% & 0.5990 & 37\% & 40\% & 359 & 965 & 662\\
trade-off, $w^*$ & 19.1\% & 0.7546 & 22\% & 24\% & 383 & 823 & 603\\
\end{tabular}
\end{ruledtabular}
\end{table}

\subsection{\texorpdfstring{A Generator Aligned with the Logical Flip}{A Generator Aligned with the Logical Flip}}

\begin{figure}[tbp]
\includegraphics[width=0.72\textwidth,height=0.42\textheight,keepaspectratio]{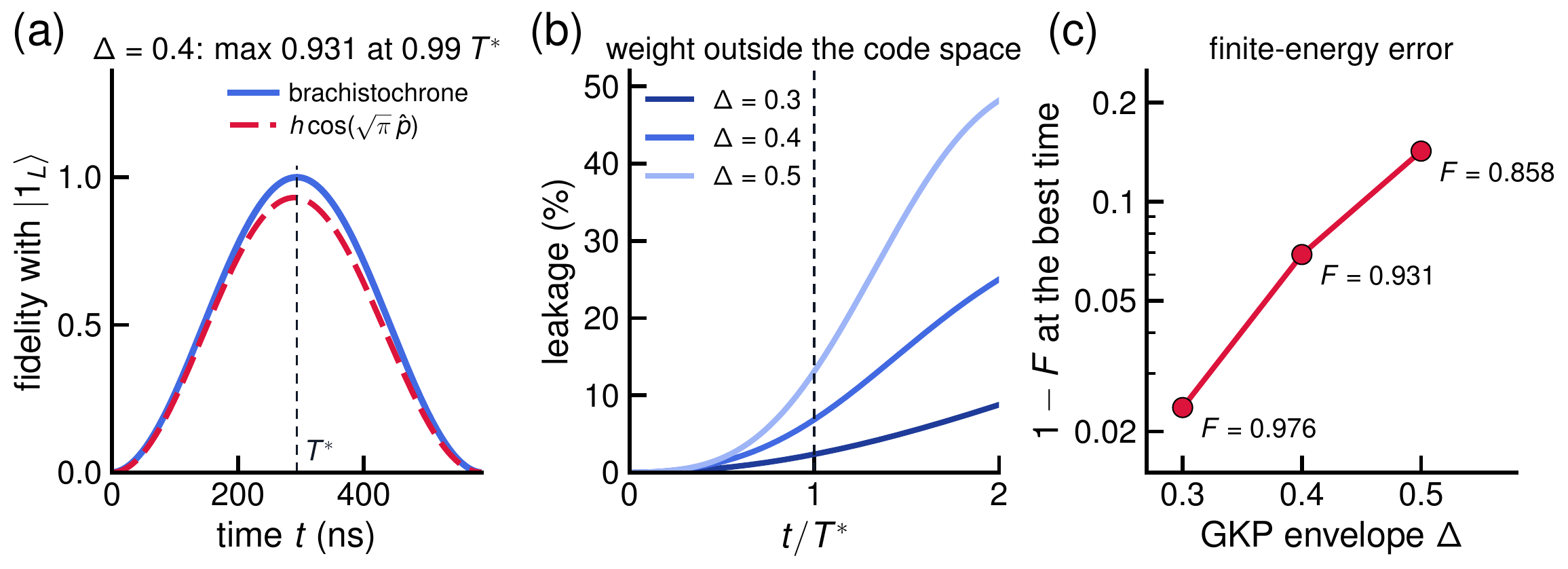}
\caption{The periodic potential $h\cos(\sqrt\pi\hat p)$ as a generator of the flip. Panels show (a)~the fidelity with $\ket{1_L}$ against time ($\Delta=0.4$) with the brachistochrone for comparison, (b)~the weight outside the code space, and (c)~the infidelity at the best time against $\Delta$.}
\label{fig:S06}
\end{figure}

On the ideal code, the periodic potential $h\cos(\sqrt\pi\hat p)$ acts as $hX_L$~\cite{S-Brooks2013}, a rotation about the logical $X$ axis rather than $Y$. At finite energy, with $h$ set by $\Delta H=\omega$ on $\ket{0_L}$, it swaps the logical states with a best fidelity of only 0.931 for $\Delta=0.4$, at $0.99\,T^*$, which falls from 0.976 to 0.858 as $\Delta$ grows from 0.3 to 0.5 (Supplementary Fig.~\ref{fig:S06}). The finite-energy envelope, not the speed, limits this generator.

The photon-optimal path is thus the right reference when the input is known. When it is not, the brachistochrone is the fastest flip and, among flips that keep the qubit encoded, the most photon and error efficient.

\section{\texorpdfstring{Conditional Squeezing and the Even-Order Operator Hierarchy}{Conditional Squeezing and the Even-Order Operator Hierarchy}}
\label{app:cs}

We choose conditional squeezing as the gate family for state preparation because its quadratic structure naturally preserves photon-number parity while providing a compact set of operations for coherently reshaping the oscillator state. The generators change the photon number only by two, making the resulting gate sequence compatible with the even-parity structure of the GKP states considered here. In the benchmarks, the conditional-squeezing drive is optimized to follow the unrestricted quantum brachistochrone, which serves as a time-optimal reference (Fig.~4, Table~I, and Sec.~\ref{app:numerics}), and, as a check, for the target-state fidelity alone (Supplementary Table~\ref{tab:capability}). In this section, we derive how such a sequence, whose native interaction is only quadratic, generates the higher-order, even-photon operators that the non-Gaussian code states require. Throughout, $\hbar=1$ and all angles are in radians.

The symbols $j$, $G_j$, $\zeta_j$, $\beta_j$, $\phi_j$, $\vartheta$, $\theta$, $\alpha$, $M$, $m$, and $z$ are local to this section. Here $j$ labels a gate (not a qubit branch, as in Sec.~\ref{app:capacity}), $\zeta_j$ is a control amplitude (not a lattice point $\zeta_{m,l}$), $\beta_j$ a gate strength (not the lattice vector $\beta$), $\vartheta$ a squeezing phase (not the arclength), $\theta$ and $\alpha$ a qubit rotation angle and axis (not the Bures angle and the lattice vector), $M$ the number of gates (not $P\nop P$), $m$ a power (not the lattice index), and $z$ the argument of the squeezing operator $S$ (not a coordinate of the plane of Sec.~\ref{app:capacity}). The squeezing amplitude $\xi$ is that of $H_{\rm CS}$ (a rate), whereas $z$ is dimensionless.

\subsection{Conventions}
The bosonic mode has annihilation operator $\aop$, $[\aop,\ad]=1$, number operator $\nop=\ad\aop$ and Fock states $\ket n$, $n=0,1,2,\dots$, with $\ad\ket n=\sqrt{n+1}\,\ket{n+1}$ and $\aop\ket n=\sqrt n\,\ket{n-1}$. The quadratures are $\hat q=(\aop+\ad)/\sqrt2$ and $\hat p=(\aop-\ad)/(i\sqrt2)$, so that $[\hat q,\hat p]=i$. In numerics the Fock space is truncated at dimension $D$ (states $\ket0,\dots,\ket{D-1}$).

The ancilla qubit has basis states $\ket{\mathrm g},\ket{\mathrm e}$ and Pauli operators $\sigma_{x,y,z}$ with $\sigma_z\ket{\mathrm g}=+\ket{\mathrm g}$ and $\sigma_z\ket{\mathrm e}=-\ket{\mathrm e}$. The projectors onto the $\sigma_z$ eigenstates are $P_\pm=(I\pm\sigma_z)/2$, that is, $P_+=\ket{\mathrm g}\bra{\mathrm g}$ and $P_-=\ket{\mathrm e}\bra{\mathrm e}$. A qubit rotation by angle $\theta$ about the equatorial axis at azimuth $\alpha$ is
\begin{align}
  R(\theta,\alpha) & =\exp\!\Bigl[-\tfrac{i\theta}{2}\bigl(\cos\alpha\,\sigma_x+\sin\alpha\,\sigma_y\bigr)\Bigr], \\
  R_y(\theta)      & \equiv R(\theta,\pi/2)=e^{-i\theta\sigma_y/2}.
\end{align}
Operators acting on the oscillator and on the qubit commute with each other. In particular every oscillator operator commutes with $\sigma_{x,y,z}$. Throughout, $[X,Y]=XY-YX$, $\operatorname{ad}_X Y\equiv[X,Y]$, and $\operatorname{ad}_X^{k}Y=[X,\operatorname{ad}_X^{k-1}Y]$ with $\operatorname{ad}_X^{0}Y=Y$. Here $M$ is the number of conditional squeezing gates, labelled by $j=1,\dots,M$, and $m$ denotes a power in $G^m$ (see the hierarchy of even-order terms below).

\subsection{\texorpdfstring{The Conditional Squeezing Gate}{The Conditional Squeezing Gate}}
The native interaction is a qubit-controlled squeezing, which we call a conditional squeezing gate. The qubit state determines the sense (sign) of a quadratic squeezing generator acting on the oscillator. For the $j$-th gate,
\begin{align}
  U_j     & =e^{G_j\sigma_z},\label{eq:Uj}                                \\
  G_j     & =\zeta_j\,\hat a^{\dagger2}-\zeta_j^{*}\,\aop^{2},\label{eq:Gj} \\
  \zeta_j & =\beta_j\,e^{i\phi_j}.\label{eq:zeta}
\end{align}
Here $\zeta_j\in\mathbb C$ is the complex control amplitude of the gate, $\beta_j=|\zeta_j|\ge0$ is its magnitude, a dimensionless squeezing strength (interaction strength times duration), and $\phi_j=\arg\zeta_j\in(-\pi,\pi]$ is its phase. The phase fixes the orientation of the squeezing axis in phase space. Changing $\phi_j$ rotates the squeezed quadrature without changing the strength. Written out, $G_j=\beta_j\bigl(e^{i\phi_j}\hat a^{\dagger2}-e^{-i\phi_j}\aop^{2}\bigr)$. Since $G_j^\dagger=\zeta_j^*\aop^2-\zeta_j\hat a^{\dagger2}=-G_j$, the operator $G_j$ is anti-Hermitian and $U_j$ is unitary. The gate is the conditional-squeezing Hamiltonian $H_{\rm CS}$ of Eq.~(10) with opposite branch amplitudes, $\xi_{\rm g}=-\xi_{\rm e}\equiv\xi$, held for a time $\Delta t$. Since $\ket{\mathrm g}\bra{\mathrm g}-\ket{\mathrm e}\bra{\mathrm e}=\sigma_z$, one has $e^{-iH_{\rm CS}\Delta t}=e^{G\sigma_z}$ with $\zeta=-\xi\Delta t/2$, and the bound $|\xi|\leq g$ gives $\beta\leq g\Delta t/2$. Using $\sigma_z=P_+-P_-$, $P_\pm^2=P_\pm$ and $P_+P_-=0$,
\begin{align}
  U_j & =P_+\,e^{G_j}+P_-\,e^{-G_j},
  \label{eq:branch}
\end{align}
so the oscillator undergoes the transformation $e^{G_j}$ if the qubit is in $\ket{\mathrm g}$ and the inverse transformation $e^{-G_j}$ if it is in $\ket{\mathrm e}$. With the squeezing operator
\begin{align}
  S(z) & =\exp\!\Bigl[\tfrac12\bigl(z^*\aop^2-z\hat a^{\dagger2}\bigr)\Bigr],\qquad z\in\mathbb C,
\end{align}
where $z=r e^{i\vartheta}$ has squeezing magnitude $r=|z|$ and phase $\vartheta=\arg z$, comparison of exponents gives $e^{\pm G_j}=S(\mp2\zeta_j)$. Hence the squeezing magnitude is $r=2\beta_j$, and the squeezing phase is $\vartheta=\phi_j$ (for $\ket{\mathrm e}$) or $\phi_j+\pi$ (for $\ket{\mathrm g}$). The two qubit branches squeeze the same quadrature in opposite senses. [The branch $\ket{\mathrm g}$ of $H_{\rm CS}$ held for $\Delta t$ is $S(\xi_{\rm g}\Delta t)$, so $r=|\xi_{\rm g}|\Delta t$, the conditional squeeze $r=gt$ of Supplementary Table~\ref{tab:params}.] For a qubit in $c_{\rm g}\ket{\mathrm g}+c_{\rm e}\ket{\mathrm e}$,
\begin{align}
  U_j\bigl[\ket\psi\otimes(c_{\rm g}\ket{\mathrm g}+c_{\rm e}\ket{\mathrm e})\bigr]
   & =c_{\rm g}\,e^{G_j}\ket\psi\otimes\ket{\mathrm g}+c_{\rm e}\,e^{-G_j}\ket\psi\otimes\ket{\mathrm e},
\end{align}
so a single conditional squeezing gate entangles the qubit with the oscillator whenever the qubit is not in a $\sigma_z$ eigenstate and the oscillator state is changed by the squeezing.

\subsection{\texorpdfstring{Commutator Algebra}{Commutator Algebra}}
The three operators
\begin{align}
  K_+ & =\tfrac12\,\hat a^{\dagger2},\qquad K_-=\tfrac12\,\aop^{2},\qquad K_0=\tfrac12\bigl(\nop+\tfrac12\bigr)
\end{align}
satisfy $[K_0,K_\pm]=\pm K_\pm$ and $[K_-,K_+]=2K_0$, that is, they close under commutation and form an $\mathrm{su}(1,1)$ algebra. Every generator is a linear combination, $G_j=2(\zeta_jK_+-\zeta_j^*K_-)$, hence all $G_j$ lie in this algebra. They change the photon number by two, $\hat a^{\dagger2}\ket n=\sqrt{(n+1)(n+2)}\,\ket{n+2}$ and $\aop^{2}\ket n=\sqrt{n(n-1)}\,\ket{n-2}$.

The commutators below are derived from three elementary identities,
\begin{align}
  [\aop^2,\hat a^{\dagger2}] & =4\nop+2,\label{eq:elem} \\
  [\nop,\hat a^{\dagger2}]   & =2\hat a^{\dagger2},\nonumber \\
  [\nop,\aop^{2}]            & =-2\aop^{2}.\nonumber
\end{align}
The first follows from $[\aop^2,\hat a^{\dagger2}]=\aop[\aop,\hat a^{\dagger2}]+[\aop,\hat a^{\dagger2}]\aop$ and $[\aop,\hat a^{\dagger2}]=2\ad$, which give $2\aop\ad+2\ad\aop=4\nop+2$. Let $G_j,G_l$ be two generators with parameters $(\beta_j,\phi_j)$ and $(\beta_l,\phi_l)$. Since $[\hat a^{\dagger2},\hat a^{\dagger2}]=[\aop^2,\aop^2]=0$, only the cross terms survive,
\begin{align}
  [G_j,G_l]
   & =-\zeta_j\zeta_l^*[\hat a^{\dagger2},\aop^{2}]-\zeta_j^*\zeta_l[\aop^{2},\hat a^{\dagger2}]\nonumber \\
   & =\bigl(\zeta_j\zeta_l^*-\zeta_j^*\zeta_l\bigr)[\aop^2,\hat a^{\dagger2}]\nonumber                 \\
   & =2i\,\mathrm{Im}\bigl(\zeta_j\zeta_l^*\bigr)\,(4\nop+2).
\end{align}
Because $\zeta_j\zeta_l^*=\beta_j\beta_l\,e^{i(\phi_j-\phi_l)}$, one has $\mathrm{Im}(\zeta_j\zeta_l^*)=\beta_j\beta_l\sin(\phi_j-\phi_l)$ and therefore
\begin{align}
  [G_j,G_l] & =8i\,\beta_j\beta_l\sin(\phi_j-\phi_l)\bigl(\nop+\tfrac12\bigr).
  \label{eq:GjGl}
\end{align}
The factor $\sin(\phi_j-\phi_l)$ arises because only the imaginary part of $\zeta_j\zeta_l^*$ (the antisymmetric combination $\zeta_j\zeta_l^*-\zeta_j^*\zeta_l$) survives in the commutator. It depends only on the relative orientation of the two squeezing axes. Two generators with $\phi_j-\phi_l=0$ or $\pi$ (the same quadrature, $\zeta_l$ a real multiple of $\zeta_j$) commute, and the commutator is largest, $|\sin(\phi_j-\phi_l)|=1$, for $\phi_j-\phi_l=\pm\pi/2$, that is, for squeezing axes $\pi/4$ apart. Noncommutativity of conditional squeezings therefore requires at least two different phases. Similarly, using Eq.~\eqref{eq:elem},
\begin{align}
  [G_j,\nop] & =\zeta_j[\hat a^{\dagger2},\nop]-\zeta_j^*[\aop^2,\nop]=-2\zeta_j\hat a^{\dagger2}-2\zeta_j^*\aop^{2}.
  \label{eq:Gn}
\end{align}

Two generators of particular use are
\begin{align}
  G_A & =\tfrac12\bigl(\aop^2-\hat a^{\dagger2}\bigr),\qquad G_B=\tfrac{i}{2}\bigl(\aop^2+\hat a^{\dagger2}\bigr).
\end{align}
They are the special cases of $G$ with $\zeta=-\tfrac12$ ($\beta=\tfrac12$, $\phi=\pi$) and $\zeta=\tfrac i2$ ($\beta=\tfrac12$, $\phi=\pi/2$), respectively. Their phases differ by $\pi/2$, so Eq.~\eqref{eq:GjGl} gives the maximal value
\begin{align}
  [G_A,G_B] & =8i\cdot\tfrac14\cdot\sin\tfrac\pi2\,\bigl(\nop+\tfrac12\bigr)=2i\bigl(\nop+\tfrac12\bigr).
\end{align}
In quadrature form, $G_A=\tfrac i2(\hat q\hat p+\hat p\hat q)$ and $G_B=\tfrac i2(\hat q^2-\hat p^2)$. They are the Hermitian generators of the tangential capacity up to a factor, $K_A=iG_A$ and $K_B=-iG_B$ (Sec.~\ref{app:capacity}).

\subsection{\texorpdfstring{Gaussian Character and the Role of the Qubit Rotations}{Gaussian Character and the Role of the Qubit Rotations}}
A Gaussian unitary is one generated by a Hamiltonian at most quadratic in $\hat q,\hat p$. It maps Gaussian states to Gaussian states. Each $e^{\pm G_j}$ is of this type. Without qubit rotations all $U_j$ are diagonal in the $\sigma_z$ basis, so a qubit initially in $\ket{\mathrm g}$ remains in $\ket{\mathrm g}$ and the oscillator receives a product of squeezings, which is a Gaussian unitary. Applied to the vacuum it yields a Gaussian state, whereas a finite-energy GKP state is non-Gaussian and cannot be reached in this way. This result is a special case of Proposition~\ref{thm:nogo}. (For an initial superposition, each qubit branch is Gaussian, but the qubit--oscillator state is a coherent sum of two branches and the oscillator alone is generally mixed.)

Non-Gaussianity therefore requires qubit rotations between the conditional squeezing gates. Rotations change the $\sigma_z$-basis content of the qubit,
\begin{align}
  R_y(\theta)\ket{\mathrm g} & =\cos\tfrac\theta2\,\ket{\mathrm g}+\sin\tfrac\theta2\,\ket{\mathrm e},  \\
  R_y(\theta)\ket{\mathrm e} & =-\sin\tfrac\theta2\,\ket{\mathrm g}+\cos\tfrac\theta2\,\ket{\mathrm e},
\end{align}
so successive conditional squeezings act on coherent superpositions of earlier conditional histories. Equivalently, since $G$ commutes with all qubit operators,
\begin{align}
  R_y(\theta)\,e^{G\sigma_z}R_y^\dagger(\theta) & =\exp\!\bigl[G\,(\cos\theta\,\sigma_z+\sin\theta\,\sigma_x)\bigr],
\end{align}
which shows that the rotation changes the qubit axis along which the next squeezing is conditioned. For $M$ conditional squeezing gates and $M-1$ rotations $R_y(\theta_j)$ the full sequence, in time order from right to left, is
\begin{align}
  U_{\mathrm{tot}} & =U_M\,R_y(\theta_{M-1})\,U_{M-1}\cdots R_y(\theta_1)\,U_1 .
  \label{eq:fullsequence}
\end{align}
Moving all rotations to the left with $e^{X}R=R\,e^{R^\dagger XR}$ gives
\begin{align}
  U_{\mathrm{tot}} & =Q_{M-1}\,\prod_{j=M}^{1}e^{\tilde G_j},\qquad Q_j=R_y(\theta_j)\cdots R_y(\theta_1),\qquad Q_0=I,\qquad
  \tilde G_j=G_j\,Q_{j-1}^\dagger\sigma_zQ_{j-1},
\end{align}
where the product is ordered with $j=M$ leftmost. Each $\tilde G_j$ has a quadratic oscillator part $G_j$ but a qubit part $Q_{j-1}^\dagger\sigma_zQ_{j-1}$ that points along a different direction for different $j$. The noncommutativity of these dressed generators generates higher-order bosonic operators.

\subsection{\texorpdfstring{One Rotation Followed by One Conditional Squeezing Gate}{One Rotation Followed by One Conditional Squeezing Gate}}
Consider $U=e^{\mathcal X}e^{\mathcal Y}$ with $\mathcal X=G\sigma_z$ (a single conditional squeezing gate, $G=\zeta\hat a^{\dagger2}-\zeta^*\aop^2$) and $\mathcal Y=-\tfrac{i\theta}{2}\sigma_y$ (a rotation $R_y(\theta)$). The Baker--Campbell--Hausdorff (BCH) expansion reads
\begin{align}
  \log(e^{\mathcal X}e^{\mathcal Y}) & =\mathcal X+\mathcal Y+\tfrac12[\mathcal X,\mathcal Y]
  +\tfrac1{12}[\mathcal X,[\mathcal X,\mathcal Y]]+\tfrac1{12}[\mathcal Y,[\mathcal Y,\mathcal X]]
  -\tfrac1{24}[\mathcal Y,[\mathcal X,[\mathcal X,\mathcal Y]]]+\cdots .
\end{align}
Since $G$ commutes with the Pauli matrices and the Pauli matrices anticommute, so that $[\sigma_z,\sigma_y]=-2i\sigma_x$ and $[\sigma_z,\sigma_x]=2i\sigma_y$, repeated commutation with $\mathcal X$ produces powers of $G$ with alternating qubit operators,
\begin{align}
  [\mathcal X,\mathcal Y]                    & =-\theta\,G\sigma_x,\qquad
  [\mathcal X,[\mathcal X,\mathcal Y]]         =-2i\theta\,G^2\sigma_y,\nonumber   \\
  \operatorname{ad}_{\mathcal X}^3\mathcal Y & =-4\theta\,G^3\sigma_x,\qquad
  \operatorname{ad}_{\mathcal X}^4\mathcal Y   =-8i\theta\,G^4\sigma_y.
  \label{eq:nested}
\end{align}
In general,
\begin{align}
  \operatorname{ad}_{\mathcal X}^k\mathcal Y & =
  \begin{cases}
    -2^{k-1}\theta\,G^k\sigma_x,  & k\ \text{odd},  \\
    -2^{k-1}i\theta\,G^k\sigma_y, & k\ \text{even}.
  \end{cases}
\end{align}
Terms with an outer $[\mathcal Y,\cdot\,]$ do not raise the oscillator order. For example $[\mathcal Y,[\mathcal X,[\mathcal X,\mathcal Y]]]=0$ because $\mathcal Y\propto\sigma_y$, and $[\mathcal Y,[\mathcal Y,\mathcal X]]=-\theta^2G\sigma_z$.

\subsection{\texorpdfstring{Photon-Order Content of \texorpdfstring{$G^m$}{G\textasciicircum m}}{Photon-Order Content of G\textasciicircum m}}
Using $\hat a^{\dagger2}\aop^2=\nop(\nop-1)$, $\aop^2\hat a^{\dagger2}=(\nop+1)(\nop+2)$ and $h(\nop)\,\aop^{2}=\aop^{2}h(\nop-2)$ for any function $h$, one finds
\begin{align}
  G^2 & =\zeta^2\hat a^{\dagger4}+(\zeta^*)^2\aop^4-2|\zeta|^2(\nop^2+\nop+1),\nonumber                         \\[4pt]
  G^3 & =\zeta^3\hat a^{\dagger6}-\zeta^2\zeta^*\hat a^{\dagger2}(3\nop^2+9\nop+14)
        +\zeta(\zeta^*)^2\aop^{2}(3\nop^2-3\nop+8)-(\zeta^*)^3\aop^6,\nonumber                                     \\[4pt]
  G^4 & =\zeta^4\hat a^{\dagger8}-4\zeta^3\zeta^*\hat a^{\dagger4}(\nop^2+5\nop+11)
        +|\zeta|^4(6\nop^4+12\nop^3+58\nop^2+52\nop+28)\nonumber                                                  \\
      & \quad-4\zeta(\zeta^*)^3\aop^4(\nop^2-3\nop+7)+(\zeta^*)^4\aop^8 .
\end{align}
In general, for a positive integer $m$,
\begin{align}
  G^m & =\zeta^m\,\hat a^{\dagger2m}+(-1)^m(\zeta^*)^m\,\aop^{2m}+\mathcal R_m,
  \label{eq:Gm}
\end{align}
where $\mathcal R_m$ collects all terms of lower photon order, including the number-conserving ones. The leading photon-number-changing order of $G^m$ is therefore
\begin{equation}
  \delta n_m=2m=2,4,6,8,\ldots
\end{equation}
Combining Eqs.~\eqref{eq:nested} and \eqref{eq:Gm}, the leading oscillator terms of $\operatorname{ad}_{\mathcal X}^k\mathcal Y$ for $k=1,\dots,4$ are, up to the numerical prefactors in Eq.~\eqref{eq:nested},
\begin{align}
  \bigl(\zeta\hat a^{\dagger2}-\zeta^*\aop^2\bigr)\sigma_x
   & \to\bigl(\zeta^2\hat a^{\dagger4}+(\zeta^*)^2\aop^4\bigr)\sigma_y
     \to\bigl(\zeta^3\hat a^{\dagger6}-(\zeta^*)^3\aop^6\bigr)\sigma_x
     \to\bigl(\zeta^4\hat a^{\dagger8}+(\zeta^*)^4\aop^8\bigr)\sigma_y,
\end{align}
that is, quadratic, quartic, sextic and octic processes. The evenness is structural. Each elementary generator changes the photon number by $\pm2$.

\subsection{\texorpdfstring{Several Gates and Limits of the Interpretation}{Several Gates and Limits of the Interpretation}}
For different generators $G_j$, Eq.~\eqref{eq:GjGl} shows that cross commutators $[G_j,G_l]$ produce number-dependent (number-conserving) terms, and Eq.~\eqref{eq:Gn} shows that $[G_j,\nop]$ returns quadratic operators. The complete effective generator $\Omega_{\mathrm{eff}}$, defined by $U_{\mathrm{tot}}=e^{\Omega_{\mathrm{eff}}}$, thus contains (i) the single-generator hierarchy above, (ii) nested cross commutators such as $[G_j,[G_{j'},G_{j''}]]$, and (iii) their qubit-dressed versions, with the dependence on the qubit axis supplied by the rotations $\theta_j$.

This hierarchy should not be interpreted too literally. The exact operation is the product~\eqref{eq:fullsequence}. The BCH logarithm is only a representation of it and requires convergence for the chosen parameters. The statement that $2,4,6,8,\ldots$-photon processes are generated refers to the photon-number-changing hierarchy produced by the quadratic elementary operators. It does not imply that any BCH term is a pure $2m$-photon Hamiltonian, nor that a hardware nonlinearity of order higher than two is required. The native interaction remains quadratic, and the higher-order structure arises solely from the composition of conditional squeezings and qubit rotations. Gate sequences of this type are therefore well suited to the preparation of even-parity bosonic codes. The brachistochrone Hamiltonian of such a preparation also changes the photon number only by even amounts, and it requires terms of high order (Supplementary Table~\ref{tab:order}). Every identity of this section was verified numerically in a truncated Fock space to machine precision.

\section{\texorpdfstring{Closeness, Entanglement, and Tangential Capacity along a Path}{Closeness, Entanglement, and Tangential Capacity along a Path}}
\label{app:capacity}

In this section, we define how closely a pulse follows a path, show that every gate set of this work must entangle qubit and oscillator to create a non-Gaussian state, and derive the tangential capacity of Eq.~(9), the largest speed at which a gate set pushes an unentangled state along a path. The minimal time of each gate set is a time-optimal control problem~\cite{S-Khaneja2001,S-Arenz2017} without a known solution here, so we compare the gate sets against one reference, the brachistochrone at the same vacuum speed. The capacity is the support function of the set of accessible velocities~\cite{S-Agrachev2004}.

\subsection{\texorpdfstring{Closeness and the Need for Entanglement}{Closeness and the Need for Entanglement}}
\label{sec:following}

A path prescribes at every instant a state $\ket{p(t)}$ and a unit tangent $\ket{\dot p}=d\ket p/d\vartheta$, with $\langle p|\dot p\rangle=0$, and a path of length $L$ traversed in $T$ advances at the speed $v_p=L/T$. The closeness $F(t)=\langle p(t)|\rho(t)|p(t)\rangle$ of the reduced oscillator state $\rho(t)$ is the probability of finding the oscillator in the path state, and $C$ is its minimum over the pulse. It is a diagnostic, whereas the figure of merit remains the target fidelity $\mathcal F_T$. For a pure path state of even parity,
\begin{equation}
F\leq\min\{\lambda_{\max}(\rho),\,P_{\rm in},\,1-P_{\rm odd}\},
\label{eq:Fbounds}
\end{equation}
with $\lambda_{\max}(\rho)$ the largest eigenvalue of $\rho$, $P_{\rm in}$ its weight in any subspace that contains $\ket p$, and $P_{\rm odd}$ its odd-parity weight. For a brachistochrone, the relevant subspace is the plane of $\ket{\psi_i}$ and $\ket\chi$. With
\begin{equation}
\begin{gathered}
x=2\,\mathrm{Re}\langle\psi_i|\rho|\chi\rangle,\qquad z=\langle\psi_i|\rho|\psi_i\rangle-\langle\chi|\rho|\chi\rangle,\\
P_{\rm in}=\langle\psi_i|\rho|\psi_i\rangle+\langle\chi|\rho|\chi\rangle,
\end{gathered}
\label{eq:plane}
\end{equation}
the closeness to $\cos\vartheta\ket{\psi_i}+\sin\vartheta\ket\chi$ is $F=\tfrac12(P_{\rm in}+x\sin2\vartheta+z\cos2\vartheta)$, so pure real states of the plane lie on the unit circle [Figs.~3(b) and 4(a)]. With $(x,z)=r_\rho(\sin2\vartheta_\rho,\cos2\vartheta_\rho)$, $1-F$ splits into three non-negative parts,
\begin{equation}
1-F=(1-P_{\rm in})+\tfrac12(P_{\rm in}-r_\rho)+r_\rho\sin^2(\vartheta-\vartheta_\rho),
\label{eq:Fsplit}
\end{equation}
namely weight outside the plane, mixedness or relative phase within it, and a lag or lead along it. A pulse that does nothing loses closeness by lag alone, $C=\cos^2\theta=0.54$ for GKP preparation. For GKP preparation the mean lag term at the minimum closeness is at most 0.003 for conditional squeezing and 0.081 for conditional displacements, so these gate sets lose closeness by departing from the path, not by lagging behind it.

\begin{proposition}[product states evolve by Gaussian unitaries]\label{thm:nogo}
Let $H(t)=\sum_kA_k(t)\otimes O_k(t)$ with Hermitian qubit operators $A_k$ and Hermitian oscillator operators $O_k$ of degree $\leq2$ in $\aop,\ad$. If $\ket{\Psi(t)}=\ket{Q(t)}\otimes\ket{\psi(t)}$ for all $t\in[0,T]$, with differentiable factors, then $\ket{\psi(T)}=U_G\ket{\psi(0)}$ up to a global phase, with $U_G$ Gaussian.
\end{proposition}
\begin{proof}
Applying $\bra{Q(t)}\otimes1$ to $i\partial_t\ket\Psi=H\ket\Psi$ gives $i\partial_t\ket\psi+i\langle Q|\partial_tQ\rangle\ket\psi=\sum_k\langle Q|A_k|Q\rangle O_k\ket\psi$. Both $i\langle Q|\partial_tQ\rangle$ and $\langle Q|A_k|Q\rangle$ are real, so $\ket\psi$ evolves, up to a phase, under the quadratic Hamiltonian $\sum_k\langle A_k\rangle O_k$, whose propagator is Gaussian.
\end{proof}
For a gate the hypothesis must hold for every encoded input, so a single Gaussian unitary acts on $\mathcal C$, the case of the numerical bound in Sec.~\ref{app:flip}. All three gate sets of Eq.~(10) are of this form, apart from the self-Kerr of the device, which is not Gaussian but changes the fidelity of a fixed pulse by less than $10^{-4}$ (Sec.~\ref{sec:kerr-absent}). A pulse that creates a non-Gaussian state from the vacuum must therefore entangle qubit and oscillator, which lowers $\lambda_{\max}(\rho)$ in Eq.~\eqref{eq:Fbounds}. Section~\ref{app:cs} shows how qubit rotations lift this restriction for conditional squeezing, whose generators conserve parity, so that it can control each parity sector, and GKP preparation needs only the even one.

\subsection{\texorpdfstring{Tangential Capacity}{Tangential Capacity}}
\label{sec:tangential}

The capacity quantifies how much of the motion along a path a gate set can supply without entangling qubit and oscillator. Write the oscillator generator of qubit branch $j$ as $G_j=\bm u_j\cdot\bm K$, with real amplitudes $\bm u_j$, $|\bm u_j|\leq g$, and Hermitian generators $\bm K$. For CS, $\bm u_j=(\mathrm{Re}\,\xi_j,\mathrm{Im}\,\xi_j)$, $K_A=\tfrac i2(\aop^2-\hat a^{\dagger2})$ and $K_B=\tfrac12(\aop^2+\hat a^{\dagger2})$.
\begin{proposition}[tangential capacity]\label{prop:cap}
Let $H=H_q\otimes1+\sum_j\ket j\bra j\otimes G_j$, with $H_q$ the qubit drive. (i)~For any joint state $\ket\Psi$, the reduced oscillator state obeys $\partial_t\rho=-i\sum_j[G_j,\varrho_j]$, with $\varrho_j=\langle j|\Psi\rangle\langle\Psi|j\rangle$. For a product state $\ket\Psi=\ket Q\otimes\ket p$ this is, at that instant, the evolution of $\ket p$ under the qubit-averaged generator $G_{\rm eff}=\bm u_{\rm eff}\cdot\bm K$, $\bm u_{\rm eff}=\sum_jP_j\bm u_j$, $P_j=|\langle j|Q\rangle|^2$. (ii)~For $\ket p$ on the path, its velocity along the path, $\mathrm{Re}\,\langle\dot p|{-i}G_{\rm eff}|p\rangle=\bm u_{\rm eff}\cdot\bm\tau$ with $\tau_l=\mathrm{Re}\,\langle\dot p|{-i}K_l|p\rangle$, is at most $v=g|\bm\tau|$, Eq.~(9), attained when every branch points along $\bm\tau$. (iii)~If $\ket{\Psi(t)}=\ket Q\otimes\ket{p(t)}$ and the path state advances at the speed $v_p$, then, with $\ket{w_j}=(1-\ket p\bra p)(-iG_j)\ket p$,
\begin{equation}
1-F(t+dt)=dt^2\sum_jP_j\big\|\ket{w_j}-v_p\ket{\dot p}\big\|^2+O(dt^3),
\label{eq:closeness-budget}
\end{equation}
and $\sum_jP_j\|\ket{w_j}-v_p\ket{\dot p}\|^2\geq\max(0,v_p-v)^2$.
\end{proposition}
\begin{proof}
(i)~$\mathrm{Tr}_q[(H_q\otimes1)X-X(H_q\otimes1)]=0$ for every joint operator $X$, and $\mathrm{Tr}_q[(\ket j\bra j\otimes G_j)\ket\Psi\bra\Psi]=G_j\varrho_j$, with $\varrho_j=P_j\ket p\bra p$ for a product state. (ii)~$\bm u_{\rm eff}$ is a convex combination of vectors of length $\leq g$, and Cauchy--Schwarz gives $\bm u_{\rm eff}\cdot\bm\tau\leq g|\bm\tau|$. (iii)~$1-F=\|(1-1\otimes\ket{p_+}\bra{p_+})\ket{\Psi(t+dt)}\|^2$ with $\ket{p_+}=\ket{p(t+dt)}=\ket p+v_p\,dt\ket{\dot p}+O(dt^2)$. The qubit drive acts on $\ket Q$ alone and contributes at $O(dt^2)$, so the vector equals $dt\sum_j\langle j|Q\rangle\ket j\otimes(\ket{w_j}-v_p\ket{\dot p})+O(dt^2)$. The component of $\ket{w_j}-v_p\ket{\dot p}$ along $\ket{\dot p}$ has the real part $\bm u_j\cdot\bm\tau-v_p$, so $\|\ket{w_j}-v_p\ket{\dot p}\|^2\geq(\bm u_j\cdot\bm\tau-v_p)^2$. Convexity, $\sum_jP_j(\bm u_j\cdot\bm\tau-v_p)^2\geq(\bm u_{\rm eff}\cdot\bm\tau-v_p)^2$, and (ii) give the bound.
\end{proof}
Part (iii) ties the capacity to the tracking objective (Sec.~\ref{app:numerics}). Along a path faster than $v$, a direct push loses closeness at second order in time, whatever the controls. Keeping $F=1$ to this order requires all populated branches to supply exactly $v_p$ along the tangent, so the state stays a product, which by Proposition~\ref{thm:nogo} is possible over a finite interval only within one Gaussian orbit. The capacity therefore bounds only the direct part of the motion, not the duration of a pulse, which may entangle. On a brachistochrone, where $v$ is constant, $T_{\tan}=L/v$ is the duration of a direct push at the capacity. Supplementary Figure~\ref{fig:capacity}(a--c) draws this construction for GKP preparation at 412~ns, where the path requires $\theta/T=0.34\,\omega$ and the capacity is $v=0.51\,\omega$.

\subsection{\texorpdfstring{Capacity along the Brachistochrones}{Capacity along the Brachistochrones}}

\begin{figure}[tbp]
\includegraphics[width=0.72\textwidth,height=0.42\textheight,keepaspectratio]{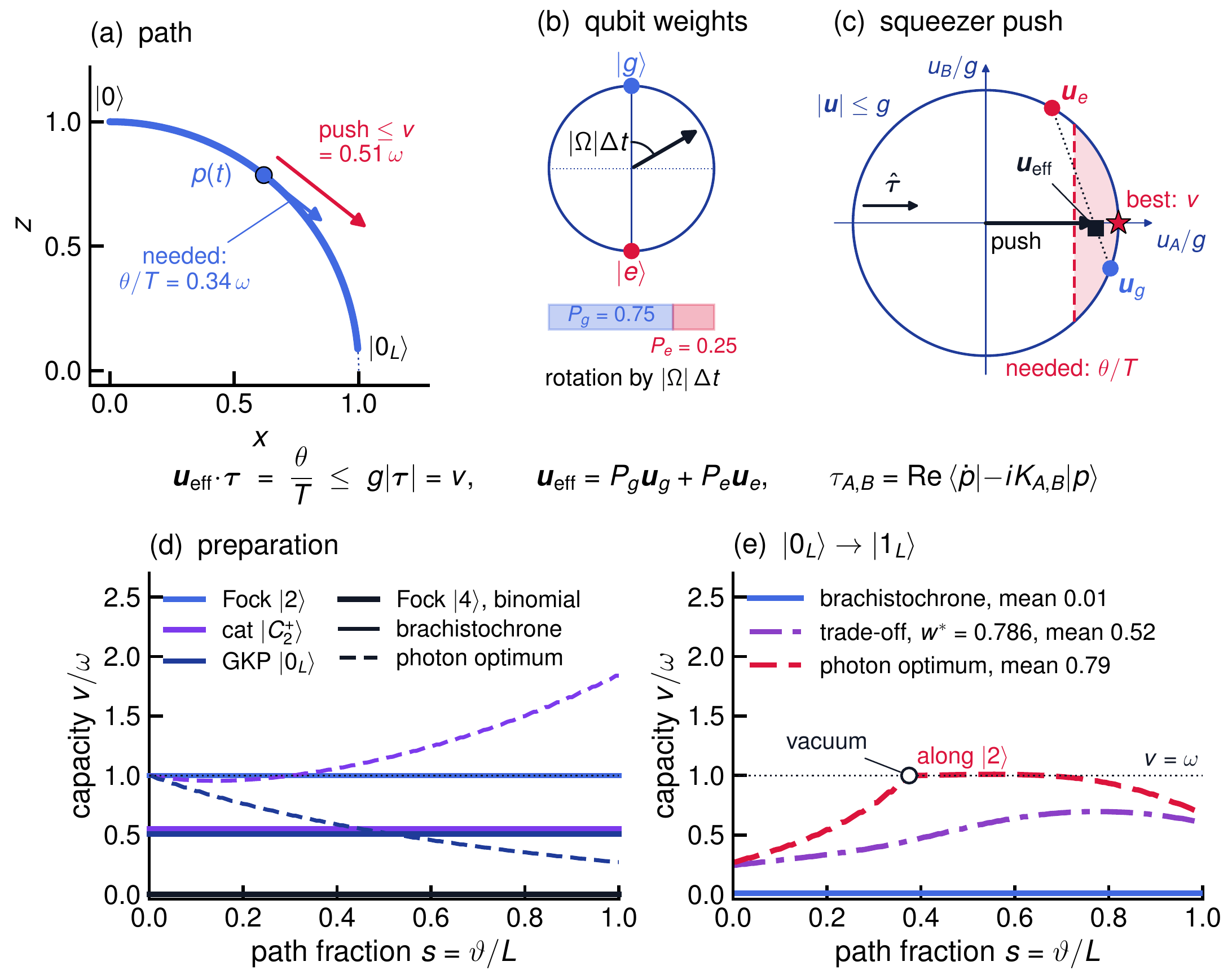}
\caption{How conditional squeezing pushes the state along a path, for GKP preparation at $T=412$~ns. (a)~The brachistochrone in the plane of Eq.~\eqref{eq:plane}, with the speed $\theta/T=0.34\,\omega$ that the path requires (blue arrow) and the largest push of the squeezers, $v=0.51\,\omega$ (crimson arrow). (b)~A qubit rotation by $60^\circ$ from $\ket{\mathrm g}$ sets the branch populations $P_{\rm g}=0.75$ and $P_{\rm e}=0.25$. (c)~Branch controls $\bm u_{\rm g}$, $\bm u_{\rm e}$ in units of $g$ and their population-weighted average $\bm u_{\rm eff}$ (square), whose push along the path is $\bm u_{\rm eff}\cdot\bm\tau$ (arrow). The star marks the largest push $v=g|\bm\tau|$, and controls in the shaded region can keep pace with the path. (d,~e)~Tangential capacity $v/\omega$, Eq.~(9), against the path fraction $s$, for preparation (solid, brachistochrone, dashed, photon optimum) and for $\ket{0_L}\to\ket{1_L}$ (trade-off member $w^*$). The dotted line marks $v=\omega$, and an open circle the vacuum.}
\label{fig:capacity}
\end{figure}

Since $-iK_l$ is anti-Hermitian, $\mathrm{Re}\langle\dot p|{-i}K_l|p\rangle=\mathrm{Re}\langle\chi|{-i}K_l|\psi_i\rangle$ at every $\vartheta$ along $\ket p=\cos\vartheta\ket{\psi_i}+\sin\vartheta\ket\chi$, so $v$ is constant along any brachistochrone. For real states only $-iK_A=G_A=\tfrac12(\aop^2-\hat a^{\dagger2})$ of Sec.~\ref{app:cs} contributes, since $-iK_B$ is imaginary. From the vacuum, $\langle\chi|G_A|0\rangle=-\langle\chi|2\rangle/\sqrt2$ and $v=g|\langle2|\chi\rangle|/\sqrt2$. This gives $v/g=0.707$, 0.390, and 0.360 for Fock $\ket2$, the cat, and GKP preparation, 0 for Fock $\ket4$ and the binomial-code state, and, from $\ket{0_L}$, 0.0061 for the flip ($T_{\tan}=296$, 471, and 275~ns, and $34~\mu$s). For displacements $\langle\dot p|(\aop\pm\ad)|p\rangle=0$ on every parity-even path, because $\aop\pm\ad$ changes the parity, and for SNAP the fixed dispersive term $-\chi_d\nop$ adds no tangential component on the real paths of this work, since $-i\nop$ is imaginary. Both therefore have $v=0$, and by Proposition~\ref{prop:cap}(iii) a direct push loses closeness at the full rate. For displacements every $\ket{w_j}$ has odd parity. Along the photon-optimal and trade-off paths the capacity varies [Supplementary Fig.~\ref{fig:capacity}(d,~e)]. The photon-optimal transfer, for example, reaches the vacuum at $s=0.375$ and leaves it along Fock $\ket2$, where $v\approx\omega$.

\subsection{\texorpdfstring{What the Capacity Predicts}{What the Capacity Predicts}}

\begin{figure}[tbp]
\includegraphics[width=0.72\textwidth,height=0.42\textheight,keepaspectratio]{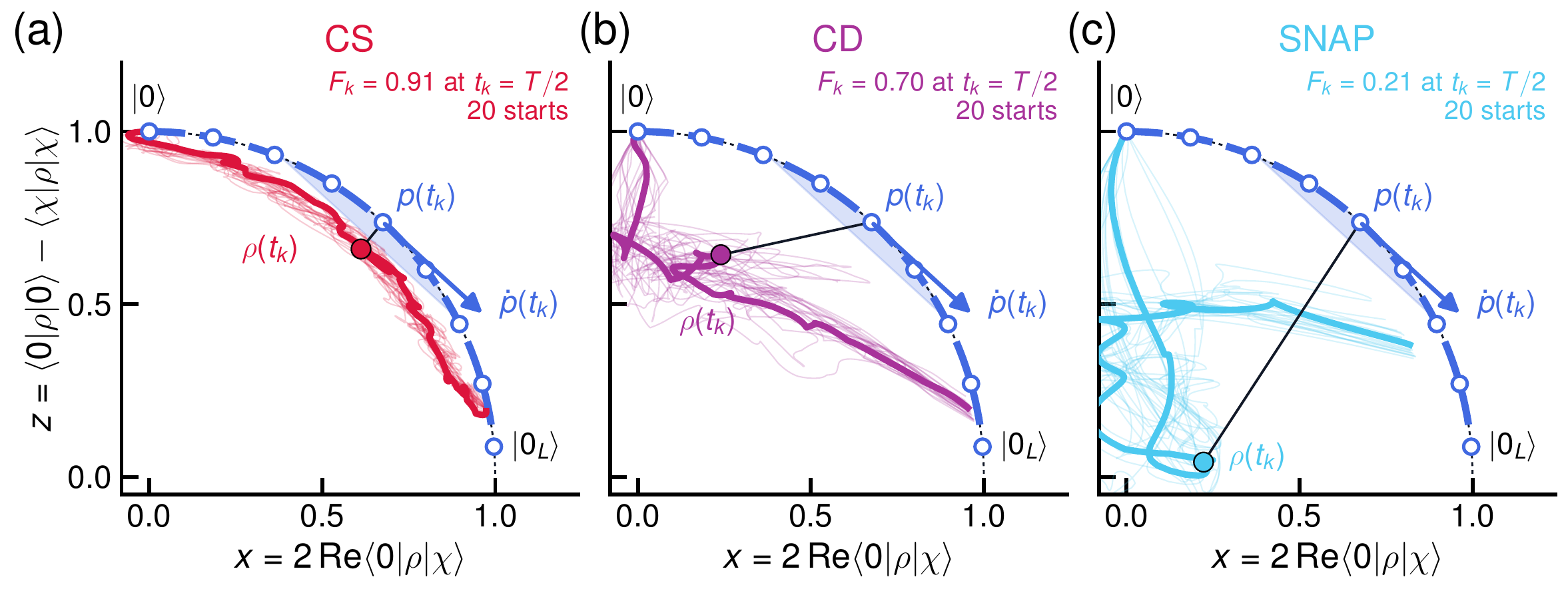}
\caption{The tracking protocol for CS (a), CD (b) and SNAP (c) in the plane of Eq.~\eqref{eq:plane}, for GKP preparation at 824~ns. Open circles mark the path states $p(t_k)$ at $t_k=kT/8$, the shaded area marks the region $F_k\geq0.97$ at $t_k=T/2$ for states of the plane, thin lines show every start, and the thick line shows the start closest to the mean, whose state at $t_k=T/2$ is joined to $p(t_k)$, with its closeness $F_k$.}
\label{fig:S08}
\end{figure}

Supplementary Figure~\ref{fig:S08} draws the tracking protocol in the plane of Eq.~\eqref{eq:plane} for GKP preparation at 824~ns. The capacity separates the gate sets by parity, and the benchmark bears this out. Conditional squeezing never populates odd photon numbers in closed evolution ($P_{\rm odd}<10^{-26}$), and at 824~ns its shortfall $1-C=0.103$ splits by Eq.~\eqref{eq:Fsplit} into 0.086 outside the plane, 0.016 within it, and a lag of 0.001 (means over 20 starts, closed evolution). Odd weight makes up 0.259 of the shortfall 0.508 of conditional displacements and 0.480 of the shortfall 0.812 of SNAP, and 66\% on average over all 185 runs of these gate sets that leave the vacuum, and at least half for conditional displacements at every duration. The capacity does not, however, rank paths for one gate set. Conditional squeezing follows the binomial-code path, along which $v=0$, as closely as any other preparation path (Supplementary Table~\ref{tab:prep}), and every gate set follows the photon-saving transfer paths more closely than the flip brachistochrone (Supplementary Table~\ref{tab:flip}).

\section{\texorpdfstring{Numerical Methods}{Numerical Methods}}
\label{app:numerics}

In this section, we give the pulse parametrization, the tracking objective and its exact gradient, the optimization schedule and weight selection, and the convergence and validation checks.

\emph{Controls.} The controls of Eq.~(10), the qubit drive $\Omega$ and the branch amplitudes $\xi_j$ (CS) or $\varepsilon_j$ (CD), or the cavity drive $\varepsilon$ of SNAP at fixed $\chi_d$, are piecewise constant on $N$ slices ($N=50$, or $N=100$ at 2133~ns and longer) of duration $\Delta t=T/N$. Six variables per slice, $(a_\Omega,f_\Omega,a_{\rm g},f_{\rm g},a_{\rm e},f_{\rm e})$, define $\Omega=10g\,a_\Omega e^{if_\Omega}$ and the amplitude $g\,a_je^{if_j}$ of branch $j$, with $a_\Omega,a_j\in[0,1]$.

\emph{Objective.} The tracking pulse minimizes
\begin{equation}
\begin{split}
\mathcal L={}&W\textstyle\sum_{k<N}\big[\max(0,\mathrm{thr}-F_k)\big]^2+W_{\rm fin}\big[\max(0,\mathrm{thr}_{\rm fin}-F_N)\big]^2\\
&+\tfrac{W_{\rm top}}{N}\textstyle\sum_{k}\langle\Psi_k|\Pi_{\rm top}|\Psi_k\rangle ,
\end{split}
\label{eq:loss}
\end{equation}
with the thresholds and weights of Supplementary Table~\ref{tab:schedule}. Here $F_k=\sum_j|\langle p(s_k),j|\Psi_k\rangle|^2$ is the closeness of the joint state at the end of slice $k$ to the path state at $s_k=k/N$ (in stage D, $s_k=\sum_{l\leq k}d_l/\sum_ld_l$ with free $d_l\geq10^{-3}$), and $W_{\rm top}=10^3$ penalizes the population of the Fock states $n\geq n_c$ ($n_c=40$ at cutoff 60, 44 at cutoff 80). We call this construction, evaluated in closed evolution, the tracking protocol. It prescribes where the state should be, not how it gets there. Its one-sided penalty vanishes while $F_k\geq\mathrm{thr}$, so the pulse moves freely within a tube around the path, directly or through qubit--oscillator entanglement, whereas the last slice demands arrival. Reported closenesses are evaluated under Eq.~\eqref{eq:lindblad}, except those of Supplementary Fig.~\ref{fig:S23} and $C_{\min}$ of Supplementary Table~\ref{tab:gate2} (closed evolution, cutoff 60), and closed and noisy mean closenesses differ by at most 0.01.

\emph{Propagation and exact gradient.} In slice $k$ the joint state evolves by $U_k=e^{-iH_k\Delta t}$ with $H_k=H_0+\sum_lc_{kl}H_l$, where the drift $H_0$ holds the self-Kerr of Sec.~\ref{app:noise} and, for SNAP, the dispersive shift. The exact gradient of $\mathcal L$ follows from one forward sweep of the states and one backward sweep of a costate, with $\partial U_k/\partial c_{kl}$ evaluated in the eigenbasis of $H_k$~\cite{S-Khaneja2005,S-deFouquieres2011}. It agrees with central finite differences to $10^{-8}$, and the controls are updated by L-BFGS-B~\cite{S-Byrd1995} within their bounds. This is gradient-ascent pulse engineering (GRAPE)~\cite{S-Khaneja2005}, applied in the same way to all three gate sets. The conditional displacements of CD are therefore free piecewise-constant controls, not the fixed echoed sequences of the conditional-displacement gates of Ref.~\cite{S-Eickbusch2022}. Noisy evolution is integrated in sub-steps of at most 2~ns, each the exact slice unitary followed by the exact channel of every dissipator of Eq.~\eqref{eq:lindblad}, with a splitting error below $3\times10^{-5}$ in every fidelity.

\emph{Schedule and weight selection.} Every run executes the same minimizations (Supplementary Table~\ref{tab:schedule}) from a smooth random start (low-frequency Fourier noise in every control). The tracking weight is then chosen on the ladder $\{10,20,40,70,150,300,600,1000,2000\}$. Each rung polishes the previous one, its pulse is evaluated with the Lindblad equation, and the largest $W$ with $\mathcal F_T\geq0.99$ is kept and refined by two bisections. If no rung reaches 0.99, the rung with the best noisy fidelity is kept.

\begin{table*}[tbp]
\caption{Stage schedule. The number of sub-stages is given in parentheses. The columns give the duration $T_{\rm st}$ of the stage problem, the closeness thresholds thr and thr$_{\rm fin}$, the weights, and the L-BFGS-B iterations.}
\label{tab:schedule}
\begin{ruledtabular}
\begin{tabular}{lcccccc}
stage & $T_{\rm st}$ & thr & $W$ & $W_{\rm fin}$ & thr$_{\rm fin}$ & iter.\\\hline
A (4) & $0.625\,T$ & 0.6--0.955 & 10 & 200 & 0.9999 & 350\\
B (2) & $T$ & 0.9, 0.955 & 10 & 200 & 0.9999 & 400\\
C (5) & $T$ & 0.80--0.95 & 10 & $2\times10^4$ & 0.99995 & 300\\
D (1), free progress & $T$ & 0.90 & 10 & $2\times10^4$ & 0.99995 & 300\\
E (3) & $T$ & 0.97--0.999 & 1000 & 1000 & 0.99995 & 300\\
polish (each rung) & $T$ & 0.97 & $W$ & $10^5$ & 0.99995 & 300\\
\end{tabular}
\end{ruledtabular}
\end{table*}

\emph{The tracking weight and the trade-off.} Following the path and arriving at the target compete (Supplementary Fig.~\ref{fig:S23}), so every run polishes its pulse at each weight of the ladder. At every weight and duration conditional squeezing follows more closely and arrives with a higher fidelity than the other gate sets, and every conditional-squeezing start follows more closely than the best start of the other gate sets at the same weight. Figure~4 shows the pulse that the weight-selection rule keeps, so $\mathcal F_T\approx0.990$ for CS at 412 and 824~ns is the floor of this rule, not a result, and the ranking does not depend on it.

\begin{figure}[tbp]
\includegraphics[width=0.72\textwidth,height=0.42\textheight,keepaspectratio]{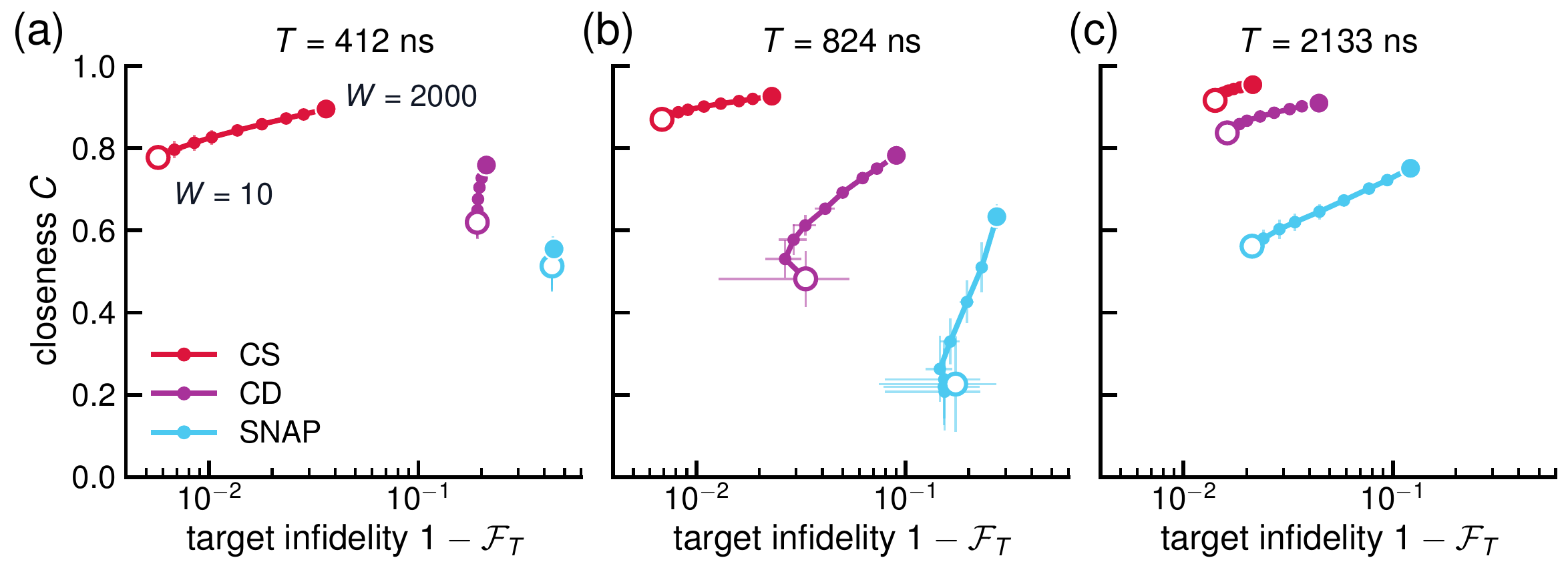}
\caption{Following against arriving, without a fidelity threshold. Minimum closeness $C$ (closed evolution) against target infidelity $1-\mathcal F_T$ (Lindblad) of GKP preparation along the brachistochrone, for every tracking weight $W=10,\dots,2000$ (mean $\pm$ s.d. over the 20 starts of Fig.~4, cutoff 60). The open marker shows $W=10$ and the filled marker $W=2000$.}
\label{fig:S23}
\end{figure}

\emph{The gate objective for logical operations.} For the flip as a gate, we minimize the mean of Eq.~\eqref{eq:loss} over the inputs $\ket{0_L}$, $\ket{1_L}$, and $\ket{+_L}$, each with the ancilla in $\ket{\mathrm g}$ and with its own reference path, its image under the flip brachistochrone. The superposition fixes the relative phase and penalizes an ancilla left entangled with the logical state.

\emph{Cutoffs and convergence.} Stages and ladder run at Fock cutoff 60, followed by a polish at cutoff 80 (none for the gate pulses of Supplementary Table~\ref{tab:gate2}), and every final pulse is evaluated at cutoff 90. The polish applies the threshold $\mathcal F_T\geq0.99$ of the weight selection at cutoff 80, so a kept pulse can lie slightly below 0.99 at cutoff 90. Between cutoffs 90, 100, and 110 the 783 pulses checked agree to $\leq3.7\times10^{-4}$ in $\mathcal F_T$ and $\leq1.8\times10^{-3}$ in $C$, except 53 of the 105 CS transfer pulses (up to $4.4\times10^{-3}$ and $9.7\times10^{-3}$) and one CS cat pulse at 2133~ns ($5.9\times10^{-3}$ in $\mathcal F_T$). The gate pulses of Supplementary Table~\ref{tab:gate2} agree between cutoffs 90 and 120 to $3.2\times10^{-4}$ when optimized for the gate alone and to $2.3\times10^{-3}$ with tracking.

\emph{Durations and statistics.} Durations are also given as multiples $T/T^*$ of the minimal time $T^*$ of the path followed. The durations 412, 824, and 2133~ns are $T/T^*=2.9$, 5.9, and 15.3 for GKP preparation and $T/T^*=1.40$, 2.81, and 7.27 for the flip brachistochrone. Bands are mean $\pm$ sample s.d. (normalized by $n-1$) over the starts, every start included, and ``best'' is the maximum. Thick lines and Wigner panels show one run, the start whose $(x,z,F)$ traces have the smallest mean-squared distance from the averages over the starts. An average over starts would be a mixture of different pulses and smear the Wigner function.

\emph{Runs that never leave the vacuum.} At zero cavity drive $\mathcal L$ is stationary for the displacement-based gate sets, because a displacement creates odd-parity components, orthogonal to every even path state and target, and the qubit drive and the dispersive term do not act on the vacuum. Once a stage drives the amplitudes to zero, their lower bound, L-BFGS-B cannot leave this point. Only SNAP runs end there (34 of 217, and none of the 221 CD and 345 CS runs), and they are kept in every statistic.

\emph{Capability without tracking.} Supplementary Table~\ref{tab:capability} uses the stages $W=0$, $W_{\rm fin}=200$, $2\times10^4$ and $10^5$ (800, 800 and 1200 iterations) at cutoff 60 from 8 starts independent of the benchmark starts, and evaluates every pulse at cutoff 90, closed and under Eq.~\eqref{eq:lindblad}. The cutoff-60 and cutoff-90 closed fidelities agree to $1.4\times10^{-6}$.

\emph{Validation.} Six stored pulses, re-integrated by an independent code with the exact Lindblad generator on each slice, reproduce the stored closed evolution to $6\times10^{-12}$ and the noisy $\mathcal F_T$, closeness traces, and $C$ to $\leq5.4\times10^{-6}$. Every closed-form expression of Secs.~\ref{app:derivation}--\ref{app:logical} agrees with a direct evaluation on the state vectors to $10^{-13}$ or better. With the Kerr term switched off, the code reproduces three stored pulses of the version without it, one per gate set at 412~ns, bit for bit from the same random starts (Sec.~\ref{app:kerr}).

\section{\texorpdfstring{Robustness of State Preparation and Further Targets}{Robustness of State Preparation and Further Targets}}
\label{app:robust}
In this section, we test whether the GKP ranking depends on the figure of merit, the resources, the noise model, or the form of the conditional-squeezing gate, extend the benchmark to further targets, and state the assumptions that bound the comparison.

\emph{Quality of the prepared GKP states.} Experiments judge GKP states by the decoded logical fidelity and the effective squeezing~\cite{S-CampagneIbarcq2020,S-Sivak2023RealTime,S-PhysRevLett.132.150607,S-deNeeve2022}, so Supplementary Table~\ref{tab:stab} also gives the stabilizers and the logical displacement $\langle\hat D(\beta)\rangle$ of the final states. They confirm the ranking of $\mathcal F_T$. At 412~ns conditional squeezing falls short of the target by 0.010 in $\langle\hat D(\beta)\rangle$ and 0.020 in $\langle S_Z\rangle$, whereas conditional displacements reach comparable values only at 2133~ns and SNAP at none of the three durations. The states with $\mathcal F_T\geq0.98$ have $\bar n={}$2.0--2.4, below the 2.44 of the target, so they are lower-energy approximants of the logical state.

\begin{table}[h]
\caption{Final states of GKP preparation (mean $\pm$ s.d. over the 20 starts of Fig.~4).}
\label{tab:stab}
\begin{ruledtabular}
\begin{tabular}{lccccc}
 & $\mathcal F_T$ & $\langle S_X\rangle$ & $\langle S_Z\rangle$ & $\langle\hat D(\beta)\rangle$ & $\bar n$\\\hline
target $\ket{0_L}$ & 1 & 0.589 & 0.606 & 0.882 & 2.44\\
vacuum & 0.544 & 0.043 & 0.043 & 0.456 & 0\\\hline
CS, 412~ns & 0.990 & $0.517(4)$ & $0.586(5)$ & $0.872(2)$ & 2.01\\
CS, 824~ns & 0.990 & $0.528(5)$ & $0.590(3)$ & $0.873(1)$ & 2.11\\
CS, 2133~ns & 0.986 & $0.550(5)$ & $0.591(3)$ & $0.872(2)$ & 2.33\\
CD, 412~ns & 0.809 & $0.168(27)$ & $0.375(7)$ & $0.763(8)$ & 0.78\\
CD, 824~ns & 0.977 & $0.516(7)$ & $0.520(17)$ & $0.847(8)$ & 1.95\\
CD, 2133~ns & 0.984 & $0.532(4)$ & $0.579(4)$ & $0.870(2)$ & 2.16\\
SNAP, 412~ns & 0.566 & $0.042(3)$ & $0.072(51)$ & $0.496(71)$ & 0.05\\
SNAP, 824~ns & 0.864 & $0.329(50)$ & $0.404(19)$ & $0.772(11)$ & 1.32\\
SNAP, 2133~ns & 0.979 & $0.522(2)$ & $0.543(4)$ & $0.853(1)$ & 1.97\\
\end{tabular}
\end{ruledtabular}
\end{table}

\emph{Energy uncertainty and photon number.} The amplitude bounds of Eq.~(10) equalize the speed at which each gate set moves the vacuum, not its speed along a pulse or in hardware. In hardware, the echoed conditional-displacement gates of Refs.~\cite{S-Eickbusch2022,S-Sivak2023RealTime} displace each branch at $3.1~\mu$s$^{-1}$ and at most $3.8~\mu$s$^{-1}$, below $\omega=5.3~\mu$s$^{-1}$, so the common bound favors CD. Conditional squeezing uses 1.2--2.3 times the energy uncertainty of conditional displacements (Supplementary Table~\ref{tab:resources}), since that of a squeezer grows with the photon number. Every pulse runs several times faster than the $0.34\,\omega$ that the path needs at 412~ns, because the paths set the destination and the pace of progress, not the instantaneous speed.

\begin{table*}[tbp]
\caption{Oscillator energy uncertainty (time average and maximum, in units of $\omega=g/\sqrt2$) and photon number of the representative pulses (closed evolution). The last column gives the population at $n\geq70$ at the photon-number peak.}
\label{tab:resources}
\begin{ruledtabular}
\begin{tabular}{lccccc}
pulse & $\overline{\Delta H}/\omega$ & $\max\Delta H/\omega$ & $\max\bar n$ & $\overline{\bar n}$ & $P_{n\geq70}$\\\hline
CS, GKP, 412~ns & 2.42 & 4.32 & 2.00 & 1.08 & $2\times10^{-6}$\\
CD, GKP, 412~ns & 1.19 & 2.06 & 0.79 & 0.40 & $<10^{-8}$\\
SNAP, GKP, 412~ns$^a$ & 1.12 & 1.37 & 0.93 & 0.47 & $<10^{-8}$\\
CS, GKP, 824~ns & 1.93 & 3.33 & 2.20 & 1.16 & $9\times10^{-5}$\\
CD, GKP, 824~ns & 1.56 & 2.29 & 1.93 & 0.96 & $<10^{-8}$\\
SNAP, GKP, 824~ns & 2.93 & 5.65 & 2.99 & 1.66 & $<10^{-8}$\\
CS, transfer, 824~ns & 4.02 & 6.17 & 3.22 & 2.70 & $5\times10^{-4}$\\
CD, transfer, 824~ns & 1.72 & 2.65 & 2.89 & 2.58 & $2\times10^{-8}$\\
SNAP, transfer, 824~ns & 4.11 & 5.81 & 3.81 & 3.03 & $3\times10^{-8}$\\
\end{tabular}
\end{ruledtabular}
\raggedright\footnotesize $^a$the representative of the 5 non-trivial starts.
\end{table*}

\emph{Conditional squeezing at half rate.} Halving the squeezing rate and the qubit-drive bound maps every conditional-squeezing pulse of duration $T$ onto one of duration $2T$, up to the self-Kerr of the device, which is not rescaled and changes no reported value. The 412-ns pulses of Fig.~4 thus become 824-ns pulses, the representative one with a time-averaged energy uncertainty of $1.21\,\omega$, below the $1.56\,\omega$ of the representative 824-ns CD pulse. They reach $\mathcal F_T=0.9880\pm0.0002$ under noise, against $0.977\pm0.005$ for conditional displacements at the full rate, and a closeness $C=0.82\pm0.02$ in closed evolution, above the at most 0.79 of conditional displacements at every tracking weight. The ranking therefore does not rest on the larger energy uncertainty of the squeezer.

\emph{Conditional squeezing as proposed.} The controlled-squeeze gate of Ref.~\cite{S-DelGrosso2025} squeezes one qubit branch only, whereas Eq.~(10) lets both branches squeeze, as two drive tones would. Restricted to one branch, with the same target-only optimization (8 starts, Supplementary Table~\ref{tab:capability}), it reaches at best 0.966 at 412~ns and 0.995 at 824~ns, against 0.998 and 0.995 with two branches and 0.82 and 0.986 for conditional displacements. The proposed circuit also rotates the idle branch, at $2\pi\times16$~MHz per photon, and has a Kerr term of about $2\pi\times0.43$~MHz. With both terms modeled, the single-branch gate reaches $0.996\pm0.002$ at 412~ns and $0.993\pm0.001$ at 824~ns, whereas SNAP given the same dispersive shift reaches $0.843\pm0.015$ and $0.947\pm0.015$. The optimization thus absorbs these terms.

\emph{Detectable errors.} The conditional-squeezing Hamiltonian, cavity dephasing, and both ancilla dissipators conserve photon-number parity, and an ancilla error applies the squeezer of the wrong branch, which also conserves it. Since the vacuum and every target are parity even, only photon loss changes the parity under conditional squeezing, and a parity measurement after the pulse~\cite{S-Sun2014} flags every trajectory with an odd number of loss events. The flag is exact at any drive amplitude or detuning and survives the dispersive and Kerr terms of the proposed circuit. It assumes an ideal parity measurement and a zero-temperature oscillator. For the pulses of Fig.~4, the odd-parity population is 0 to machine precision in closed evolution and $0.0007$, 0.0016, and 0.0041 at 412, 824, and 2133~ns under noise. To first order it is the probability $\kappa\int\bar n\,dt$ of a loss event, about one third of the noise-induced error. The flag misses an even number of losses, cavity dephasing, and ancilla errors, and we do not claim that the pulses are error transparent~\cite{S-Vy2013,S-Kapit2018,S-Ma2020ET} or path independent~\cite{S-Ma2020PI}. Under conditional displacements and SNAP the oscillator leaves the even sector, so the flag also fires without a loss. The distinction of conditional squeezing is therefore an exact flag, which makes a loss a heralded error as in erasure conversion~\cite{S-Wu2022,S-Kubica2023,S-Teoh2023}.

\emph{Noise models.} Supplementary Table~\ref{tab:proposal} evaluates the pulses of Fig.~4 and of the transfer along the flip brachistochrone with the three sets of rates of Sec.~\ref{app:noise}. With the same dephasing time for every gate set, the ranking is the same with every set of rates. Conditional squeezing leads for GKP preparation at every duration, and conditional displacements lead for the transfer along the flip brachistochrone in 17 of their 20 starts, which exceed every conditional-squeezing start under noise, whereas the other three stall at the best plain displacement and lower their mean below that of conditional squeezing. At the coherence times of the proposal, ancilla dephasing dominates and no pulse reaches 0.99.

\begin{table*}[tbp]
\caption{Target fidelity of the stored pulses (mean $\pm$ s.d. over 20 starts, cutoff 90) under closed evolution, the rates of Ref.~\cite{S-Sivak2023RealTime} with the echo-derived ($T_{\phi q}=414~\mu$s) and Ramsey-derived (70~$\mu$s, a Markovian bracket) ancilla dephasing, and the rates assumed in the conditional-squeezing proposal~\cite{S-DelGrosso2025}.}
\label{tab:proposal}
\begin{ruledtabular}
\begin{tabular}{lcccc}
pulses & closed & echo & Ramsey & proposal\\\hline
GKP, CS, 412~ns & 0.9927(3) & 0.9903(2) & 0.9892(2) & 0.9806(5)\\
GKP, CS, 824~ns & 0.9951(5) & 0.9901(4) & 0.9881(4) & 0.9722(12)\\
GKP, CS, 2133~ns & 0.9986(3) & 0.9857(6) & 0.9805(8) & 0.9408(36)\\
GKP, CD, 412~ns & 0.810(9) & 0.809(9) & 0.808(9) & 0.799(9)\\
GKP, CD, 824~ns & 0.981(5) & 0.977(5) & 0.973(5) & 0.948(5)\\
GKP, CD, 2133~ns & 0.9968(2) & 0.9839(10) & 0.9775(11) & 0.929(4)\\
GKP, SNAP, 412~ns & 0.566(40) & 0.566(39) & 0.566(39) & 0.565(37)\\
GKP, SNAP, 824~ns & 0.868(22) & 0.864(22) & 0.862(22) & 0.840(21)\\
GKP, SNAP, 2133~ns & 0.9921(7) & 0.9789(10) & 0.9712(8) & 0.9143(16)\\
transfer, CS, 824~ns & 0.993(3) & 0.983(4) & 0.980(4) & 0.955(3)\\
transfer, CD, 824~ns & 0.964(80) & 0.957(79) & 0.955(78) & 0.935(73)\\
transfer, SNAP, 824~ns & 0.93(2) & 0.93(2) & 0.92(2) & 0.90(2)\\
\end{tabular}
\end{ruledtabular}
\end{table*}

\begin{figure}[tbp]
\includegraphics[width=0.72\textwidth,height=0.42\textheight,keepaspectratio]{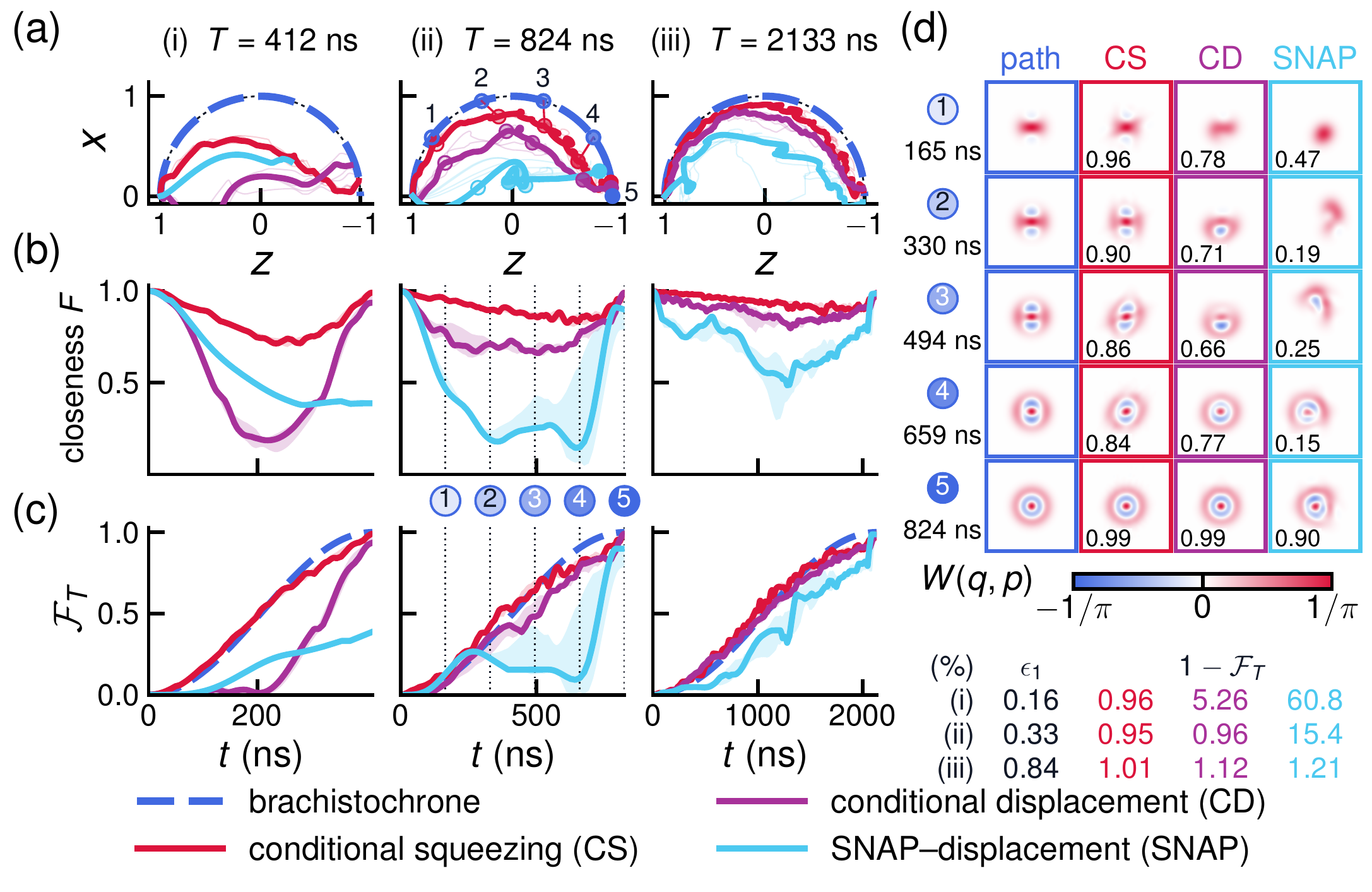}
\caption{As Fig.~4, for Fock $\ket2$ (five starts per gate set, with $z$ horizontal in a). (d), at the right, has one row per $t_k$, and the errors at $T$ ($\epsilon_1$, $1-\mathcal F_T$, in \%) are tabulated below it.}
\label{fig:S09}
\end{figure}

\begin{figure}[tbp]
\includegraphics[width=0.72\textwidth,height=0.42\textheight,keepaspectratio]{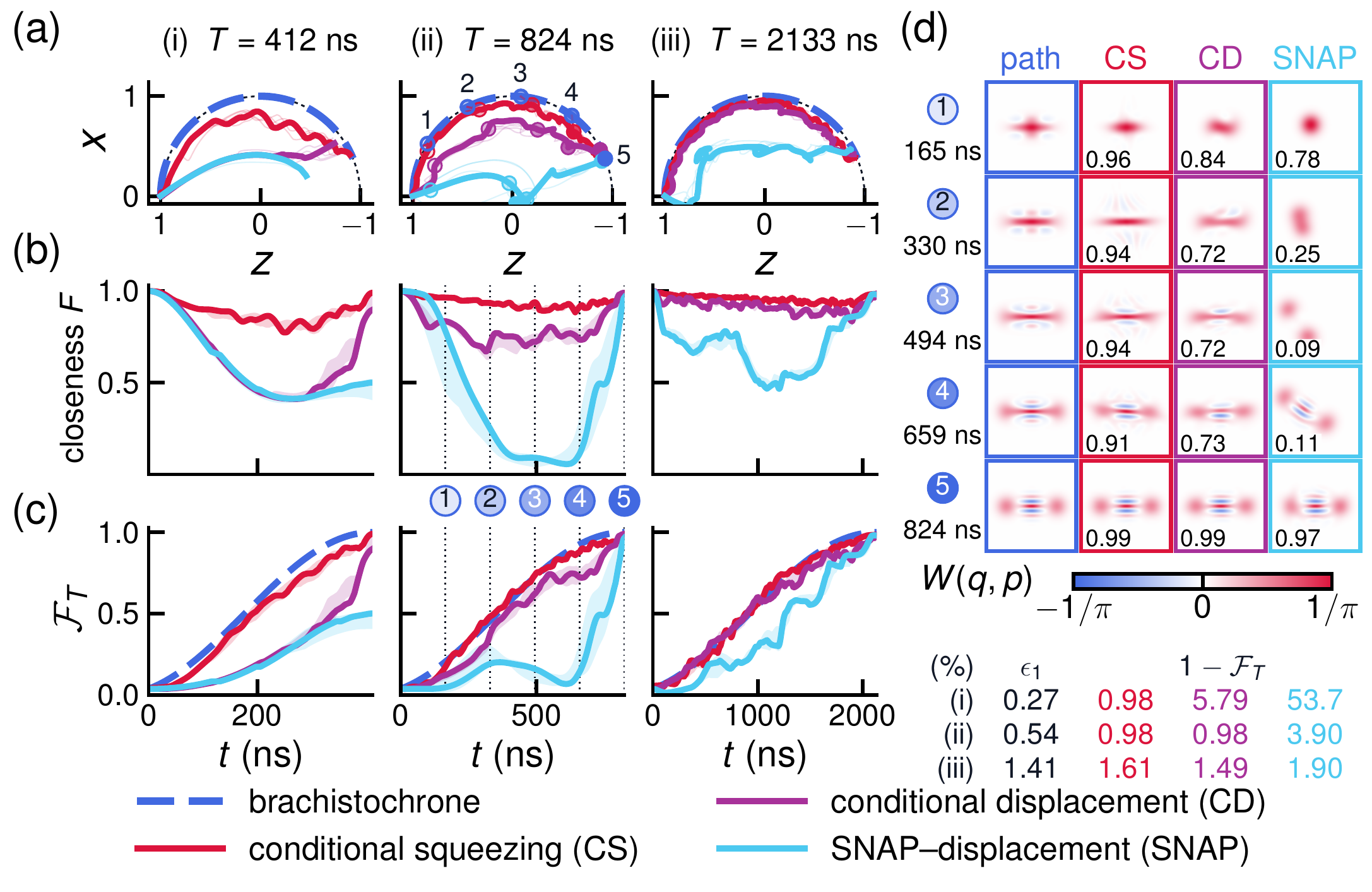}
\caption{As Fig.~4, for the cat $\ket{C_2^+}$ (five starts per gate set, with $z$ horizontal in a). (d), at the right, has one row per $t_k$, and the errors at $T$ ($\epsilon_1$, $1-\mathcal F_T$, in \%) are tabulated below it.}
\label{fig:S10}
\end{figure}

\begin{figure}[tbp]
\includegraphics[width=0.72\textwidth,height=0.42\textheight,keepaspectratio]{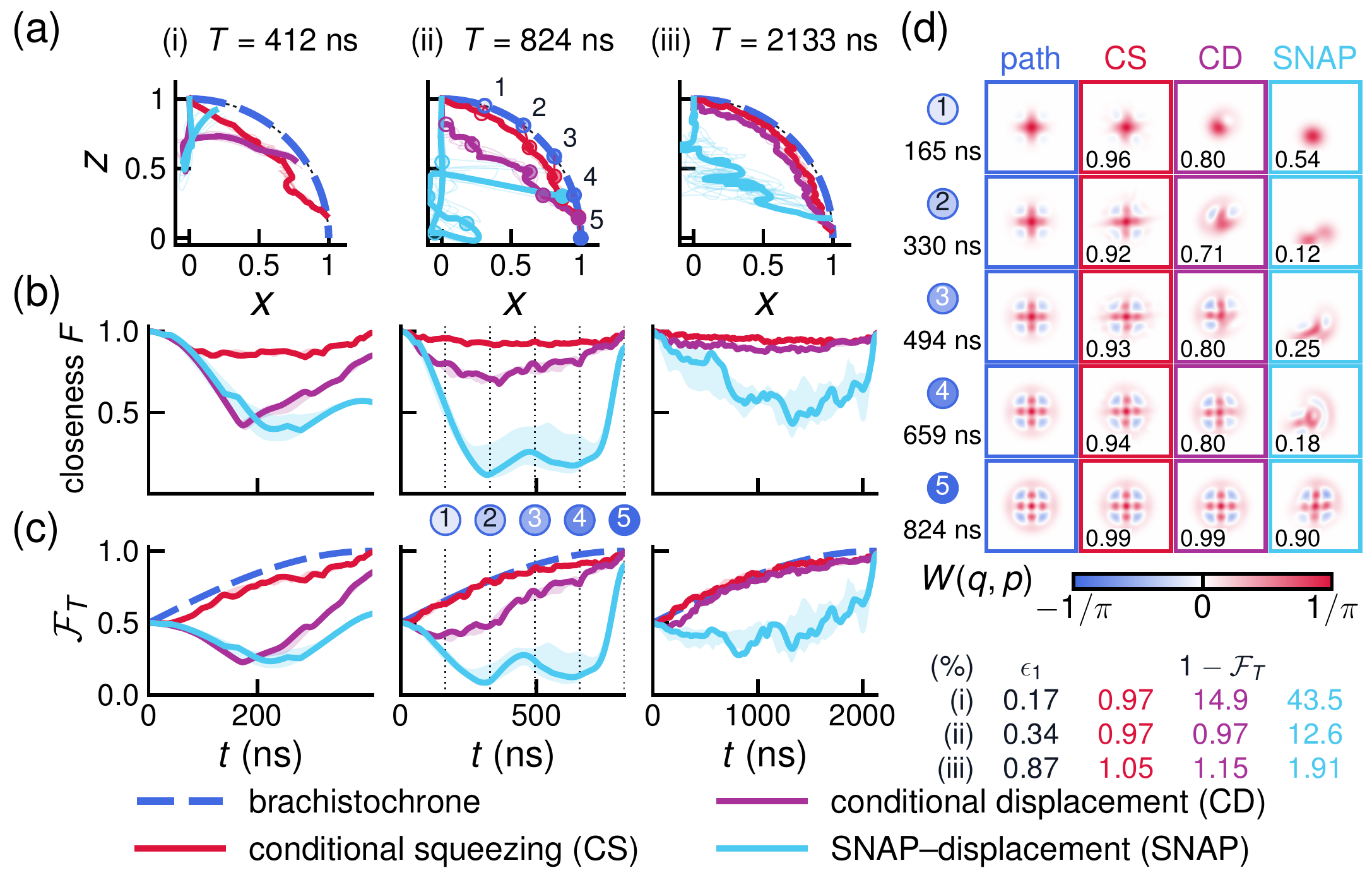}
\caption{As Fig.~4, for the binomial-code state $(\ket0+\ket4)/\sqrt2$, with five starts each for CS and CD. The eight SNAP starts are warm-started from target-only pulses, since every cold start at 412 and 824~ns stays at the zero-drive point. (d), at the right, has one row per $t_k$, and the errors at $T$ ($\epsilon_1$, $1-\mathcal F_T$, in \%) are tabulated below it.}
\label{fig:S11}
\end{figure}

\emph{Further preparation targets.} Supplementary Table~\ref{tab:prep} and Supplementary Figs.~\ref{fig:S09}--\ref{fig:S11} extend Fig.~4 to Fock $\ket2$, the cat $\ket{C_2^+}$, and the binomial-code state (five starts per gate set and duration). Conditional squeezing reaches a mean $\mathcal F_T\geq{}$0.98 at every duration and follows each path most closely. The path to the binomial-code state moves the vacuum straight toward $\ket4$, which a squeezer cannot reach in a single two-photon step, so no generator has tangential capacity along it ($v=0$, Sec.~\ref{app:capacity}). Conditional squeezing still follows this path, because qubit rotations between conditional squeezes compose two-photon steps into four-photon ones (Sec.~\ref{app:cs}), which entangles qubit and oscillator during the pulse. For the cat at 412~ns, shorter than the 471~ns that a direct push at the capacity would take, it likewise relies on this entanglement rather than on a direct push. Conditional displacements need 824~ns, and SNAP reaches 0.98--0.99 only at 2133~ns.

\begin{table*}[tbp]
\caption{Preparation by the three gate sets along the brachistochrone, with the target fidelity $\mathcal F_T$ under the Lindblad equation as mean (s.d. in units of $10^{-3}$ when $\geq5\times10^{-4}$) [best] and the mean minimum closeness $C$. The Fock $\ket2$, cat, and binomial runs use 5 starts, and the GKP runs 20 starts (Fig.~4). Every SNAP start for the binomial-code state at 412 and 824~ns stays at the zero-drive point ($\mathcal F_T=0.500$, $C=0.50$), and warm-started SNAP runs are given in the text.}
\label{tab:prep}
\begin{ruledtabular}
\begin{tabular}{llcccccc}
target & $T$ (ns) & \multicolumn{2}{c}{CS} & \multicolumn{2}{c}{CD} & \multicolumn{2}{c}{SNAP}\\
 & & $\mathcal F_T$ & $C$ & $\mathcal F_T$ & $C$ & $\mathcal F_T$ & $C$\\\hline
Fock $\ket2$ & 412 & 0.990 [0.991] & 0.70 & 0.947(25) [0.971] & 0.16 & 0.392(9) [0.407] & 0.37\\
 & 824 & 0.990 [0.991] & 0.81 & 0.990 [0.991] & 0.65 & 0.846(60) [0.899] & 0.16\\
 & 2133 & 0.990 [0.990] & 0.86 & 0.989(1) [0.990] & 0.78 & 0.988(1) [0.989] & 0.41\\\hline
cat $\ket{C_2^+}$ & 412 & 0.990 [0.991] & 0.78 & 0.942(38) [0.981] & 0.42 & 0.463(58) [0.502] & 0.39\\
 & 824 & 0.990 [0.990] & 0.88 & 0.990 [0.990] & 0.64 & 0.961(28) [0.986] & 0.05\\
 & 2133 & 0.984(4) [0.986] & 0.91 & 0.985 [0.986] & 0.84 & 0.981(1) [0.981] & 0.48\\\hline
binomial $\frac{\ket0+\ket4}{\sqrt2}$ & 412 & 0.990 [0.991] & 0.84 & 0.851(22) [0.880] & 0.42 & 0.500 [0.500] & 0.50\\
 & 824 & 0.990 [0.990] & 0.91 & 0.990 [0.991] & 0.67 & 0.500 [0.500] & 0.50\\
 & 2133 & 0.990 [0.990] & 0.92 & 0.988 [0.989] & 0.87 & 0.983 [0.983] & 0.53\\\hline
GKP $\ket{0_L}$ & 412 & 0.990 [0.991] & 0.82 & 0.809(9) [0.830] & 0.62 & 0.566(39) [0.640] & 0.51\\
 & 824 & 0.990 [0.991] & 0.90 & 0.977(5) [0.984] & 0.49 & 0.864(22) [0.885] & 0.19\\
 & 2133 & 0.986(1) [0.987] & 0.91 & 0.984(1) [0.986] & 0.84 & 0.979(1) [0.980] & 0.56\\
\end{tabular}
\end{ruledtabular}
\end{table*}

\emph{SNAP without the zero-drive trap.} Every SNAP start for the binomial-code state at 412 and 824~ns, and 15 of the 20 GKP starts at 412~ns, stays at the zero-drive point. Started from target-only pulses (8 starts per case), SNAP leaves the vacuum but by a route far from the path, with $C={}$0.38, 0.17, and 0.35 for the binomial-code state (Supplementary Fig.~\ref{fig:S11}), below the 0.50 of a pulse that does nothing, and 0.37 and 0.22 for GKP. At 412~ns SNAP is limited also without the path, as one cycle of number-selective phases takes $2\pi/\chi_d=838$~ns.

\emph{Scope.} (i)~Conditional squeezing is modeled at the rate of a hardware proposal~\cite{S-DelGrosso2025} that has not yet been realized. Its idle-branch rotation and Kerr term are absorbed by the optimization, whereas pump-induced dephasing and heating, cross-Stark shifts, flux noise, and the experimental overheads of echoed conditional-displacement and SNAP gates~\cite{S-Eickbusch2022,S-Heeres2015} are not modeled. (ii)~Ancilla errors act during the conditional operations and are part of every reported fidelity. The model is at zero temperature, and the ancilla starts in $\ket{\mathrm g}$. (iii)~The benchmark uses $\Delta=0.4$, and  Sec.~\ref{app:kerr} repeats it at $\Delta=0.3$ and 0.5, where conditional squeezing again reaches the highest $\mathcal F_T$ at 412 and 824~ns, although at $\Delta=0.3$ it follows the path less closely than at $\Delta=0.4$. (iv)~The optimization is closed-system and local and tests three durations. At 2133~ns conditional squeezing and displacements lie within 0.003 in $\mathcal F_T$ in closed evolution and at the rates of Ref.~\cite{S-Sivak2023RealTime}, and within 0.013 at the rates of the proposal, with conditional squeezing ahead (Supplementary Table~\ref{tab:proposal}). At 412~ns the ranking rests on 20 tracking starts per gate set (Fig.~4 and Table~I) and 8 target-only starts (Supplementary Table~\ref{tab:capability}), which cannot exclude better pulses. The best tracking run of conditional displacements reaches 0.830, above its target-only best of 0.817, and that of SNAP reaches 0.640, against 0.644, so these values are lower bounds, with a margin far below the gap to conditional squeezing (0.990 along the path and 0.998 without it). (v)~The pulses are open-loop optima of the model Hamiltonian. A 1\% amplitude error lowers the mean closed-evolution $\mathcal F_T$ of the stored preparation pulses by at most 0.0040 and leaves the ranking unchanged, whereas static terms of a device, such as a detuning or a Kerr term, must enter the optimization, as the self-Kerr of the device does in the benchmark of this work (Sec.~\ref{app:kerr}).

\section{\texorpdfstring{Logical Flip and State-to-State Transfer}{Logical Flip and State-to-State Transfer}}
\label{app:flip}

In this section, we benchmark the logical flip as a gate and then the state-to-state transfer between the logical states, whose pulses we also test on the whole code space.

\emph{The logical flip as a gate.} We benchmark the flip as a gate (Supplementary Table~\ref{tab:gate2}) with the gate objective of Sec.~\ref{app:numerics} and the gate metric of Sec.~\ref{app:logical} under noise (8 starts per gate set and duration). Optimized for the gate alone, conditional squeezing implements the flip with $0.987\pm0.001$ at 824~ns, consistently over all starts, and $0.95\pm0.05$ at 412~ns. It does so through qubit--oscillator entanglement (Sec.~\ref{app:pulses}), since without entanglement the oscillator evolves by a Gaussian unitary (Proposition~\ref{thm:nogo}), up to the negligible self-Kerr of the device, and a numerical search finds no Gaussian unitary with a gate fidelity above 0.780, that of the best plain displacement. Conditional displacements reach 0.991 in one start at 824~ns, whereas the other seven stall at 0.772--0.774, the best plain displacement reduced by noise. This stall is a stationary point of the optimization, not a demonstrated limit of the gate set. SNAP reaches at most 0.805. Read through an ideal decoder, the stalled pulses act as an exact logical $X$, but under noise idling scores the same decoded value against the identity, so the decoded metric cannot rank noisy pulses, and we score against the swap $X_L$, which keeps the qubit encoded.

Tracking the brachistochrone images changes the picture for conditional displacements. Their mean rises from 0.80 to 0.91 at $W=70$ (best 0.983), because the tracking term steers six of the eight starts away from the stall. For conditional squeezing, tracking costs fidelity, and no gate set stays close to the images of all three inputs, the smallest closeness over them staying below 0.40 in every start and at every weight. The flip brachistochrone is therefore a reference for the duration and photon cost of a flip that keeps the qubit encoded, not a path that these gate sets follow in our local searches.

\begin{table*}[tbp]
\caption{The logical flip as a gate. Gate fidelity with $X_L$ (Lindblad, cutoff 90) of pulses optimized for the gate alone and with the gate-tracking objective (mean $\pm$ s.d. over 8 starts, best start in brackets). The rows $C_{\min}$ give the smallest closed-evolution closeness over the three inputs of the tracking pulses (the mean over starts at $W=70$, and the largest value over all starts, weights, and both durations, at cutoff 60).}
\label{tab:gate2}
\begin{ruledtabular}
\begin{tabular}{lccc}
 & CS & CD & SNAP\\\hline
gate only, 412~ns & $0.945\pm0.050$ [0.983] & $0.681\pm0.190$ [0.777] & $0.697\pm0.009$ [0.707]\\
gate only, 824~ns & $0.987\pm0.001$ [0.989] & $0.801\pm0.077$ [0.991] & $0.739\pm0.054$ [0.805]\\
tracking, 824~ns, $W=10$ & $0.965\pm0.009$ [0.979] & $0.817\pm0.153$ [0.989] & $0.672\pm0.028$ [0.742]\\
tracking, 824~ns, $W=70$ & $0.954\pm0.005$ [0.961] & $0.914\pm0.116$ [0.983] & $0.671\pm0.026$ [0.735]\\
$C_{\min}$, 824~ns, $W=70$ & 0.27 & 0.10 & 0.05\\
$C_{\min}$, largest & 0.39 & 0.27 & 0.10\\
\end{tabular}
\end{ruledtabular}
\end{table*}

\begin{figure}[tbp]
\includegraphics[width=0.72\textwidth,height=0.42\textheight,keepaspectratio]{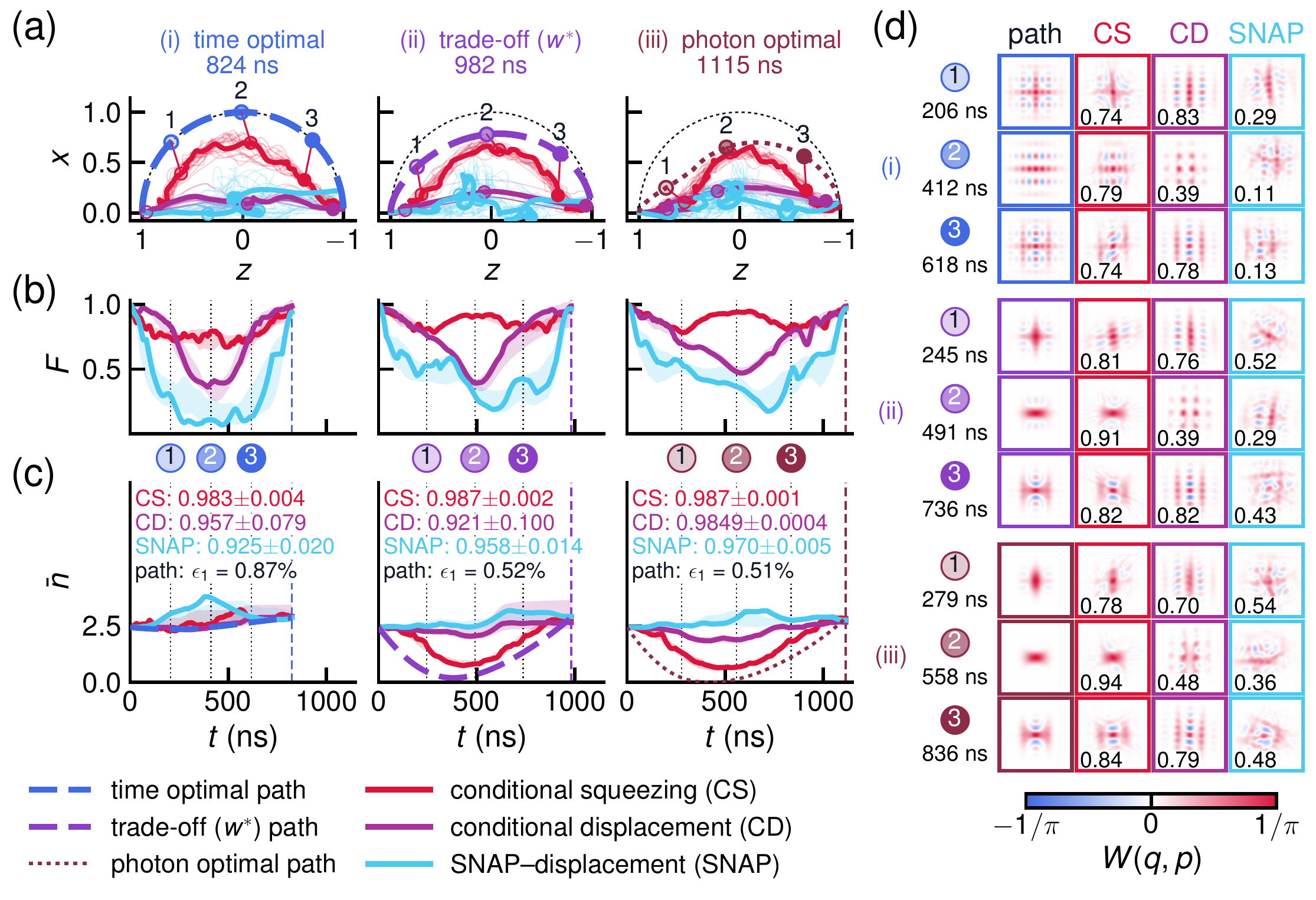}
\caption{The state-to-state transfer $\ket{0_L}\to\ket{1_L}$ (pulses optimized from $\ket{0_L}$ alone) along the time-optimal (i), trade-off (ii) and photon-optimal (iii) paths, each at 2.81 times its minimal time, with CS, CD and SNAP (20 starts each). Panels as in Fig.~4, except that (a) has $z$ horizontal, (c) shows the photon number with $\mathcal F_T$ (mean $\pm$ s.d.) and the path's $\epsilon_1$, and (d), at the right, has one row per time $t_k=kT/4$, $k=1,2,3$, of each path.}
\label{fig:flip}
\end{figure}

\begin{table*}[tbp]
\caption{State-to-state transfer $\ket{0_L}\to\ket{1_L}$ along the three paths at multiples $T/T^*$ of their minimal times $T^*$. The table gives the target fidelity (mean, s.d. in units of $10^{-3}$ when $\geq5\times10^{-4}$) and mean minimum closeness of the three gate sets, and the exposure of the CS pulses (photon~ns).}
\label{tab:flip}
\begin{ruledtabular}
\begin{tabular}{llcccccccc}
$T/T^*$ & path & $T$ (ns) & CS $\mathcal F_T$ & CS $C$ & CS exposure & CD $\mathcal F_T$ & CD $C$ & SNAP $\mathcal F_T$ & SNAP $C$\\\hline
1.17 & time optimal & 344 & 0.988(2) & 0.35 & 1033 & 0.777 & 0.19 & 0.609(146) & 0.17\\
 & trade-off & 410 & 0.990(1) & 0.70 & 807 & 0.777 & 0.31 & 0.644(117) & 0.30\\
 & photon optimal & 466 & 0.989(1) & 0.74 & 861 & 0.776 & 0.34 & 0.799(4) & 0.22\\\hline
1.40 & time optimal & 412 & 0.989(1) & 0.46 & 1171 & 0.777 & 0.19 & 0.677(90) & 0.12\\
 & trade-off & 491 & 0.989(1) & 0.72 & 931 & 0.776 & 0.31 & 0.735(106) & 0.23\\
 & photon optimal & 558 & 0.987(3) & 0.77 & 998 & 0.776 & 0.34 & 0.830(25) & 0.26\\\hline
2.81 & time optimal & 824 & 0.983(4) & 0.63 & 2262 & 0.957(79) & 0.30 & 0.925(20) & 0.08\\
 & trade-off & 982 & 0.987(2) & 0.76 & 1732 & 0.921(100) & 0.37 & 0.958(14) & 0.17\\
 & photon optimal & 1115 & 0.987(1) & 0.79 & 1784 & 0.985 & 0.48 & 0.970(5) & 0.22\\\hline
7.27 & time optimal & 2133 & 0.967(3) & 0.79 & 5917 & 0.975(1) & 0.60 & 0.973(2) & 0.32\\
 & trade-off & 2542 & 0.975(2) & 0.82 & 4135 & 0.975(2) & 0.72 & 0.971 & 0.42\\
 & photon optimal & 2887 & 0.976(2) & 0.82 & 4289 & 0.975(1) & 0.72 & 0.969 & 0.45\\
\end{tabular}
\end{ruledtabular}
\end{table*}

\begin{figure}[tbp]
\includegraphics[width=0.72\textwidth,height=0.42\textheight,keepaspectratio]{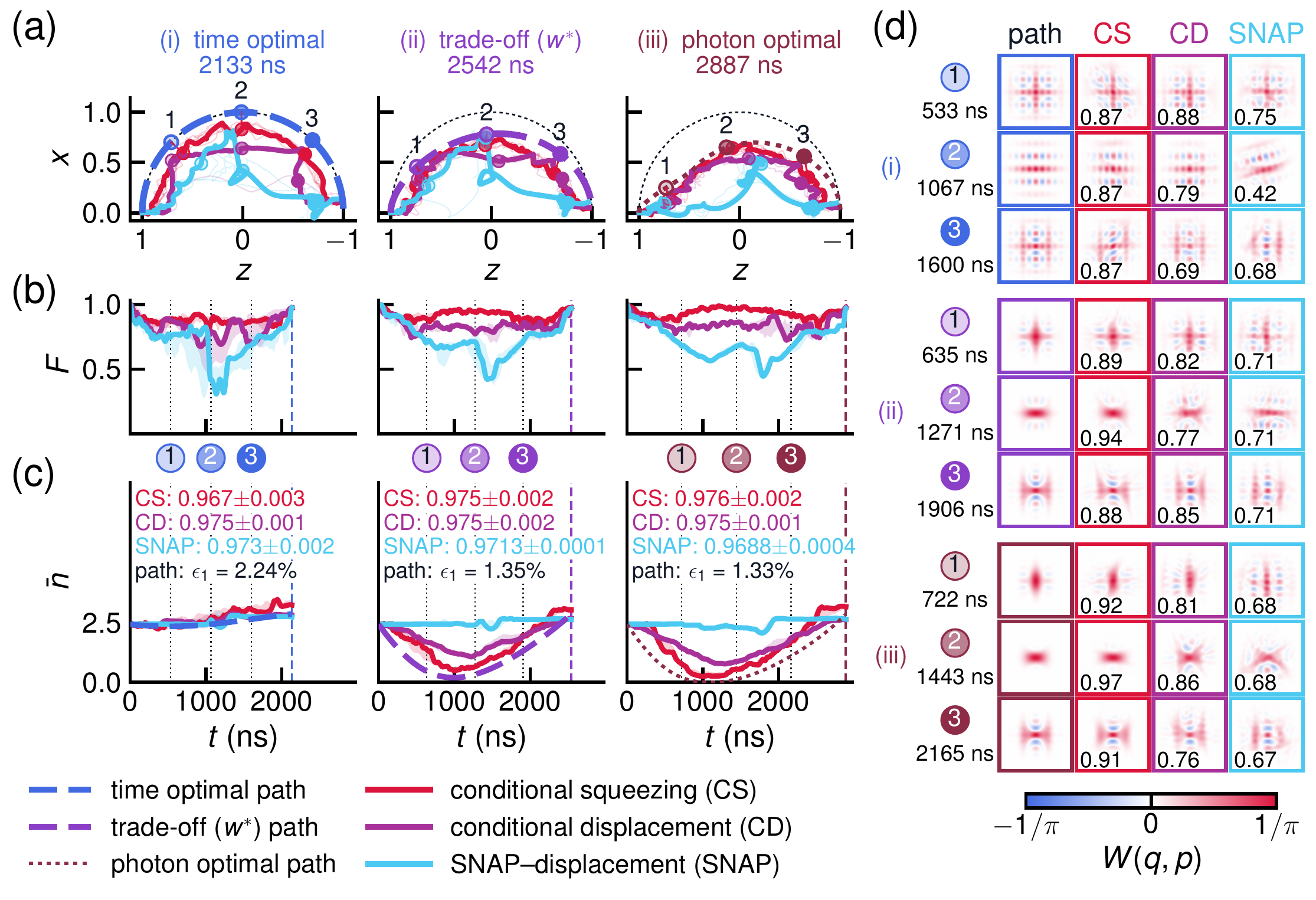}
\caption{As Supplementary Fig.~\ref{fig:flip}, at 7.27 times the minimal time of each path (5 starts for CS and 3 for CD and SNAP).}
\label{fig:S22}
\end{figure}

\begin{table*}[tbp]
\caption{The transfer pulses at $T/T^*=2.81$ applied to the code space (Lindblad, cutoff 90, mean $\pm$ s.d. over 20 starts). The table gives the fidelity of the two basis transfers and the gate fidelity with $X_L$ after the best logical frame update, with the best start in brackets.}
\label{tab:gate}
\begin{ruledtabular}
\begin{tabular}{llccc}
path & gate set & $\ket{0_L}\to\ket{1_L}$ & $\ket{1_L}\to\ket{0_L}$ & gate fidelity\\\hline
brachistochrone & CS & 0.983(4) & 0.12(5) & 0.44(5) [0.53]\\
 & CD & 0.957(79) & 0.949(76) & 0.953(77) [0.987]\\
 & SNAP & 0.925(20) & 0.28(10) & 0.55(6) [0.66]\\
trade-off & CS & 0.987(2) & 0.13(10) & 0.47(8) [0.68]\\
 & CD & 0.921(100) & 0.69(12) & 0.80(6) [0.92]\\
 & SNAP & 0.958(14) & 0.27(10) & 0.54(6) [0.67]\\
photon optimal & CS & 0.987(1) & 0.10(5) & 0.44(4) [0.53]\\
 & CD & 0.985 & 0.41(21) & 0.65(14) [0.85]\\
 & SNAP & 0.970(5) & 0.27(8) & 0.55(5) [0.63]\\\hline
$H_{\rm opt}$, 824~ns & & 0.993 & 0.993 & 0.993\\
\end{tabular}
\end{ruledtabular}
\end{table*}

\emph{The state-to-state transfer between the logical states.} Supplementary Figure~\ref{fig:flip} and Supplementary Table~\ref{tab:flip} give the transfer $\ket{0_L}\to\ket{1_L}$, with pulses optimized from $\ket{0_L}$ alone, along the three paths at 1.17 to 7.27 times their minimal times. Conditional squeezing follows the photon-saving paths more closely than the brachistochrone at every duration, with 15--30\% less exposure of the pulse, consistent with its capacity along them (Sec.~\ref{app:capacity}). The displacement-based gate sets, with $v=0$ on all three paths, also follow the photon-saving paths more closely, so the route through the vacuum is easier to follow in general. Conditional displacements complete the transfer by translating the code and follow none of the paths for $T/T^*\leq2.81$ ($C\leq{}$0.48), nor does SNAP ($C\leq{}$0.31). At $T/T^*=7.27$ (Supplementary Fig.~\ref{fig:S22}) loss and dephasing limit every gate set ($\mathcal F_T={}$0.97--0.98). Applied to the six cardinal states (Supplementary Table~\ref{tab:gate}), only the conditional-displacement pulses along the brachistochrone act as the gate (0.982--0.987 in 17 of 20 starts, with no frame update needed, whereas the other three stall at the best plain displacement). Every conditional-squeezing and SNAP pulse, and most displacement pulses along the photon-saving paths, perform the transfer only, at cutoffs 90 and 120 alike.

\section{\texorpdfstring{Optimized Pulses}{Optimized Pulses}}
\label{app:pulses}

\begin{figure}[tbp]
\includegraphics[width=0.72\textwidth,height=0.42\textheight,keepaspectratio]{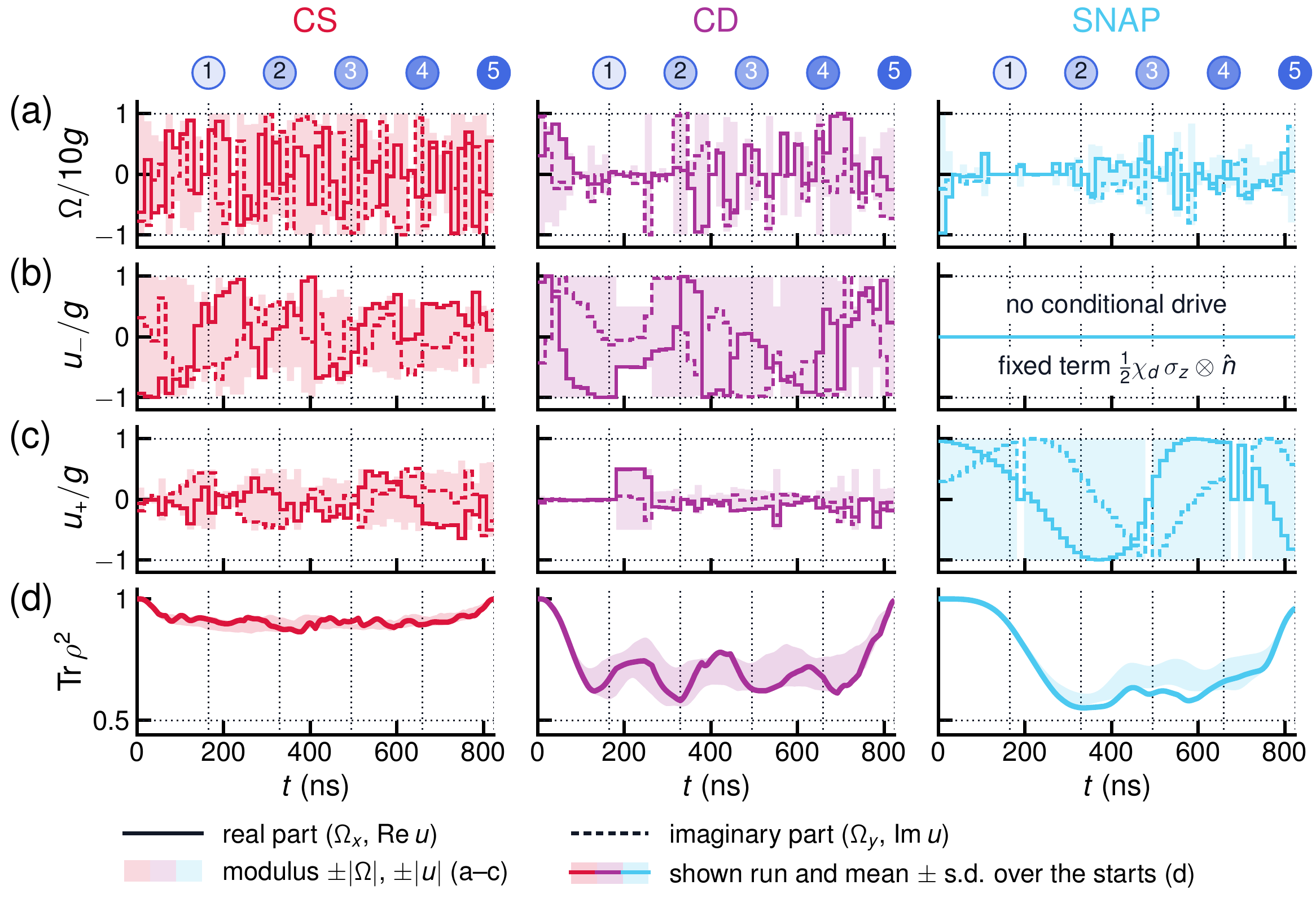}
\caption{Pulses of the GKP preparation runs of Fig.~4(d) at $T=824$~ns, one representative start per gate set, on $N=50$ slices of 16.48~ns and relative to their bounds (dotted lines at $\pm1$), with the real part solid, the imaginary part dashed, and the modulus shaded. (a)~Qubit drive $\Omega$, $|\Omega|\leq10g$. (b,~c)~Conditional part $u_-$ and common part $u_+$ of the branch amplitudes [Eq.~\eqref{eq:upm}]. For SNAP, $u_+=\varepsilon$, $u_-=0$, and the only conditional term is the dispersive coupling, $\chi_d=g$. (d)~Oscillator purity in closed evolution for the shown run (line) and as mean $\pm$ s.d. over the 20 starts (band), with $1/2$ for maximal entanglement (dotted line). Numbered markers give the times $t_k$ of Fig.~4(d). The shown runs reach $\mathcal F_T={}$0.991 (CS), 0.972 (CD), and 0.878 (SNAP) under noise.}
\label{fig:pulses-tracking}
\end{figure}

\begin{figure}[tbp]
\includegraphics[width=0.72\textwidth,height=0.42\textheight,keepaspectratio]{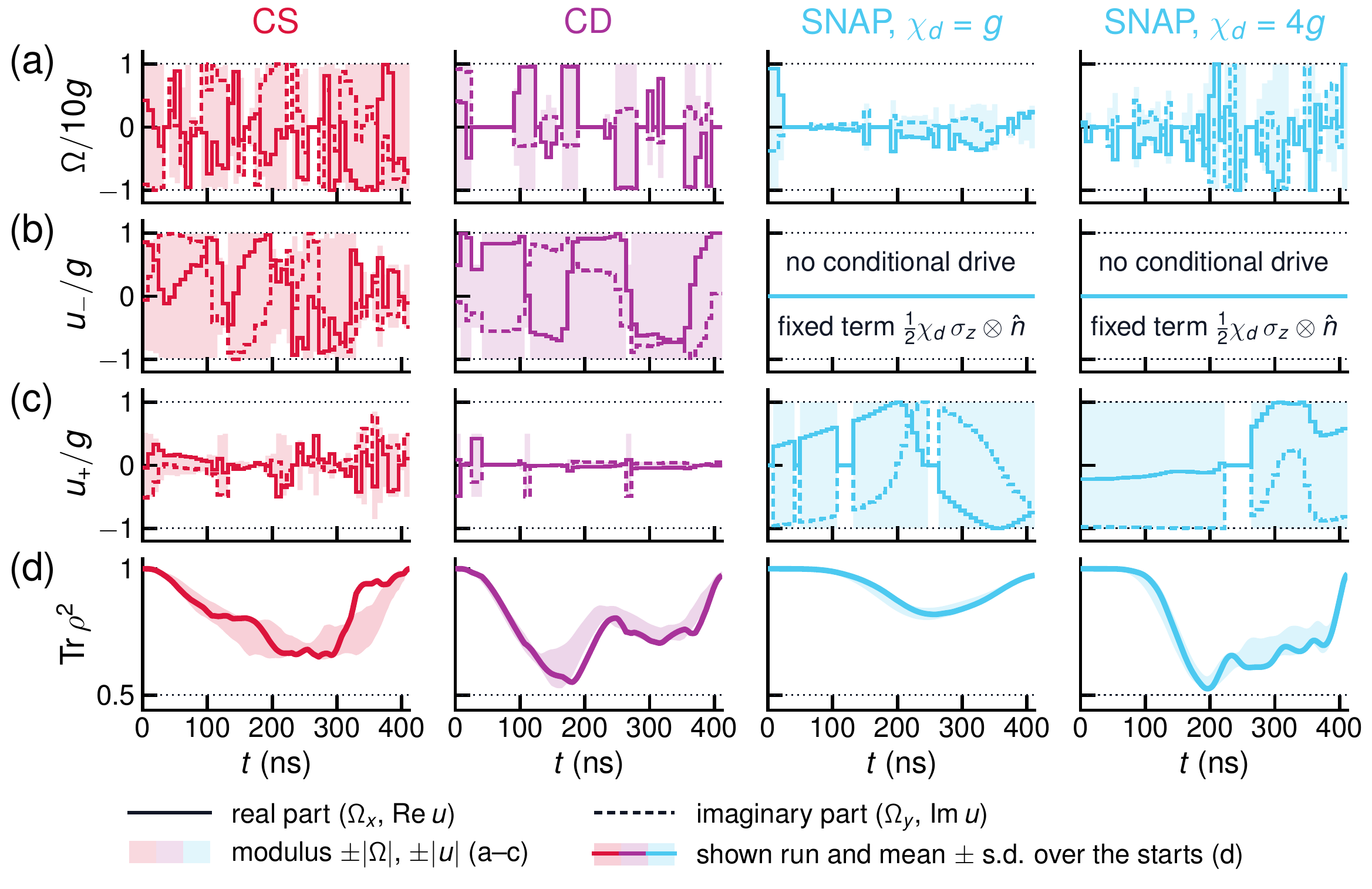}
\caption{Pulses of the GKP preparation of Supplementary Table~\ref{tab:capability} at $T=412$~ns, optimized for the final fidelity alone, for CS, CD, and SNAP with $\chi_d=g$ and $4g$. Each column shows the start whose fidelity under noise is closest to the mean over its 8 starts. Panels as in Supplementary Fig.~\ref{fig:pulses-tracking}, with $N=50$ slices of 8.24~ns and the band in (d) over the 8 starts. The shown runs reach $\mathcal F_T={}$0.997 (CS), 0.807 (CD), 0.629 ($\chi_d=g$), and 0.838 ($\chi_d=4g$).}
\label{fig:pulses-target}
\end{figure}

\begin{figure}[tbp]
\includegraphics[width=0.72\textwidth,height=0.42\textheight,keepaspectratio]{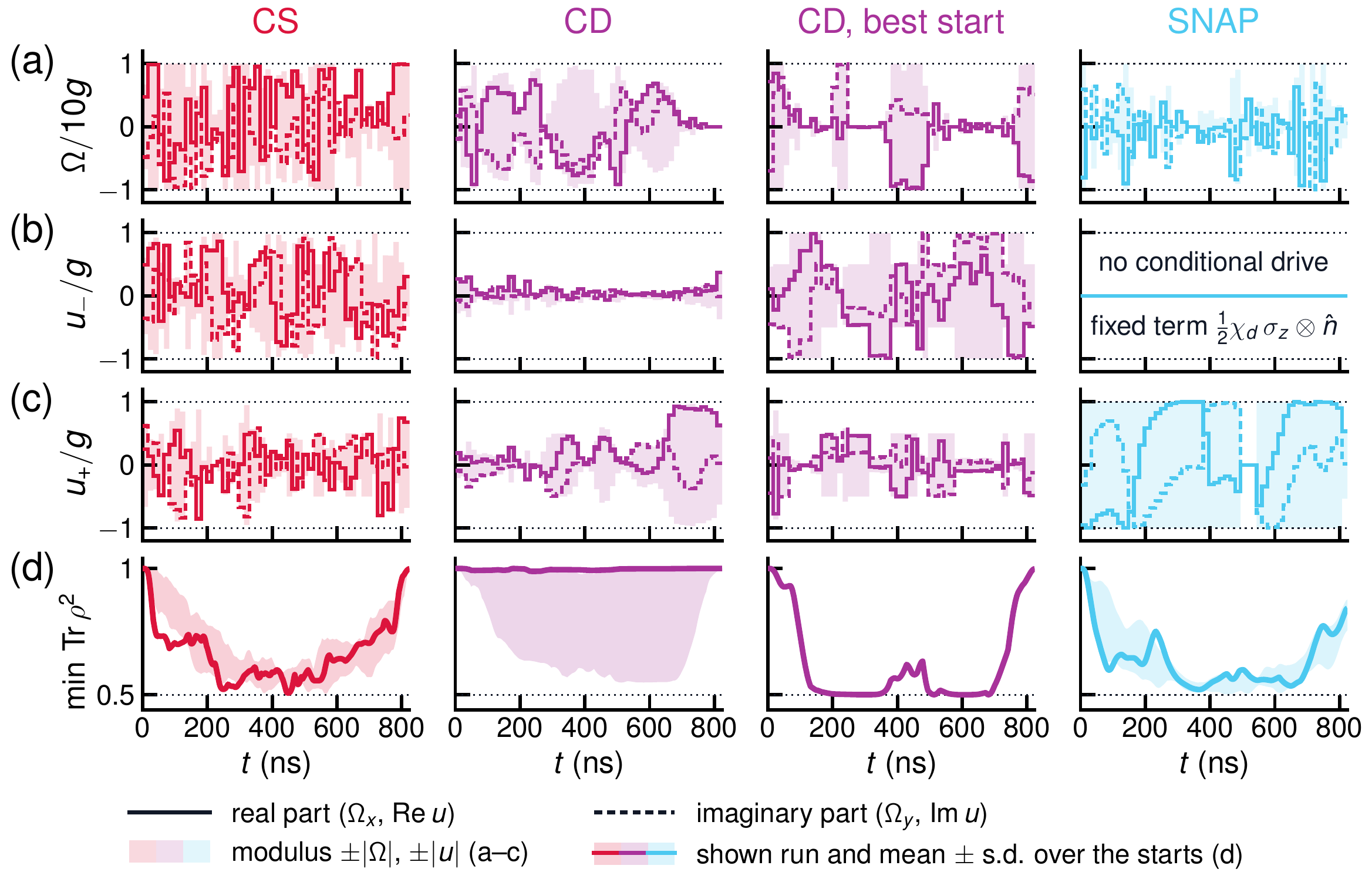}
\caption{Pulses of the logical flip optimized for the gate objective alone at $T=824$~ns (Supplementary Table~\ref{tab:gate2}). For each gate set a column shows the start whose score is closest to the mean over its 8 starts, and for CD a further column shows the best start, the only one that implements the flip. Panels as in Supplementary Fig.~\ref{fig:pulses-tracking}, except that (d) shows the smallest oscillator purity over the inputs $\ket{0_L}$, $\ket{1_L}$, and $\ket{+_L}$. The shown pulses score 0.987 (CS), 0.774 and 0.991 (CD), and 0.743 (SNAP).}
\label{fig:pulses-flip}
\end{figure}

In this section, we show the optimized pulses behind Fig.~4, Supplementary Table~\ref{tab:capability}, and Sec.~\ref{app:flip}, for GKP preparation along the brachistochrone (Supplementary Fig.~\ref{fig:pulses-tracking}), for GKP preparation optimized for the final fidelity alone (Supplementary Fig.~\ref{fig:pulses-target}), and for the logical flip optimized as a gate (Supplementary Fig.~\ref{fig:pulses-flip}). Supplementary Table~\ref{tab:pulses} gives their statistics over all starts. These pulses are reference solutions of the model of Eq.~(10) under one common amplitude budget, not hardware schedules.

\emph{Reading the pulses.} Each figure shows the controls of one representative start per gate set, normalized to their bounds. Row~(a) shows the qubit drive $\Omega$. For CS and CD, rows~(b) and (c) show the branch amplitudes $u_j=\xi_j$ or $\varepsilon_j$ through their conditional and common parts, $u_\pm=(u_{\rm g}\pm u_{\rm e})/2$, since
\begin{equation}
\sum_{j}\ket j\bra j\otimes G(u_j)=1\otimes G(u_+)+\sigma_z\otimes G(u_-),\qquad|u_+|^2+|u_-|^2=\tfrac12\big(|u_{\rm g}|^2+|u_{\rm e}|^2\big)\leq g^2,
\label{eq:upm}
\end{equation}
with $\sigma_z=\ket{\mathrm g}\bra{\mathrm g}-\ket{\mathrm e}\bra{\mathrm e}$. The common part acts on the oscillator whatever the qubit state, as a plain squeezer for CS and a plain displacement for CD, whereas the conditional part acts with opposite signs on the two qubit states. For SNAP $u_-=0$, and the only conditional term is the dispersive term. Row~(d) shows the oscillator purity $\mathrm{Tr}\,\rho^2$ in closed evolution, 1 for a product state and 1/2 for a maximally entangled one, which by Eq.~\eqref{eq:Fbounds} caps the closeness,
\begin{equation}
F\leq\lambda_{\max}=\tfrac12\big[1+\big(2\,\mathrm{Tr}\,\rho^2-1\big)^{1/2}\big].
\label{eq:purity-cap}
\end{equation}

\begin{table*}[tbp]
\caption{Statistics of the optimized pulses, mean (s.d. in units of the last digit) over all starts. The columns give the fraction of oscillator amplitudes within $10^{-3}g$ of their bound $g$ (both branches pooled for CS and CD, the cavity drive for SNAP), the conditional share $\int|u_-|^2dt/\int(|u_+|^2+|u_-|^2)\,dt$ of Eq.~\eqref{eq:upm}, the time-averaged qubit drive $|\Omega|/10g$, the smallest oscillator purity and the largest photon number (closed evolution at the slice ends, cutoff 90, for the flip the extremes over the inputs $\ket{0_L}$, $\ket{1_L}$, and $\ket{+_L}$), and the change $\Delta F$ of the fidelity under Eq.~\eqref{eq:lindblad} ($\mathcal F_T$, or the gate fidelity for the flip), in units of $10^{-3}$, when every control is smoothed with a Gaussian of standard deviation $\sigma$. A dash marks a quantity that SNAP does not have or a duration not smoothed.}
\label{tab:pulses}
\footnotesize
\setlength{\tabcolsep}{4pt}
\begin{ruledtabular}
\begin{tabular}{llcccccccc}
pulses & $T$ (ns) & at bound & cond.\ share & $|\Omega|/10g$ & min $\mathrm{Tr}\,\rho^2$ & max $\bar n$ & $\Delta F$, $\sigma=1$~ns & $\Delta F$, $\sigma=2$~ns & $\Delta F$, $\sigma=4$~ns\\\hline
\multicolumn{10}{l}{GKP preparation along the brachistochrone (Fig.~4, 20 starts)}\\
CS & 412 & 0.82(4) & 0.85(3) & 0.82(3) & 0.775(16) & 2.06(5) & $-0.3(3)$ & $-2.2(11)$ & $-17(9)$\\
 & 824 & 0.53(4) & 0.782(19) & 0.77(3) & 0.852(13) & 2.13(4) & $-0.3(4)$ & $-1.4(8)$ & $-10(3)$\\
 & 2133 & 0.12(3) & 0.67(2) & 0.63(5) & 0.873(12) & 2.31(2) & -- & -- & --\\
CD & 412 & 0.95(3) & 0.972(17) & 0.42(8) & 0.57(3) & 0.79(4) & $-0.82(11)$ & $-2.3(3)$ & $-7.8(13)$\\
 & 824 & 0.95(2) & 0.964(13) & 0.47(6) & 0.56(3) & 1.95(8) & $-0.40(11)$ & $-1.5(3)$ & $-7.3(16)$\\
 & 2133 & 0.53(3) & 0.759(19) & 0.54(3) & 0.805(15) & 2.16(2) & -- & -- & --\\
SNAP$^a$ & 412 & 0.2(4) & -- & 0.32(10) & 0.95(9) & 0.2(4) & $-0.1(2)$ & $-0.3(5)$ & $-0.7(12)$\\
 & 824 & 0.91(6) & -- & 0.23(2) & 0.58(2) & 2.5(4) & $-0.46(14)$ & $-1.3(3)$ & $-4.9(11)$\\
 & 2133 & 0.14(5) & -- & 0.191(10) & 0.597(14) & 2.04(5) & -- & -- & --\\
\hline
\multicolumn{10}{l}{GKP preparation for the final fidelity alone (Supplementary Table~\ref{tab:capability}, 8 starts)}\\
CS & 412 & 0.75(8) & 0.85(5) & 0.77(6) & 0.647(15) & 2.49(5) & $-0.32(9)$ & $-1.7(7)$ & $-12(5)$\\
CD & 412 & 0.90(7) & 0.94(4) & 0.39(3) & 0.58(5) & 0.77(6) & $-0.76(11)$ & $-2.2(3)$ & $-7.3(11)$\\
SNAP, $\chi_d=g$ & 412 & 0.85(5) & -- & 0.190(19) & 0.82(2) & 0.79(9) & $-0.38(10)$ & $-0.9(2)$ & $-2.1(6)$\\
SNAP, $\chi_d=4g$ & 412 & 0.92(5) & -- & 0.55(7) & 0.543(12) & 1.00(11) & $-0.88(19)$ & $-4.0(12)$ & $-28(9)$\\
\hline
\multicolumn{10}{l}{Logical flip optimized as a gate, gate objective alone (Supplementary Table~\ref{tab:gate2}, 8 starts)}\\
CS & 824 & 0.50(8) & 0.73(4) & 0.73(4) & 0.520(15) & 5.4(14) & $-0.38(8)$ & $-2.0(6)$ & $-15(5)$\\
CD & 824 & 0.3(3) & 0.5(3) & 0.48(10) & 0.7(2) & 4.2(5) & $-0.12(7)$ & $-0.5(3)$ & $-2.8(17)$\\
SNAP & 824 & 0.93(5) & -- & 0.36(2) & 0.513(11) & 4.6(10) & $-0.1(6)$ & $-0.9(12)$ & $-8(4)$\\
\end{tabular}
\end{ruledtabular}
\raggedright\footnotesize $^a$15 of the 20 SNAP starts at 412~ns stay at the vacuum (Sec.~\ref{app:numerics}) and are included. Over the other starts the smallest purity is 0.809(18).
\end{table*}

\emph{Preparation along the brachistochrone.} Three features of Supplementary Fig.~\ref{fig:pulses-tracking} hold across the starts (Supplementary Table~\ref{tab:pulses}). First, the shortest pulses spend nearly the whole oscillator budget, with the oscillator amplitudes of CS and CD at their bound in 0.82 and 0.95 of all slices at 412~ns. Second, the conditional part carries most of the oscillator drive in every start. Third, every start that leaves the vacuum entangles qubit and oscillator during the pulse, as Proposition~\ref{thm:nogo} requires, and returns close to a product state by the end. CS entangles less than CD at every duration. By Eq.~\eqref{eq:purity-cap} the smallest purity caps the minimum closeness, on average at 0.92 for CS and 0.67 for CD at 824~ns, against the reached $C=0.90$ and 0.49, so entanglement alone forces most of their loss of closeness. In the CS pulses at 412 and 824~ns, the qubit drive and the conditional part are both at least half their bound in 0.78 and 0.67 of the slices. The alternation of conditional squeezing gates and qubit rotations in Sec.~\ref{app:cs} thus describes the operators these pulses generate, not their shape.

\emph{Preparation for the final fidelity alone.} Supplementary Table~\ref{tab:capability} gives the fidelities that the gate sets reach without the path. Free of it, the CS pulses of Supplementary Fig.~\ref{fig:pulses-target} entangle more deeply than along it (smallest purity 0.65 against 0.77) and reach 0.998. CD entangles even more deeply (0.58) yet reaches only 0.80, so entanglement is necessary but not sufficient. SNAP entangles only through its dispersive term, weakly at $\chi_d=g$ (0.82, $\mathcal F_T=0.63$) and nearly maximally at $\chi_d=4g$ (0.54, $\mathcal F_T=0.84$).

\begin{table}[tbp]
\caption{GKP preparation optimized for the target alone, without a path. Fidelity with $\ket{0_L}$ under noise (Lindblad, Fock cutoff 90), as mean (s.d.) [best] over 8 starts.}
\label{tab:capability}
\begin{ruledtabular}
\begin{tabular}{lcc}
gate set & 412~ns & 824~ns\\\hline
CS & 0.9975(1) [0.998] & 0.9947(5) [0.995]\\
CD & 0.799(18) [0.817] & 0.979(14) [0.986]\\
SNAP, $\chi_d=g$ & 0.628(9) [0.644] & 0.845(25) [0.872]\\
SNAP, $\chi_d=4g$ & 0.840(21) [0.876] & 0.988(1) [0.989]\\
\end{tabular}
\end{ruledtabular}
\end{table}

\emph{Logical flip.} Every CS flip pulse of Supplementary Fig.~\ref{fig:pulses-flip} passes close to a maximally entangled state (smallest purity $0.520\pm0.015$) and reaches $\bar n=5.4\pm1.4$, against $0.852\pm0.013$ and $2.13\pm0.04$ for the CS preparation pulses along the path at the same duration. The flip pulses thus entangle far more deeply, which supports the conclusion of Sec.~\ref{app:flip} that conditional squeezing implements the flip through entanglement. The representative CD pulse is a stalled one, which stays close to a product state and acts as a plain displacement would, whereas the one CD start that implements the flip entangles maximally.

\emph{Finite bandwidth.} The pulses switch instantly between slices. Smoothing every control of the stored pulses at 412 and 824~ns with a normalized Gaussian of standard deviation $\sigma$, without re-optimization, keeps every control within its bound. At $\sigma=1$~ns (a half-power bandwidth of 133~MHz) the mean fidelity of every gate set and protocol changes by less than $10^{-3}$, and at $\sigma=4$~ns (33~MHz) by up to 0.03, with the same order of the gate sets at every $\sigma$.

\emph{Reproduction.} The 180 tracking pulses of Fig.~4 and the 144 pulses of the flip as a gate, re-simulated from their stored slice variables, reproduce the stored results bit for bit, as does the target-only optimization of Supplementary Table~\ref{tab:capability} at 412~ns, repeated from the same seeds.

\section{\texorpdfstring{Photon-Optimal Logical Flips That Keep the Qubit Encoded}{Photon-Optimal Logical Flips That Keep the Qubit Encoded}}
\label{app:flipopt}

In this section, we show that the photon-optimal flip among those that keep the qubit encoded is the time-optimal one, exactly for the averaged exposure and to first order in the noise rates for the averaged error, and contrast it with the transfer of one known input and with flips that leave the code space.

\begin{figure}[tbp]
\includegraphics[width=0.72\textwidth,height=0.42\textheight,keepaspectratio]{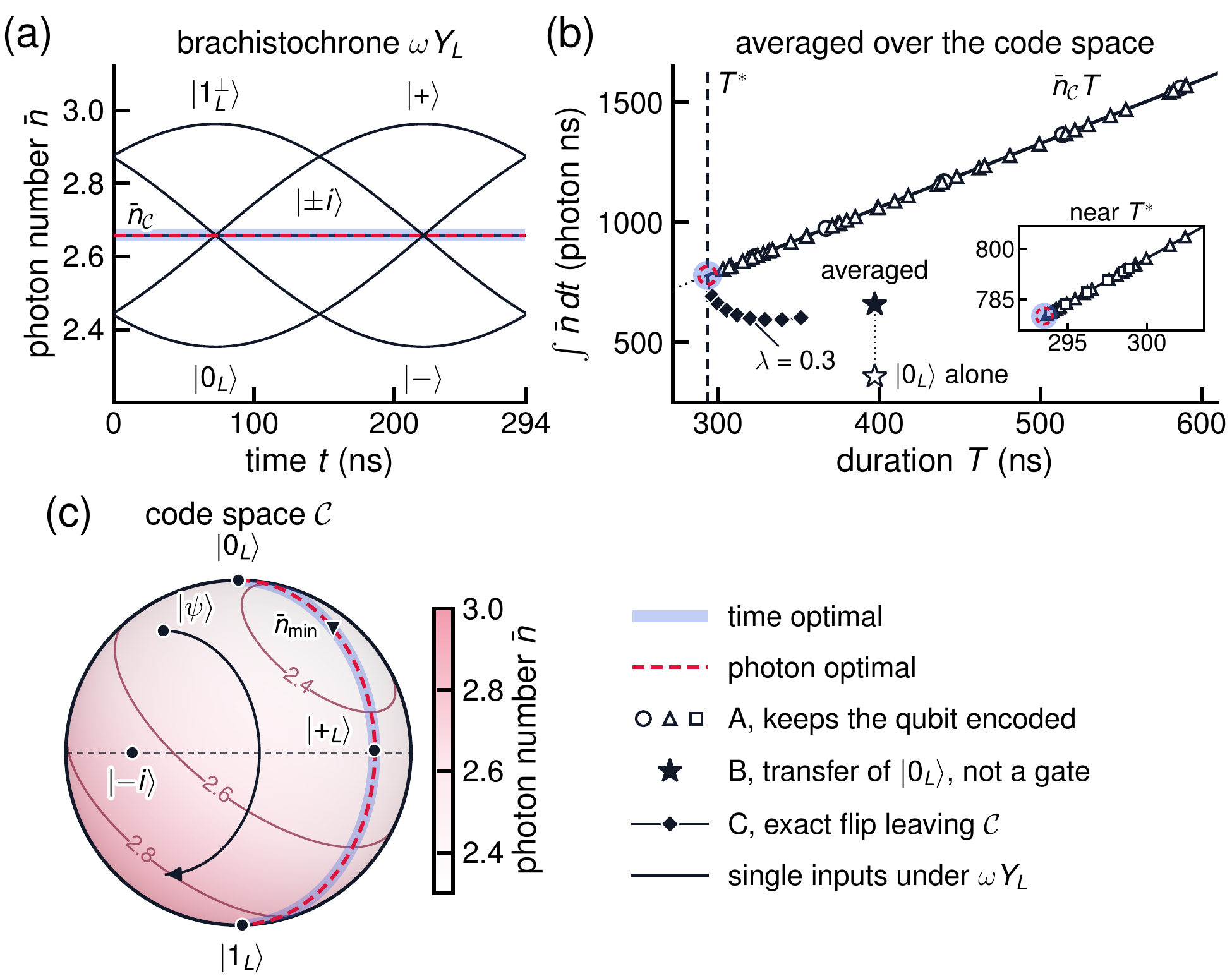}
\caption{Photon-optimal logical flips that keep the qubit encoded ($\Delta=0.4$, Fock cutoff 80). (a)~Photon number along the flip brachistochrone $\omega Y_L$ for the six states of an orthonormal octahedron of the code space (thin lines) and their mean (crimson dashed), which stays within $10^{-14}$ of $\bar n_{\mathcal C}=2.658$ (blue band), the average over the code space. (b)~Exposure $\int\bar n\,dt$ averaged over the code space against the duration $T$. Flips of case A (circles, slower brachistochrones, triangles, random smooth evolutions within the code space, squares, two-leg detours) lie on the line $\bar n_{\mathcal C}T$ (inset, within 3\% of $T^*$), with the brachistochrone (blue disk) at their common minimum. Tilted flips of case C (diamonds, $\lambda=0.1$ to 0.45) fall below the line. The generators of the photon-optimal transfer of case B (stars) give $\ket{0_L}$ alone 359~photon~ns (open star) and 659~photon~ns averaged over the code space (filled star), but are not a gate. (c)~Bloch sphere of the code space shaded by $\bar n$. The brachistochrone carries $\ket{0_L}$ along a great circle (blue band) through the state with the fewest photons (triangle) and through $\ket{+_L}$ at $T^*/2$, and the photon optimum of $\ket{0_L}$ alone within the code space (crimson dashed) coincides with it.}
\label{fig:flipopt}
\end{figure}

\emph{Three cases.} (A)~A logical flip that keeps the qubit encoded is a gate whose evolution satisfies $U(t)\mathcal C=\mathcal C$ at all times. (B)~A state-to-state transfer $\ket{0_L}\to\ket{1_L}$ only has to steer the one known input $\ket{0_L}$ to its target~\cite{S-Carlini2006,S-Carlini2007,S-Khaneja2005}, and it is not a gate. (C)~An exact flip that leaves the code space is a gate whose evolution leaves $\mathcal C$ during the operation and returns at $T$.

\emph{The problem for case A.} A gate acts on unknown inputs, so we measure its photon cost by the exposure averaged uniformly over the code space,
\begin{equation}
E[U]=\int_0^T\langle\bar n\rangle_{\mathcal C}(t)\,dt,\qquad\langle\bar n\rangle_{\mathcal C}(t)=\int\bra\psi U(t)^\dagger\nop\,U(t)\ket\psi\,d\psi,
\label{eq:flipopt-E}
\end{equation}
which sets the average loss error $\kappa E$ to first order (Lemma~\ref{lem:loss}), together with the average first-order error $\bar\epsilon_1$ of Proposition~\ref{thm:code}. We call a flip admissible if its oscillator Hamiltonian $H(t)$, $t\in[0,T]$, generates an evolution that keeps the qubit encoded and implements the flip, $U(T)=X_Le^{-i\varphi Z_L/2}$ on $\mathcal C$ for some $\varphi$, with $\Delta H\leq\omega$ in the state that starts as $\ket{0_L}$. Over the admissible flips we minimize the duration $T$, the exposure $E$, or the error $\bar\epsilon_1$.

\begin{proposition}[photon-optimal flips that keep the qubit encoded]\label{prop:flipopt}
(i) With $U(0)=1$, the evolution keeps the qubit encoded if and only if $[H(t),\Pi]=0$ at all times. (ii) Every admissible flip has $E=\bar n_{\mathcal C}T$ and $\bar\epsilon_1=T(\Gamma_q+\kappa\bar n_{\mathcal C}+\kappa_\phi V_{\mathcal C})$, whatever its Hamiltonian, so minimizing $E$ or $\bar\epsilon_1$ over the admissible flips is the same problem as minimizing $T$, and this remains true under any further constraint on $H(t)$. (iii) Every admissible flip takes $T\geq T^*=\theta_L/\omega$, and every flip with $T=T^*$ moves $\ket{0_L}$ along the great circle of the flip brachistochrone at speed $\omega$, which fixes $U(t)$ on $\mathcal C$ up to phases that change no cost. The flip brachistochrone therefore solves all three problems, with $E=\bar n_{\mathcal C}T^*=780$~photon~ns.
\end{proposition}
\begin{proof}
(i) $P(t)=U(t)\Pi U(t)^\dagger$ obeys $\dot P=-i[H,P]$ with $P(0)=\Pi$. If $[H,\Pi]=0$, the constant $P=\Pi$ is the unique solution, and conversely $P(t)=\Pi$ at all times gives $0=\dot P=-i[H,\Pi]$. (ii) By (i), $U(t)$ restricted to $\mathcal C$ is a unitary of $\mathcal C$, and Proposition~\ref{thm:code} gives $\langle\bar n\rangle_{\mathcal C}(t)=\tfrac12\mathrm{Tr}(U\Pi U^\dagger\nop)=\tfrac12\mathrm{Tr}(\Pi\nop)=\bar n_{\mathcal C}$ and the constant average variance $V_{\mathcal C}$ at every $t$, which we integrate over $[0,T]$. (iii) A flip maps $\ket{0_L}$ to $X_Le^{-i\varphi Z_L/2}\ket{0_L}=e^{-i\varphi/2}\ket{1_L}$, at Fubini--Study distance $\theta_L$, and the length $\int_0^T\Delta H\,dt\leq\omega T$ of the trajectory of $\ket{0_L}$ cannot be shorter than that distance~\cite{S-Anandan1990,S-MandelstamTamm1945}, so $\theta_L\leq\omega T$ (Corollary~\ref{cor:flip}). Equality forces the trajectory to be a shortest path, which for $0<\theta_L<\pi/2$ is the unique great-circle arc in $\mathcal C$, traversed at speed $\omega$. Since $U(t)$ is unitary on $\mathcal C$, it then maps the state of $\mathcal C$ orthogonal to $\ket{0_L}$ to the state orthogonal to the arc, each up to a phase. These phases change neither the averages nor the logical action beyond the frame update, and the Hamiltonian on the complement of $\mathcal C$ changes no cost. The flip brachistochrone $\omega Y_L$ keeps the qubit encoded, has $\Delta H\leq\omega$ on every code state, and implements $X_LZ_L$ at $T^*$ (Sec.~\ref{sec:flip-hopt}).
\end{proof}

Within case A the photon cost depends on the duration only, so the time optimization solves the photon optimization. This concerns the uniform average over inputs, which sets the average gate fidelity. The exposure of single inputs ranges from 723 to 838~photon~ns along the brachistochrone, and only their average is constant [Supplementary Fig.~\ref{fig:flipopt}(a)]. All 104 flips of case A that we propagated lie on the line $E=\bar n_{\mathcal C}T$ to a relative $10^{-15}$, and none is faster than $T^*$ [Supplementary Fig.~\ref{fig:flipopt}(b)].

\emph{Case B.} The photon optimum of the transfer of $\ket{0_L}$ alone runs through the vacuum, lasts 397~ns, and exposes $\ket{0_L}$ to 359~photon~ns, about half of its 739~photon~ns along the brachistochrone, but its generators map $\ket{1_L}$ to $\ket{0_L}$ with fidelity 0.60 only (Sec.~\ref{app:logical}). Confined to the code space, the photon optimum of $\ket{0_L}$ alone is the great circle of the brachistochrone itself, a local minimum that a numerical search does not improve [Supplementary Fig.~\ref{fig:flipopt}(c)]. The whole saving of the transfer therefore comes from leaving the code space.

\emph{Case C.} If the evolution leaves $\mathcal C$, part (i) fails and the averaged photon number is no longer conserved. The tilted flips of Sec.~\ref{sec:exactflip} lower the averaged exposure by 11\%, 19\%, and 23\% at 1\%, 4\%, and 9\% longer duration and fall below the line $\bar n_{\mathcal C}T$ [Supplementary Fig.~\ref{fig:flipopt}(b)]. Lower photon costs for gates therefore require leaving the code space, for example by moving the code plane through regions with fewer photons or by holding the logical state in the control qubit while the oscillator passes through the vacuum.

\section{\texorpdfstring{Cavity Self-Kerr and the Envelope Width}{Cavity Self-Kerr and the Envelope Width}}
\label{app:kerr}

In this section, we include the self-Kerr of the cavity in every benchmark and quantify its effect, and we repeat the benchmark at other envelope widths. Unless stated otherwise, the protocol, bounds, noise rates, cutoffs, and random starts are those of Fig.~4 (Sec.~\ref{app:numerics}).

\subsection{\texorpdfstring{Self-Kerr of the Device in Every Benchmark}{Self-Kerr of the Device in Every Benchmark}}
\label{sec:kerr-device}
Every pulse optimization and evaluation of the main text and of Secs.~\ref{app:numerics}--\ref{app:pulses} includes $H_K=\tfrac K2\hat a^{\dagger2}\aop^2$ with $K/2\pi=-4.8$~Hz (Sec.~\ref{app:noise}), except the runs with the Hamiltonian of the CS proposal in Sec.~\ref{app:robust}, which carry the far larger Kerr term of that circuit. With $K=0$, the code reproduces the stored pulses of the version without the Kerr term bit for bit from the same random starts (three runs at 412~ns, every slice variable and every reported quantity identical). With the Kerr term, L-BFGS-B follows a different route from the same start and ends in another local optimum, as it does when the random start is shifted by only $10^{-12}$, so single pulses change while the means over the 20 starts agree within 2.1 standard errors for every gate set and duration (Supplementary Table~\ref{tab:kerrpaired}). The Kerr term itself changes the target fidelity of a fixed pulse by at most $3\times10^{-5}$ under the Lindblad equation (Sec.~\ref{sec:kerr-absent}), far below the spread over the starts. For CS, the means agree to $10^{-4}$ in $\mathcal F_T$ and to 0.002 in $C$, although single starts move by up to 0.062 in $C$.

\begin{table*}[tbp]
\caption{GKP preparation along the brachistochrone (Fig.~4) without and with the self-Kerr of the device, from the same 20 random starts. Target fidelity $\mathcal F_T$ and minimum closeness $C$ under the Lindblad equation (cutoff 90), mean (s.d.), and the difference of the means with its standard error.}
\label{tab:kerrpaired}
\begin{ruledtabular}
\begin{tabular}{lccccccc}
gate set & $T$ (ns) & $\mathcal F_T$, $K=0$ & $\mathcal F_T$, $K/2\pi=-4.8$~Hz & $C$, $K=0$ & $C$, $K/2\pi=-4.8$~Hz & $\delta\mathcal F_T$ & $\delta C$\\\hline
CS & 412 & 0.9903(2) & 0.9903(2) & 0.83(2) & 0.82(2) & $0.0000(1)$ & $-0.002(6)$\\
CS & 824 & 0.9901(2) & 0.9901(4) & 0.90(1) & 0.90(1) & $0.0000(1)$ & $0.000(3)$\\
CS & 2133 & 0.9857(6) & 0.9857(6) & 0.91(1) & 0.91(1) & $0.0000(2)$ & $0.000(3)$\\
CD & 412 & 0.8073(94) & 0.8093(90) & 0.62(3) & 0.62(4) & $+0.0019(29)$ & $-0.001(12)$\\
CD & 824 & 0.9795(33) & 0.9768(48) & 0.51(6) & 0.49(5) & $-0.0027(13)$ & $-0.020(18)$\\
CD & 2133 & 0.9841(10) & 0.9839(10) & 0.84(1) & 0.84(1) & $-0.0002(3)$ & $-0.006(4)$\\
SNAP & 412 & 0.5762(452) & 0.5662(394) & 0.48(9) & 0.51(7) & $-0.0100(134)$ & $+0.021(26)$\\
SNAP & 824 & 0.8344(1008) & 0.8644(220) & 0.19(12) & 0.19(6) & $+0.0300(231)$ & $-0.005(31)$\\
SNAP & 2133 & 0.9787(12) & 0.9789(10) & 0.55(3) & 0.56(2) & $+0.0002(3)$ & $+0.014(8)$\\
\end{tabular}
\end{ruledtabular}
\end{table*}

\subsection{\texorpdfstring{A Kerr Term Absent from the Model}{A Kerr Term Absent from the Model}}
\label{sec:kerr-absent}
A Kerr term that the optimization does not include, or a miscalibration $\delta K$ of a modeled one, changes the fidelity of a pulse. Supplementary Table~\ref{tab:kerrsweep} and Supplementary Fig.~\ref{fig:kerr}(c) re-evaluate the pulses optimized without the Kerr term with $K$ up to $10^4$ times the device value. At the device value the change is of either sign and of first order in $K$, because the pulses do not reach their targets exactly, and it stays below $3\times10^{-5}$ for all 427 pulses, below the cutoff uncertainty of the pulses (Sec.~\ref{app:numerics}). Between $K/2\pi=-0.48$ and $-4.8$~kHz the mean loss of each group of Supplementary Table~\ref{tab:kerrsweep} grows 20 to 99 times, faster than linearly and at most quadratically in $K$, except for the CD and SNAP pulses at 412~ns, which still gain at $-0.48$~kHz, and it grows with the duration of the pulse and with the photon number along it. At $K/2\pi=-4.8$~kHz the 412-ns CS pulses lose 0.012, whereas the 2133-ns pulses of every gate set lose 0.15--0.17, so the short pulses are the least exposed. The CD and SNAP pulses at 412~ns change least because they hold few photons. Between the logical states, CS pulses that follow the trade-off or the photon-optimal path lose 0.067 and 0.069, against 0.092 along the brachistochrone, although they last longer, whereas CD and SNAP pulses, which stay far from every path, lose more along these paths. For preparation, CS pulses along the photon-optimal path lose 0.013, 0.039, and 0.14 at their three durations, against 0.012, 0.048, and 0.17 along the brachistochrone.

\begin{table*}[tbp]
\caption{Change of the target fidelity of the stored pulses, optimized without a Kerr term, when the self-Kerr $\tfrac K2\hat a^{\dagger2}\aop^2$ is switched on, $K/2\pi=-4.8$~Hz (device), $-0.48$, $-4.8$, and $-48$~kHz (Lindblad equation at the rates of Ref.~\cite{S-Sivak2023RealTime}, cutoff 90, mean over 20 starts). The flip runs are those at $T/T^*=2.81$, so their durations differ between the paths.}
\label{tab:kerrsweep}
\begin{ruledtabular}
\begin{tabular}{llcccccccc}
transfer & path & gate set & $T$ (ns) & $\mathcal F_T(K=0)$ & $-4.8$~Hz & $-0.48$~kHz & $-4.8$~kHz & $-48$~kHz\\\hline
GKP & brach. & CS & 412 & 0.9903 & $-9.1\times10^{-8}$ & $-1.4\times10^{-4}$ & $-1.2\times10^{-2}$ & $-3.2\times10^{-1}$\\
GKP & brach. & CS & 824 & 0.9901 & $-1.1\times10^{-7}$ & $-7.8\times10^{-4}$ & $-4.8\times10^{-2}$ & $-4.3\times10^{-1}$\\
GKP & brach. & CS & 2133 & 0.9857 & $-5.8\times10^{-7}$ & $-3.5\times10^{-3}$ & $-1.7\times10^{-1}$ & $-4.9\times10^{-1}$\\
GKP & brach. & CD & 412 & 0.8073 & $3.6\times10^{-8}$ & $2.0\times10^{-6}$ & $-1.2\times10^{-4}$ & $-1.5\times10^{-2}$\\
GKP & brach. & CD & 824 & 0.9795 & $-1.5\times10^{-8}$ & $-9.4\times10^{-5}$ & $-9.3\times10^{-3}$ & $-4.5\times10^{-1}$\\
GKP & brach. & CD & 2133 & 0.9841 & $-4.9\times10^{-7}$ & $-2.2\times10^{-3}$ & $-1.5\times10^{-1}$ & $-4.9\times10^{-1}$\\
GKP & brach. & SNAP & 412 & 0.5762 & $1.5\times10^{-8}$ & $1.3\times10^{-6}$ & $-8.0\times10^{-6}$ & $-2.0\times10^{-3}$\\
GKP & brach. & SNAP & 824 & 0.8344 & $-4.9\times10^{-7}$ & $-2.1\times10^{-4}$ & $-1.6\times10^{-2}$ & $-3.4\times10^{-1}$\\
GKP & brach. & SNAP & 2133 & 0.9787 & $-5.6\times10^{-7}$ & $-2.9\times10^{-3}$ & $-1.7\times10^{-1}$ & $-6.0\times10^{-1}$\\
GKP & photon-opt. & CS & 439 & 0.9904 & $1.0\times10^{-7}$ & $-1.4\times10^{-4}$ & $-1.3\times10^{-2}$ & $-3.2\times10^{-1}$\\
GKP & photon-opt. & CS & 879 & 0.9901 & $3.9\times10^{-7}$ & $-5.7\times10^{-4}$ & $-3.9\times10^{-2}$ & $-4.4\times10^{-1}$\\
GKP & photon-opt. & CS & 2275 & 0.9858 & $-1.8\times10^{-7}$ & $-3.0\times10^{-3}$ & $-1.4\times10^{-1}$ & $-5.1\times10^{-1}$\\
flip & brach. & CS & 824 & 0.9835 & $5.8\times10^{-7}$ & $-2.4\times10^{-3}$ & $-9.2\times10^{-2}$ & $-6.6\times10^{-1}$\\
flip & brach. & CD & 824 & 0.9890 & $1.5\times10^{-7}$ & $-1.5\times10^{-3}$ & $-8.9\times10^{-2}$ & $-5.1\times10^{-1}$\\
flip & brach. & SNAP & 824 & 0.9139 & $1.3\times10^{-6}$ & $-1.8\times10^{-3}$ & $-1.0\times10^{-1}$ & $-5.6\times10^{-1}$\\
flip & trade-off & CS & 982 & 0.9873 & $1.1\times10^{-6}$ & $-2.6\times10^{-3}$ & $-6.7\times10^{-2}$ & $-4.3\times10^{-1}$\\
flip & trade-off & CD & 982 & 0.9121 & $-4.7\times10^{-7}$ & $-2.3\times10^{-3}$ & $-1.1\times10^{-1}$ & $-5.1\times10^{-1}$\\
flip & trade-off & SNAP & 982 & 0.9592 & $1.8\times10^{-7}$ & $-2.5\times10^{-3}$ & $-1.3\times10^{-1}$ & $-6.0\times10^{-1}$\\
flip & photon-opt. & CS & 1115 & 0.9868 & $-1.8\times10^{-6}$ & $-3.4\times10^{-3}$ & $-6.9\times10^{-2}$ & $-4.5\times10^{-1}$\\
flip & photon-opt. & CD & 1115 & 0.9847 & $-8.7\times10^{-7}$ & $-2.8\times10^{-3}$ & $-1.3\times10^{-1}$ & $-5.7\times10^{-1}$\\
flip & photon-opt. & SNAP & 1115 & 0.9683 & $-7.2\times10^{-8}$ & $-3.2\times10^{-3}$ & $-1.5\times10^{-1}$ & $-6.0\times10^{-1}$\\
\end{tabular}
\end{ruledtabular}
\end{table*}

\subsection{\texorpdfstring{A Kerr Term in the Model}{A Kerr Term in the Model}}
\label{sec:kerr-model}
A Kerr term that the model includes is absorbed by the optimization. Supplementary Table~\ref{tab:kerrmodel} and Supplementary Fig.~\ref{fig:kerr}(d) re-optimize starts 1--5 of Fig.~4 at 412 and 824~ns with $K/2\pi=-4.8$ and $-48$~kHz in the model, a thousand and ten thousand times the device value, and with $-0.43$~MHz, the Kerr term of the circuit of Ref.~\cite{S-DelGrosso2025} (Sec.~\ref{app:robust}), and evaluate the pulses with the same term. At $-4.8$ and $-48$~kHz, where the pulses optimized without the term lose 0.012 and 0.32 (CS, 412~ns, Sec.~\ref{sec:kerr-absent}), the re-optimized CS pulses keep $\mathcal F_T=0.990$, the threshold of the selection rule, and follow the path as closely as with the device Kerr ($C=0.83$--0.89, against 0.82--0.90). At $-0.43$~MHz they still keep $\mathcal F_T=0.990$ but follow the path less closely ($C=0.65$ and 0.64), since a Kerr term this strong rotates the phase of $\ket4$ by 6.7~rad within 412~ns, a motion that the brachistochrone of the model without it does not contain. A Kerr term is non-Gaussian by itself, so that displacements alone create non-Gaussian states, and the displacement-based gate sets gain from the strongest Kerr term. At $-0.43$~MHz, CD reaches 0.838 and 0.983 at 412 and 824~ns, against 0.810 and 0.976 with the device Kerr, SNAP reaches 0.919 at 824~ns, against 0.860, and no SNAP start stays at the zero-drive point. CS still prepares $\ket{0_L}$ with the highest fidelity at every Kerr value.

\begin{table}[tbp]
\caption{Starts 1--5 of Fig.~4 at 412 and 824~ns, re-optimized with a large self-Kerr in the model and evaluated with it, against the same starts with the device Kerr ($K/2\pi=-4.8$~Hz). Target fidelity $\mathcal F_T$ and minimum closeness $C$ under the Lindblad equation, mean (s.d.), with the s.d. in units of the last digit and omitted when it rounds to zero. $K/2\pi=-0.43$~MHz is the Kerr term of the circuit of Ref.~\cite{S-DelGrosso2025}. At 412~ns, two, five, and none of the five SNAP starts stay at the zero-drive point at $-4.8$~kHz, $-48$~kHz, and $-0.43$~MHz, against three with the device Kerr.}
\label{tab:kerrmodel}
\begin{ruledtabular}
\begin{tabular}{lcccccc}
$K/2\pi$ & gate set & $T$ (ns) & $\mathcal F_T$ & $C$ & $\mathcal F_T$, device & $C$, device\\\hline
$-4.8$~kHz & CS & 412 & 0.990 & 0.84(1) & 0.990 & 0.82(3)\\
 & CS & 824 & 0.990 & 0.89(1) & 0.990 & 0.90(1)\\
 & CD & 412 & 0.810(11) & 0.63(3) & 0.810(7) & 0.65(1)\\
 & CD & 824 & 0.981(1) & 0.52(7) & 0.976(5) & 0.52(4)\\
 & SNAP & 412 & 0.597(48) & 0.44(10) & 0.579(48) & 0.49(9)\\
 & SNAP & 824 & 0.874(8) & 0.15(4) & 0.860(20) & 0.21(5)\\
$-48$~kHz & CS & 412 & 0.990 & 0.83(1) & 0.990 & 0.82(3)\\
 & CS & 824 & 0.990 & 0.89(1) & 0.990 & 0.90(1)\\
 & CD & 412 & 0.817(7) & 0.62(2) & 0.810(7) & 0.65(1)\\
 & CD & 824 & 0.982(3) & 0.57(5) & 0.976(5) & 0.52(4)\\
 & SNAP & 412 & 0.544 & 0.54 & 0.579(48) & 0.49(9)\\
 & SNAP & 824 & 0.854(34) & 0.14(7) & 0.860(20) & 0.21(5)\\
$-0.43$~MHz & CS & 412 & 0.990 & 0.65(4) & 0.990 & 0.82(3)\\
 & CS & 824 & 0.990 & 0.64(3) & 0.990 & 0.90(1)\\
 & CD & 412 & 0.838(8) & 0.56(2) & 0.810(7) & 0.65(1)\\
 & CD & 824 & 0.983(1) & 0.44(1) & 0.976(5) & 0.52(4)\\
 & SNAP & 412 & 0.636(11) & 0.46(3) & 0.579(48) & 0.49(9)\\
 & SNAP & 824 & 0.919(25) & 0.25(3) & 0.860(20) & 0.21(5)\\
\end{tabular}
\end{ruledtabular}
\end{table}

\subsection{\texorpdfstring{The Optimal Paths under a Kerr Term}{The Optimal Paths under a Kerr Term}}
\label{sec:kerr-paths}
\begin{figure*}[tbp]
\includegraphics[width=0.72\textwidth,height=0.42\textheight,keepaspectratio]{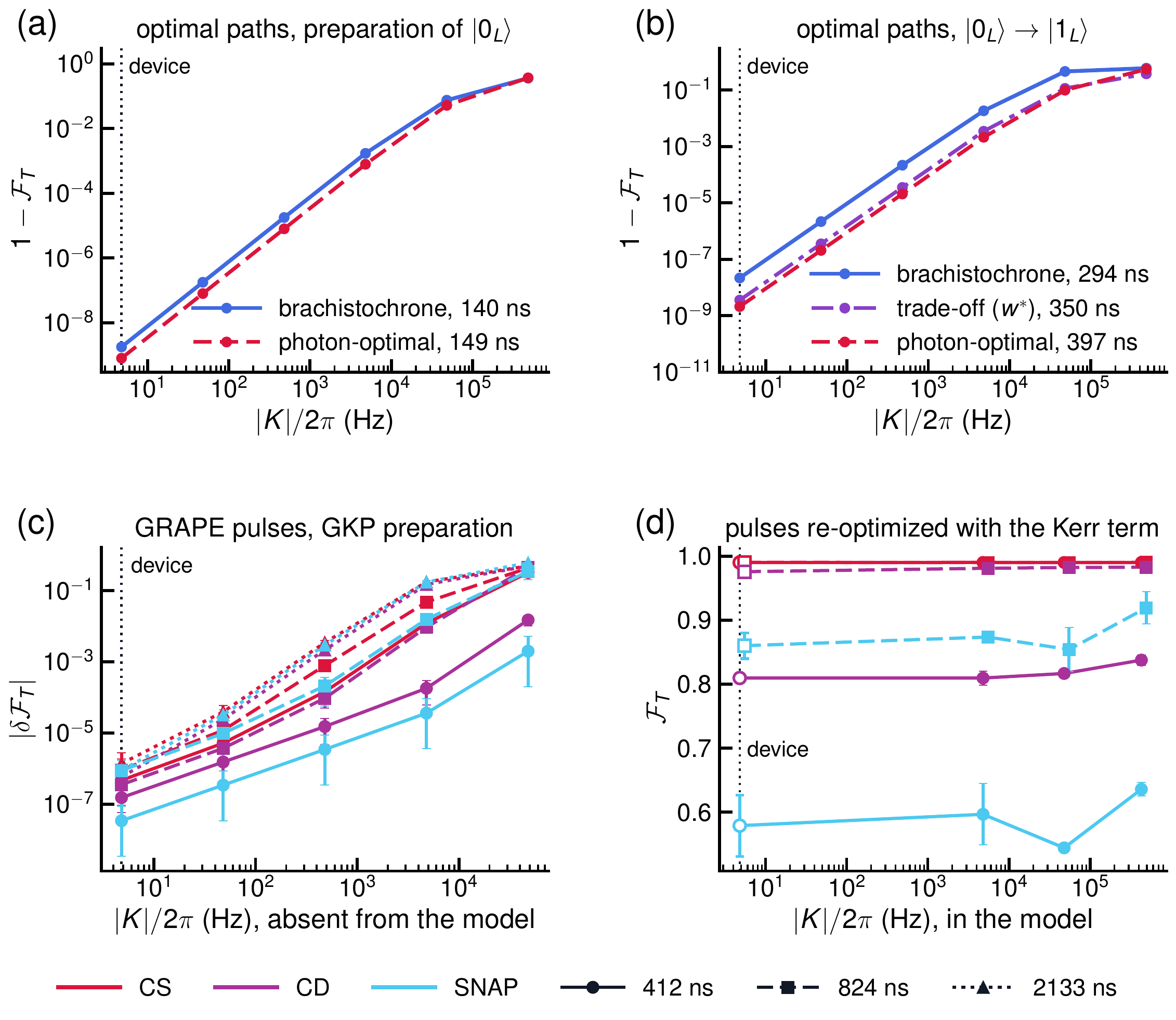}
\caption{Sensitivity to a cavity self-Kerr $\tfrac K2\hat a^{\dagger2}\aop^2$ against $|K|/2\pi$, with the device value $K/2\pi=-4.8$~Hz of Ref.~\cite{S-Sivak2023RealTime} dotted. (a)~Infidelity of the preparation brachistochrone and of the photon-optimal path generated without compensating the Kerr term (closed evolution, cutoff 90). (b)~The same for the flip brachistochrone, the trade-off path at $w^*$, and the photon-optimal transfer through the vacuum. (c)~Change $|\delta\mathcal F_T|$ of the GRAPE pulses optimized without a Kerr term from the 20 starts of Fig.~4, under the Lindblad equation, mean over the 20 starts with one standard deviation (lower bars cut where the deviation exceeds the mean), for CS, CD, and SNAP at 412, 824, and 2133~ns. (d)~Target fidelity of the pulses re-optimized with the Kerr term in the model (filled) against the same starts with the device Kerr only (open), at 412 and 824~ns, mean over the starts with one standard deviation (the 824-ns points are shifted slightly to the right).}
\label{fig:kerr}
\end{figure*}

Supplementary Table~\ref{tab:kerrpaths} and Supplementary Fig.~\ref{fig:kerr}(a,~b) generate each optimal path by its rank-two generator, as in Sec.~\ref{app:noise}, now with a self-Kerr that the generator does not compensate (closed evolution, cutoff 90). Since the paths reach their targets exactly, the error is quadratic in $K$, up to $K/2\pi\approx-0.5$~kHz, and stays below $2.2\times10^{-8}$ at the device value. A Kerr term shifts the energy of $\ket n$ by $Kn(n-1)/2$, which vanishes on the vacuum and on a single photon, so paths with fewer photons accumulate less error. In the quadratic regime, the photon-optimal preparation is 2.2 times less sensitive than the brachistochrone, although it lasts 7\% longer, and the photon-optimal transfer, which passes through the vacuum, is 10.5 times less sensitive than the flip brachistochrone, although it lasts 35\% longer. The trade-off path lies in between, 6.2 times less sensitive. Saving photons therefore also protects against a Kerr term that the control does not know, the error that remains after calibration.

\begin{table}[tbp]
\caption{Infidelity of the optimal paths at $\Delta H=\omega$, generated without compensating a self-Kerr $\tfrac K2\hat a^{\dagger2}\aop^2$ (closed evolution, cutoff 90), with the duration $T$ and the mean photon number $\langle\bar n\rangle$ along each path.}
\label{tab:kerrpaths}
\begin{ruledtabular}
\begin{tabular}{lcccccc}
 & & & \multicolumn{4}{c}{$1-\mathcal F_T$ at $K/2\pi=$}\\\cline{4-7}
path & $T$ (ns) & $\langle\bar n\rangle$ & $-4.8$~Hz & $-0.48$~kHz & $-4.8$~kHz & $-48$~kHz\\\hline
preparation, brachistochrone & 140 & 0.88 & $1.8\times10^{-9}$ & $1.8\times10^{-5}$ & $1.7\times10^{-3}$ & $7.5\times10^{-2}$\\
preparation, photon-optimal & 149 & 0.76 & $8.0\times10^{-10}$ & $8.0\times10^{-6}$ & $7.9\times10^{-4}$ & $5.2\times10^{-2}$\\
logical flip, brachistochrone & 294 & 2.52 & $2.2\times10^{-8}$ & $2.2\times10^{-4}$ & $1.8\times10^{-2}$ & $4.6\times10^{-1}$\\
transfer, trade-off ($w^*$) & 350 & 1.10 & $3.6\times10^{-9}$ & $3.5\times10^{-5}$ & $3.5\times10^{-3}$ & $1.1\times10^{-1}$\\
transfer, photon-optimal & 397 & 0.90 & $2.1\times10^{-9}$ & $2.1\times10^{-5}$ & $2.1\times10^{-3}$ & $9.9\times10^{-2}$\\
\end{tabular}
\end{ruledtabular}
\end{table}

\subsection{\texorpdfstring{Other Envelope Widths}{Other Envelope Widths}}
\label{sec:delta-bench}
Supplementary Table~\ref{tab:delta} collects the path quantities of Sec.~\ref{sec:delta} for $\Delta=0.3$--0.5, and Supplementary Table~\ref{tab:deltabench} repeats the benchmark of Fig.~4 at $\Delta=0.3$ and 0.5 for 412 and 824~ns, from starts 1--5, next to the same starts at $\Delta=0.4$. Section~\ref{sec:delta} predicts that a direct push of CS along the preparation brachistochrone needs at least 478~ns at $\Delta=0.3$, against 159~ns at $\Delta=0.5$. The minimum closeness of CS at 412~ns falls from 0.96 at $\Delta=0.5$ to 0.82 at $\Delta=0.4$ and 0.45 at $\Delta=0.3$, and at 824~ns, beyond that bound, it rises to 0.68 at $\Delta=0.3$. These are the pulses that the selection rule of Sec.~\ref{app:numerics} keeps. At $\Delta=0.3$ no tracking weight reaches $\mathcal F_T=0.99$ under noise, so for every CS start the rule keeps the weakest weight, which gives the highest fidelity, whereas at $\Delta=0.5$ it keeps the strongest. The fall persists at equal weight. On the ladder at $\Delta=0.3$ and 412~ns (closed closeness and noisy fidelity at the optimization cutoff), CS keeps $C=0.73$ at the strongest weight, where its $\mathcal F_T$ drops to 0.90, and it follows more closely than CD from $W=70$ upward. At the kept weights, however, the CD pulses, which reach only $\mathcal F_T=0.58$, stay closer to the path than those of CS ($C=0.53$ against 0.45). CS still reaches the highest $\mathcal F_T$ at every width and duration. At $\Delta=0.3$ and 412~ns it reaches 0.985, against 0.58 for CD and 0.35 for SNAP, four of whose five starts stay at the zero-drive point, and its lead over CD at 412~ns grows from 0.04 at $\Delta=0.5$ to 0.18 at $\Delta=0.4$ and 0.41 at $\Delta=0.3$. At 824~ns, CD reaches 0.909 at $\Delta=0.3$ along a route far from the path, with $C=0.15$, below the 0.33 of a pulse that does nothing. At $\Delta=0.5$, whose target has the fidelity 0.73 with the vacuum, CD reaches 0.991 at 824~ns, at the threshold of the selection rule, against 0.997 for CS. Over these widths, larger GKP states thus widen the lead of CS in fidelity, although following the path directly needs longer pulses.

\begin{table}[tbp]
\caption{GKP preparation from the vacuum along the brachistochrone at three envelope widths, from starts 1--5 of Fig.~4, with the self-Kerr of the device. Target fidelity $\mathcal F_T$, mean (s.d.) [best], and minimum closeness $C$, mean (s.d.), both under the Lindblad equation. The s.d. is in units of the last digit and is omitted when it rounds to zero. A pulse that does nothing has $C=|\langle0|0_L\rangle|^2$, which is 0.33, 0.54, and 0.73 at the three widths. At 412~ns, four, three, and two of the five SNAP starts at $\Delta=0.3$, 0.4, and 0.5 stay at the zero-drive point, with the $\mathcal F_T$ and $C$ of a pulse that does nothing. The runs at $\Delta=0.3$ use the cutoffs 90, 110, and 120 in place of 60, 80, and 90, and the truncation guard at $n\geq60$ and 61 in place of 40 and 44, since that target holds $1.2\times10^{-3}$ of its population at $n\geq40$, where the truncation guard of the cutoff-60 optimization acts.}
\label{tab:deltabench}
\begin{ruledtabular}
\begin{tabular}{lcccc}
$\Delta$ & gate set & $T$ (ns) & $\mathcal F_T$ & $C$\\\hline
0.30 & CS & 412 & 0.985(3) [0.986] & 0.45(1)\\
0.30 & CS & 824 & 0.983(1) [0.984] & 0.68(4)\\
0.30 & CD & 412 & 0.576(17) [0.592] & 0.53(1)\\
0.30 & CD & 824 & 0.909(12) [0.921] & 0.15(2)\\
0.30 & SNAP & 412 & 0.347(33) [0.406] & 0.33(1)\\
0.30 & SNAP & 824 & 0.661(30) [0.706] & 0.12(2)\\
0.40 & CS & 412 & 0.990 [0.990] & 0.82(3)\\
0.40 & CS & 824 & 0.990 [0.991] & 0.90(1)\\
0.40 & CD & 412 & 0.810(7) [0.818] & 0.65(1)\\
0.40 & CD & 824 & 0.976(5) [0.981] & 0.52(4)\\
0.40 & SNAP & 412 & 0.579(48) [0.640] & 0.49(9)\\
0.40 & SNAP & 824 & 0.860(20) [0.884] & 0.21(5)\\
0.50 & CS & 412 & 0.993(1) [0.994] & 0.96\\
0.50 & CS & 824 & 0.997 [0.997] & 0.97\\
0.50 & CD & 412 & 0.949(5) [0.956] & 0.70(2)\\
0.50 & CD & 824 & 0.991 [0.991] & 0.81(2)\\
0.50 & SNAP & 412 & 0.783(46) [0.828] & 0.57(18)\\
0.50 & SNAP & 824 & 0.950(10) [0.961] & 0.32(7)\\
\end{tabular}
\end{ruledtabular}
\end{table}

\begin{table*}[tbp]
\setlength{\tabcolsep}{5pt}
\caption{Dependence on the envelope width $\Delta$ at $\Delta H=\omega$. $\bar n$ of $\ket{0_L}$, overlap $c=\langle0_L|1_L\rangle$ of the code words, defect $1-(1-2c^2)^2$ of the flip brachistochrone, imbalance $1-\nu_-/\nu_+$ of the two Hadamard sectors (Sec.~\ref{sec:gkp-words}), durations of the preparation brachistochrone $T^*$ and of the photon-optimal preparation $T_{\rm ph}$ with the exposure saving of the latter, durations of the flip brachistochrone $T^*_L$ and of the photon-optimal transfer through the vacuum $T_{\rm vac}$ with its saving, tangential capacity $v/\omega$ of conditional squeezing at the vacuum, the speed $\theta/\omega T$ that the preparation brachistochrone needs at $T=412$~ns, and the shortest duration $T^*\omega/v$ of a direct push at the capacity.}
\label{tab:delta}
\begin{ruledtabular}
\begin{tabular}{cccccccccccccc}
$\Delta$ & $\bar n$ & $c$ & defect & $1-\frac{\nu_-}{\nu_+}$ & $T^*$ (ns) & $T_{\rm ph}$ (ns) & saving (\%) & $T^*_L$ (ns) & $T_{\rm vac}$ (ns) & saving (\%) & $v/\omega$ & $\frac{\theta}{\omega\,412~\mathrm{ns}}$ & $\frac{T^*\omega}{v}$ (ns)\\\hline
0.30 & 5.06 & $3.2\times10^{-4}$ & $4.0\times10^{-7}$ & $4.5\times10^{-4}$ & 180 & 197 & 11.4 & 296 & 474 & 51.8 & 0.38 & 0.44 & 478\\
0.35 & 3.52 & $3.2\times10^{-3}$ & $4.0\times10^{-5}$ & $4.5\times10^{-3}$ & 160 & 172 & 9.2 & 296 & 434 & 52.6 & 0.44 & 0.39 & 363\\
0.40 & 2.44 & 0.014 & $8.0\times10^{-4}$ & $2.0\times10^{-2}$ & 140 & 149 & 8.2 & 294 & 397 & 51.4 & 0.51 & 0.34 & 275\\
0.45 & 1.65 & 0.039 & $6.2\times10^{-3}$ & $5.4\times10^{-2}$ & 120 & 127 & 7.6 & 289 & 364 & 47.6 & 0.58 & 0.29 & 208\\
0.50 & 1.08 & 0.081 & $2.6\times10^{-2}$ & $1.1\times10^{-1}$ & 102 & 107 & 7.0 & 281 & 333 & 40.9 & 0.64 & 0.25 & 159\\
\end{tabular}
\end{ruledtabular}
\end{table*}

\section{\texorpdfstring{Conditional Displacements from a Squeezed Vacuum}{Conditional Displacements from a Squeezed Vacuum}}
\label{app:sqcd}

Deterministic protocols build the GKP grid from a squeezed vacuum with conditional displacements and qubit rotations~\cite{S-Hastrup2021,S-Singh2026}. Their gates belong to the gate set CD, so we run the protocol of Ref.~\cite{S-Hastrup2021} on the hardware of Fig.~4 and repeat the benchmark of Fig.~4 for CD from the initial state $\hat S(r)\ket0$, squeezed in $\hat q$ with $e^{-r}=\Delta$, the peak width of the target ($r=0.916$, 7.96~dB). It has $\bar n=1.10$ and $\mathcal F_T=0.785$, against 0 and 0.544 for the vacuum. The squeezed vacuum is given at no cost, although its preparation takes at least $r/g=122$~ns with a squeezer at the rate $g$, and Rabi interactions can prepare it as well~\cite{S-HastrupPRL2021,S-Singh2026}.

\subsection{\texorpdfstring{The Measurement-Free Protocol}{The Measurement-Free Protocol}}
\label{sec:hastrup}
The protocol of Ref.~\cite{S-Hastrup2021} starts from $\hat S(r)\ket0$ with the qubit in $\ket g$ and applies $N$ rounds of three gates, a preparation gate $e^{iu_k\hat q\hat\sigma_y}$ that shapes the envelope, a displacement gate $e^{iv_k\hat p\hat\sigma_x}$ that splits every peak in two, and a disentangling gate $e^{iw_k\hat q\hat\sigma_y}$. The lattice fixes $v_k$ and $w_k$, and the output approximates $\ket{1_L}$ with $2^N$ peaks. Since CD conditions on $\hat\sigma_z$, each gate $e^{is\hat Q\hat\sigma_n}$ runs as a conditional displacement at the bound $|\varepsilon|=g$ for the time $|s|/g$, and quarter turns of the qubit at $|\Omega|=10g$ change the frame between gates of different type. For the published cases of two and three rounds at 11.5 and 16.6~dB, the preparation gates of Ref.~\cite{S-Hastrup2021} give 0.934 and 0.993 with its grid states, against the reported 0.935 and 0.993, and preparation gates that maximize the fidelity of the ideal gate sequence raise these to 0.989 and 0.995. Supplementary Table~\ref{tab:hastrup} gives the results with such gates at $\Delta=0.4$, with the noise and the self-Kerr of Fig.~4. One round prepares the two central peaks of $\ket{1_L}$ in 337~ns with $\mathcal F_T=0.963$, and two rounds reach 0.973 in 857~ns. A third round does not help, because at its optimum its first two displacement gates merge into one and it reduces to two rounds, since the target holds almost no weight beyond the four central peaks. The remaining closed infidelity of 0.018 is mostly the mixedness of about $(\pi/32)e^{-2r}=0.016$ that the last disentangling gate leaves, because its rotation angle varies across each peak. The composite unentangling pulses of Ref.~\cite{S-Singh2026} are designed to remove this error. Our variant with two splits by $\sqrt\pi$ prepares $\ket{0_L}$, with three peaks, in 667~ns at $\mathcal F_T=0.963$ [Supplementary Fig.~\ref{fig:sqcd}(a)]. None of these routes follows the brachistochrone from $\hat S(r)\ket0$ to its target (largest $C=0.12$), since the displaced peaks pass through states far from every state of the path.

\begin{table}[tbp]
\caption{Deterministic preparation by conditional displacements from the squeezed vacuum $\hat S(r)\ket0$, $e^{-r}=\Delta=0.4$, on the hardware of Fig.~4 ($|\varepsilon|\leq g$, $|\Omega|\leq10g$, cutoff 100). Duration $T$, the sum of the gate times at the bounds, target fidelity $\mathcal F_T$ of the closed evolution and under the Lindblad equation with the self-Kerr of the device, and minimum closeness $C$ to the brachistochrone from $\hat S(r)\ket0$ to the target. The preparation gates maximize the fidelity of the ideal gate sequence. The simulated pulses fill whole slices of 1~ns and are 1.7--3.1~ns longer.}
\label{tab:hastrup}
\begin{ruledtabular}
\begin{tabular}{lcccccc}
protocol & state & rounds & $T$ (ns) & $\mathcal F_T$, closed & $\mathcal F_T$, Lindblad & $C$\\\hline
Ref.~\cite{S-Hastrup2021} & $\ket{1_L}$ & 1 & 337 & 0.965 & 0.963 & 0.12\\
Ref.~\cite{S-Hastrup2021} & $\ket{1_L}$ & 2 & 857 & 0.982 & 0.973 & 0.00\\
two splits by $\sqrt\pi$ & $\ket{0_L}$ & 2 & 667 & 0.967 & 0.963 & 0.00\\
\end{tabular}
\end{ruledtabular}
\end{table}

\subsection{\texorpdfstring{Optimized Pulses from the Squeezed Vacuum}{Optimized Pulses from the Squeezed Vacuum}}
\label{sec:sqcd}
\begin{figure*}[tbp]
\includegraphics[width=0.72\textwidth,height=0.42\textheight,keepaspectratio]{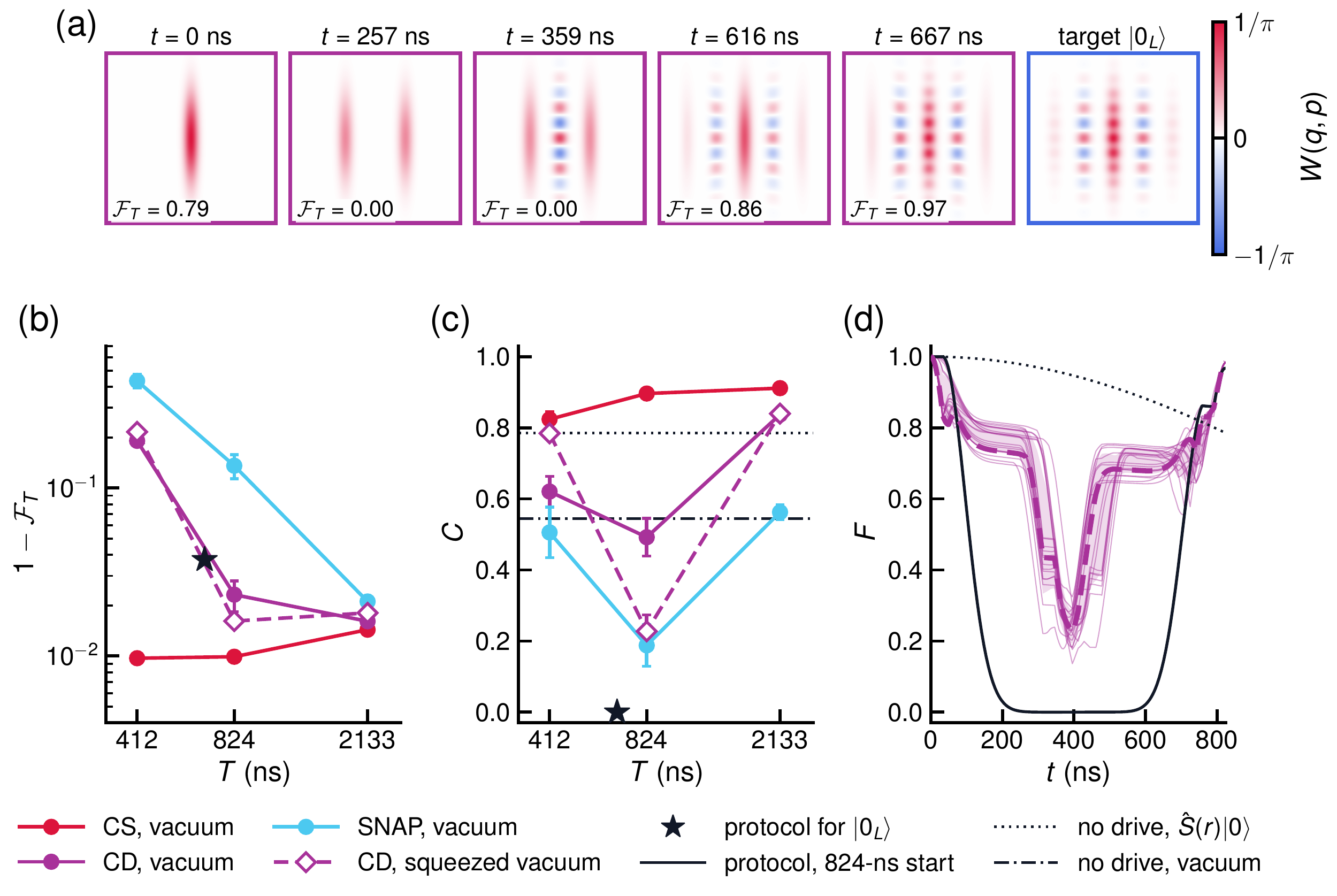}
\caption{Conditional displacements from the squeezed vacuum $\hat S(r)\ket0$, $e^{-r}=\Delta=0.4$, with the self-Kerr of the device. (a)~Wigner functions along the protocol for $\ket{0_L}$ of Supplementary Table~\ref{tab:hastrup} (closed evolution, cutoff 90) at $t=0$ and at the ends of the first split, of the merged disentangling and preparation gate, of the second split, and of the last disentangling gate, next to the target, each with its fidelity $\mathcal F_T$ with $\ket{0_L}$. (b)~Target infidelity $1-\mathcal F_T$ and (c)~minimum closeness $C$ against the duration $T$ (logarithmic axis) for CS, CD, and SNAP from the vacuum (the 20 starts of Fig.~4) and for CD from the squeezed vacuum (20 starts from the protocol), mean $\pm$ s.d. under the Lindblad equation. The star marks the protocol itself. In (c), the dotted and dash-dotted lines give the closeness of a pulse that does nothing from $\hat S(r)\ket0$ and from the vacuum in the closed evolution, and the closeness of the runs from the squeezed vacuum refers to the brachistochrone from $\hat S(r)\ket0$. (d)~Closeness $F(t)$ to that brachistochrone at 824~ns for every start from the squeezed vacuum (thin lines), the start closest to the mean (dashed), and the mean $\pm$ one standard deviation (band), all under the Lindblad equation, with the protocol laid on the 50 slices of the runs (solid) and the pulse that does nothing (dotted), both in the closed evolution.}
\label{fig:sqcd}
\end{figure*}

In the GRAPE benchmark, the reference path is the brachistochrone from $\hat S(r)\ket0$ to $\ket{0_L}$, of length 0.48~rad. Since the squeezed vacuum and the target are parity even, CD still has no tangential capacity along this path (Sec.~\ref{app:capacity}), and a conditional displacement changes neither the closeness nor the target fidelity of a parity-even oscillator to first order, so the zero-drive pulse is a stationary point of the objective. From the random starts of Fig.~4, the 16 runs at 2133~ns that finished stopped at stationary points, with the oscillator back in the squeezed vacuum or in a state with $\mathcal F_T=0.88$, and a run with three target-only warm-up stages stopped there as well. The runs of Supplementary Table~\ref{tab:sqcd} therefore start from the protocol for $\ket{0_L}$ above, laid on the slices of each duration, unchanged for start 1 and slightly perturbed for starts 2--20, and run the weight ladder of Fig.~4 directly. At 412~ns, shorter than the 667~ns that the protocol needs, its displacements reach only about half their strength, and this start has $\mathcal F_T=0.07$. Supplementary Table~\ref{tab:sqcd} and Supplementary Fig.~\ref{fig:sqcd}(b,~c) compare these runs with those from the vacuum. At 412~ns every start leaves the oscillator in the squeezed vacuum, with $\mathcal F_T=0.784$, as a pulse that does nothing. At 824~ns the pulses reach $\mathcal F_T=0.984$, close to the 0.990 of CS from the vacuum, but along a route far from the path, with $C=0.23$, below the 0.78 of a pulse that does nothing. The optimization lifts the closeness of the protocol start, which falls to zero [Supplementary Fig.~\ref{fig:sqcd}(d)], because it displaces only part of the peak, whereas the protocol splits all of it. At 2133~ns they reach 0.982 with $C=0.84$, only slightly above the 0.78 of a pulse that does nothing. In these runs, a squeezed start therefore lets CD neither follow the brachistochrone nor prepare the target faster than CS does from the vacuum.

\begin{table*}[tbp]
\caption{GKP preparation from the vacuum (CS, CD, SNAP, the 20 starts of Fig.~4) and by conditional displacements from the squeezed vacuum $\hat S(r)\ket0$, $e^{-r}=\Delta$ (20 starts, the protocol for $\ket{0_L}$ of Supplementary Table~\ref{tab:hastrup} and 19 small perturbations of it), all with the self-Kerr of the device. Target fidelity $\mathcal F_T$, mean (s.d.) [best], and minimum closeness $C$ to the brachistochrone from the initial state, mean (s.d.), both under the Lindblad equation. The s.d. is in units of the last digit and is omitted when it rounds to zero.}
\label{tab:sqcd}
\begin{ruledtabular}
\begin{tabular}{ccccccccc}
 & \multicolumn{2}{c}{CD, squeezed vacuum} & \multicolumn{2}{c}{CS, vacuum} & \multicolumn{2}{c}{CD, vacuum} & \multicolumn{2}{c}{SNAP, vacuum}\\
$T$ (ns) & $\mathcal F_T$ & $C$ & $\mathcal F_T$ & $C$ & $\mathcal F_T$ & $C$ & $\mathcal F_T$ & $C$\\\hline
412 & 0.784 [0.784] & 0.78 & 0.990 [0.991] & 0.82(2) & 0.809(9) [0.830] & 0.62(4) & 0.566(39) [0.640] & 0.51(7)\\
824 & 0.984(1) [0.985] & 0.23(5) & 0.990 [0.991] & 0.90(1) & 0.977(5) [0.984] & 0.49(5) & 0.864(22) [0.885] & 0.19(6)\\
2133 & 0.982 [0.983] & 0.84 & 0.986(1) [0.987] & 0.91(1) & 0.984(1) [0.986] & 0.84(1) & 0.979(1) [0.980] & 0.56(2)\\
\end{tabular}
\end{ruledtabular}
\end{table*}

